\documentclass[sigconf, nonacm]{acmart}

\usepackage{booktabs} % For formal tables
\usepackage{tikz}
\usepackage{listings}
\usepackage{subfigure}
\usepackage[para,online,flushleft]{threeparttable}
\usepackage{multirow}
\usepackage{multicol}
\usepackage{booktabs}
\usepackage{soul}
\usepackage{amsmath}

\usepackage{mathabx}

\usepackage[ruled, vlined, linesnumbered]{algorithm2e}
\usepackage{enumitem}
\usepackage[utf8]{inputenc}
\usepackage{latexsym}
\usepackage{xcolor,colortbl}
\usepackage{hhline}
\usepackage{cancel}
\usepackage{tikzscale}
\usepackage{soul} 
\usepackage{balance}

\renewcommand{\path}{\textsf{path}}

\newcommand{\Q}{\mathcal{Q}}
\newcommand{\E}{\mathcal{E}}
\newcommand{\C}{\mathcal{C}}

\newcommand{\V}{\mathcal{V}}

\newcommand{\qichen}[1]{{\color{black}#1}}
\newcommand{\update}[1]{{\color{black}#1}}
\newcommand{\xiao}[1]{{\color{red}#1}}

\newcommand{\dom}{\mathrm{dom}}  
\newcommand{\bm}{\mathbf}
\newcommand{\x}{e}
\newcommand{\X}{\mathcal{X}}
\newcommand{\I}{\mathcal{I}}

\newcommand{\y}{\bm{y}}

\newcommand{\T}{\mathcal{T}}

\newcommand{\con}{\E_\textsf{con}}

\newcommand{\key}{\textsf{key}}
\renewcommand{\paragraph}[1]{\medskip\noindent{\bf {#1. }}}
\setlist{leftmargin = 5mm}

\usetikzlibrary{topaths,calc}
\tikzset{fontscale/.style = {font=Large}
    }

\begin{document}
\title{Change Propagation Without Joins}

\author{Qichen Wang}
\affiliation{%
  \institution{Hong Kong Baptist University}
  \country{}
}
\email{qcwang@hkbu.edu.hk}

\author{Xiao Hu}
\affiliation{%
  \institution{University of Waterloo}
  \country{}
}
\email{xiaohu@uwaterloo.ca}

\author{Binyang Dai, Ke Yi}
\affiliation{%
  \institution{HKUST}
  \country{}
}
\email{{bdaiab, yike}@ust.hk}

\pagestyle{plain}

\begin{abstract}
We revisit the classical change propagation framework for query evaluation under updates.  The standard framework takes a query plan and materializes the intermediate views, which incurs high polynomial costs in both space and time, with the join operator being the culprit.  In this paper, we propose a new change propagation framework without joins, thus naturally avoiding this polynomial blowup.  Meanwhile, we show that the new framework still supports constant-delay enumeration of both the deltas and the full query results, the same as in the standard framework.  Furthermore, we provide a quantitative analysis of its update cost, which not only recovers many recent theoretical results on the problem, but also yields an effective approach to optimizing the query plan.  The new framework is also easy to be integrated into an existing streaming database system.  Experimental results show that our system prototype, implemented using Flink DataStream API, significantly outperforms other systems in terms of space, time, and latency.
\end{abstract}

\maketitle

\sloppy

\section{Introduction}
We study the problem of \textit{query evaluation under updates}, a.k.a. \textit{incremental view maintenance}.  Given a query $Q$, a database $D$, and a sequence of updates, where each update is either the insertion or deletion of a tuple, the goal is to maintain the query results $Q(D)$ continuously.  More precisely, there are two modes to return the updated $Q(D)$ to the user (an end user or an upper-level application): \textit{full enumeration} and \textit{delta enumeration}.  The former is pull-based, i.e., the system returns $Q(D)$ passively upon request of the user; while in the latter case, we push the delta $\Delta Q(D,t)$, i.e., the change to $Q(D)$ caused by the insertion/deletion of $t$, to the user after each update $t$.  These two modes are applicable to different scenarios. Full enumeration cannot be done too frequently if $Q(D)$ is large, and it may miss some ephemeral events in between two requests.  Delta enumeration offers real-time responses with low latency, but it requires the user to have the ability to ``consume'' the deltas in a timely fashion.  It can be considered as a stream-in-stream-out operator, where the input is a stream of updates to the base tables, while the output is a stream of updates to the query result (i.e., a stream of deltas).  If the user wishes to always have a complete and accurate $Q(D)$, it has to maintain $Q(D)$ and update it with the deltas as they are received.  If approximation is acceptable, some more %space- and time-
efficient streaming algorithms can be used instead.

\begin{figure}[t]\centering
\minipage{0.4\linewidth}
\centering
\subfigure[Old plan]{
\resizebox{1.1\linewidth}{!}{
        \begin{tikzpicture}
  \node (p1) at (0, 1.5) {$V_4 = V_2 \Join V_3$};
  \node (p2) at (0, 0) {$V_1 = R_2 \Join R_3$};
  \node (p3) at (1.5, 0.75) {$V_3 = \pi_{x_4} R_4$};
  \node (p8) at (1.5, 0) {$R_4$};
  \node (p4) at (-1.5, 0.75) {$V_2 = R_1 \Join V_1$};
  \node (p5) at (-0.5, -0.75) {$R_2$};
  \node (p6) at (0.5, -0.75) {$R_3$};
  \node (p7) at (-1.5, 0) {$R_1$};
  \begin{scope}[every path/.style={<-}]
    \draw (p1) -- (p3);
    \draw (p1) -- (p4);
    \draw (p4) -- (p2);
    \draw (p2) -- (p5);
    \draw (p2) -- (p6);
    \draw (p4) -- (p7);
      \draw (p3) -- (p8);
  \end{scope}
\end{tikzpicture}
    }    
    \label{fig:dp1}
}

\subfigure[Another old plan]{
\centering
    \resizebox{\linewidth}{!}{
        \begin{tikzpicture}
  \node (p1) at (0, 2.5) {$V_2=R_1 \Join R_2 \Join R_3 \Join V_1$};
  \node (p2) at (-1.2, 1.75) {$R_1$};
  \node (p3) at (-0.4, 1.75) {$R_2$};
  \node (p4) at (0.4, 1.75) {$R_3$};
  \node (p5) at (1.6, 1.75) {$V_1 = \pi_{x_4}R_4$};
  \begin{scope}[every path/.style={<-}]
    \draw (p1) -- (p2);
    \draw (p1) -- (p4);
    \draw (p1) -- (p3);
    \draw (p1) -- (p5);
  \end{scope}
\end{tikzpicture}
    }
    \label{fig:dp2}
}
 \endminipage 
\minipage{0.6\linewidth}
\subfigure[Our new plan]{
   \resizebox{\linewidth}{!}{
        \begin{tikzpicture}[font=\Large]
  \node (p1) at (0, 4) {$\cap : V_s([x_3])$};
  \node (p2) at (-2, 3) {$\pi : V_p(R_2)$};
  \node (p3) at (2, 3) {$\pi : V_p(R_3)$};
  \node (p4) at (-2, 2) {$\ltimes : V_s(R_2)$};
  \node (p5) at (2, 2) {$\ltimes : V_s(R_3) = R_3 \ltimes V_p(R_4)$};
  \node (p6) at (-2.5, 0) {$\pi : V_p(R_1)$};
  \node (p7) at (-1.5, 1) {$R_2$};
  \node (p8) at (1.5, 1) {$R_3$};
  \node (p9) at (2.5, 0) {$\pi : V_p(R_4)$};
  \node (p10) at (-2.5, -1) {$V_s(R_1) = R_1$};
  \node (p11) at (2.5, -1) {$V_s(R_4) = R_4$};
  \node (a1) at (-3.7, 0.4) {};
  \node (a2) at (-0.5, 0.4) {};
  \node (a3) at (-3.7, -1.3) {};
  \node (a4) at (-0.5, -1.3) {};
  \node[text=red] (a5) at (-1, -1) {$R_1$};
  \node (b1) at (-3.3, 3.3) {};
  \node (b2) at (-0.2, 3.3) {};
  \node (b3) at (-3.3, 0.7) {};
  \node (b4) at (-0.2, 0.7) {};
  \node[text=red] (b5) at (-0.7, 1) {$R_2$};
  \node (c1) at (4, 3.3) {};
  \node (c2) at (0, 3.3) {};
  \node (c3) at (4, 0.7) {};
  \node (c4) at (0, 0.7) {};
   \node[text=red] (c5) at (0.7, 1) {$R_3$};
  \node (d1) at (3.7, 0.4) {};
  \node (d2) at (0.5, 0.4) {};
  \node (d3) at (3.7, -1.3) {};
  \node (d4) at (0.5, -1.3) {};
  \node[text=red] (d5) at (1, -1) {$R_4$};
  \node (e1) at (-2.2, 4.5) {};
  \node (e2) at (2.2, 4.5) {};
  \node (e3) at (-2.2, 3.5) {};
  \node (e4) at (2.2, 3.5) {};
  \node[text=red] at (1.8, 4) {$[x_3]$};
  \begin{scope}[every path/.style={<-}]
    \draw (p1) -- (p2);
    \draw (p1) -- (p3);
    \draw (p2) -- (p4);
    \draw (p3) -- (p5);
    \draw (p4) -- (p6);
    \draw (p4) -- (p7);
    \draw (p6) -- (p10);
    \draw (p5) -- (p8);
    \draw (p5) -- (p9);
    \draw (p9) -- (p11);
  \end{scope}
  \begin{scope}
    \draw[dashed] (a1) -- (a2);
    \draw[dashed] (a1) -- (a3);
    \draw[dashed] (a2) -- (a4);
    \draw[dashed] (a3) -- (a4);
  \end{scope}
  \begin{scope}
    \draw[dashed] (b1) -- (b2);
    \draw[dashed] (b1) -- (b3);
    \draw[dashed] (b2) -- (b4);
    \draw[dashed] (b3) -- (b4);
  \end{scope}
  \begin{scope}
    \draw[dashed] (c1) -- (c2);
    \draw[dashed] (c1) -- (c3);
    \draw[dashed] (c2) -- (c4);
    \draw[dashed] (c3) -- (c4);
  \end{scope}
  \begin{scope}
    \draw[dashed] (d1) -- (d2);
    \draw[dashed] (d1) -- (d3);
    \draw[dashed] (d2) -- (d4);
    \draw[dashed] (d3) -- (d4);
  \end{scope}
  \begin{scope}
    \draw[dashed] (e1) -- (e2);
    \draw[dashed] (e1) -- (e3);
    \draw[dashed] (e2) -- (e4);
    \draw[dashed] (e3) -- (e4);    
  \end{scope}
\end{tikzpicture}
    }
    \label{fig:np}
}
\endminipage
%\vspace{-1em}
\caption{For $Q= \pi_{x_1,x_2,x_3,x_4}R_1(x_1, x_2) \Join R_2(x_2, x_3) \Join R_3(x_3, x_4) \\ \Join R_4(x_4, x_5)$, \ref{fig:dp1} and \ref{fig:dp2} are two plans under the standard change propagation framework and \ref{fig:np} is our new plan.}
\label{fig:1}
\end{figure}
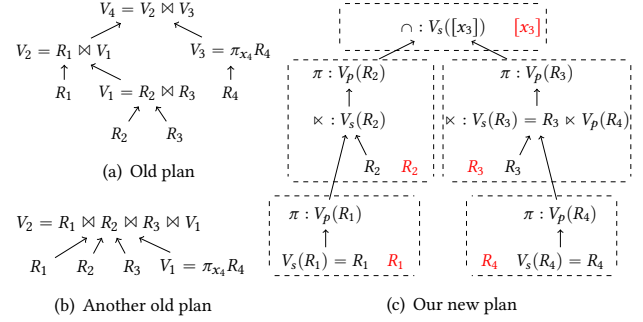

\smallskip \noindent {\bf Change propagation.}
Change propagation \cite{ross1996materialized,chirkova2012materialized,cikm01:lee} is a widely used framework in database systems for solving this problem.  It can be instantiated with any query plan, which is a tree where the leaves are the base relations and each internal node is a relational operator.  At each internal node, it maintains the results of the sub-query corresponding to the subtree at this internal node, which is often called a \textit{materialized view}.  \qichen{Figure~\ref{fig:dp1} shows a particular query plan for the query 4-Hop query from benchmark \cite{nguyen2015join}
\[Q := \pi_{x_1,x_2,x_3,x_4} R_1(x_1, x_2) \Join R_2(x_2, x_3) \Join R_3(x_3, x_4) \Join R_4(x_4, x_5).\] 
Under the standard change propagation framework, we maintain four materialized views $V_1, V_2, V_3, V_4=Q$  (if only delta enumeration is needed, then $V_4$ need not be maintained).  When a tuple $t$ is inserted or deleted in a relation, say $R_1$, it follows the leaf-to-root path to propagate the deltas to the root.  More precisely, it first computes $\Delta V_2 =  \Delta R_1 \Join V_1 = t \Join V_1$, then computes $\Delta Q  = \Delta V_4 =  \Delta V_2 \Join V_3$. Note that with the help of the materialized views, it avoids re-computing some of the sub-queries during updates.

However, the penalty is space: both $V_1$ and $V_2$ can have quadratic size in the worst case \cite{atserias2013size}.  To avoid space blowup, one can use a different query plan, say, the one shown in Figure~\ref{fig:dp2}.  This query plan does not have any materialized views (except $V_1 =\pi_{x_4} R_4$, which has at most linear size), but it has to compute a multi-way join, e.g., $R_1\Join R_2 \Join R_3 \Join t$ upon each update in $R_4$, which could take quadratic time.  Making things worse, this quadratic blowup exacerbates for queries involving more relations \cite{atserias2013size}.}

Prior work has designed advanced techniques to address this space or time blowup.  The Dynamic Yannakakis algorithm \cite{idris17:_dynam, idris2019efficient, idris2020general} has linear space and linear update time while supporting constant-delay enumeration for \textit{free-connex queries}\footnote{All technical terms in the introduction are formally defined in Section \ref{sec:preliminaries}.}; the update time further reduces to $O(1)$ amortized\footnote{All update time bounds are amortized in this paper.} for \textit{q-hierarchical queries}.  Concurrently, \citet{berkholz17:_answer} designed a different algorithm for the q-hierarchical case with the same space/time guarantees.  However, these algorithms have not been integrated into any full-fledged database or data warehouse products, possibly due to the complications of the techniques and the use of non-standard operations not routinely found in existing database systems.

\smallskip \noindent {\bf Change propagation without joins.}
The main contribution of this paper is to achieve (and improve for certain classes of queries and/or update sequences) the results above, but still under the standard change propagation framework.  Our observation is that the only relational operator that may cause a super-linear blowup is join.  Thus, if the query plan has no joins, then both space and update time will be at most linear.  To avoid joins, our high-level strategy is to replace each join in the query plan by a semi-join (or an intersection) plus a projection.  However, not every query plan is amenable to this replacement strategy.  The key technical contribution of this paper, therefore, is the construction of such a query plan for every free-connex conjunctive query.  For example, such a join-free query plan for the earlier query is shown in Figure \ref{fig:np}, which will be elaborated in Section \ref{sec:propagation}.

Since our query plan has no joins, linear space and linear update time follow straightforwardly.  Still, two technical challenges remain: (1) how to support constant-delay enumeration, and (2) how to achieve an update time better than linear.  (1) is trivial under a traditional query plan where the root corresponds to the query results $Q(D)$.  Since our query plan is join-free, no node in the plan corresponds to  $Q(D)$.  Instead, our query plan can be considered as a compact, linear-size representation of a polynomially sized $Q(D)$.  By borrowing ideas from the static case \cite{bagan2007acyclic}, we show how to enumerate $Q(D)$ with constant delay, by appropriately traversing this compact representation.  Supporting constant-delay enumeration of the delta $\Delta Q(D, t)$, on the other hand, is quite different from the static case, and we need new techniques which exploit some important properties of our query plan.

To address issue (2), \citet{wang2020maintaining} introduced the notion of \textit{enclosureness} $\lambda$ of an update sequence, which captures the hardness of the update sequence.  It is linear in the worst case, but is often a constant in many common cases, such as any first-in-first-out (FIFO) update sequence.  They also designed an algorithm with update cost $O(\lambda)$ for \textit{foreign-key acyclic queries}. %(i.e., each join is between a primary key and a foreign key, and the PK-FK references form a DAG with only one root).  
Such queries are relatively easy to handle since their result size is at most linear, so they are immune to the polynomial blowup problem caused by non-key joins, such as free-connex queries.  Indeed, we show (c.f. Theorem \ref{the:lb-non-weak-hierarchical-full}) that there is a simple free-connex query for which it is impossible to achieve $O(|D|^{1/2-\varepsilon})$ update time even over FIFO update sequences, which implies that the previous definition of $\lambda$ is not achievable for free-connex queries.  Nevertheless, we show that, after a simple relaxation of the definition, $\lambda$ is still an appropriate measure of the update complexity; in particular, we show that change propagation under our query plan achieves $O(\lambda)$ update time for every free-connex query under the new definition.  To further illustrate the usefulness of our new definition of $\lambda$, we show that for certain queries (such as q-hierarchical queries) and/or update sequences (such as FIFO or insertion-only), $\lambda$ is indeed a small constant.  For general queries, $\lambda$ also provides guidance on what would constitute a good query plan for change propagation.

\smallskip \noindent {\bf Our results.}
Specifically, this paper achieves the following results:
\begin{enumerate}
    \item We show how to construct a change propagation query plan without joins for any free-connex conjunctive query, such that the space needed by the query plan is linear and the update time is $O(\lambda)$, for an appropriately defined notion of enclosureness $\lambda$ of the update sequence.
    \item We show how to support constant-delay enumeration of both full query results and each delta in our query plan.
    \item We show that $\lambda$ is a constant for certain classes of conjunctive queries (such as q-hierarchical queries) and/or special update sequences (such as FIFO or insertion-only).  These results not only recover the prior known result of \cite{idris17:_dynam,berkholz17:_answer} on q-hierarchical queries, but also extend it to cover many other cases commonly encountered in practice.
    \item We show how our framework can handle various extensions such as selections, aggregations, and non-free-connex queries.
    \item We demonstrate the practicality of our new %change propagation 
    framework by implementing it on top of Flink and comparing it with state-of-the-art view maintenance and SQL-over-stream systems.  
\end{enumerate}

\section{Related Work}
Our new change propagation framework is inspired by several lines of research.  In the static case, the classical Yannakakis algorithm \cite{yannakakis1981algorithms} has runtime $O(|D| + |Q(D)|)$ for every free-connex query.  It consists of two stages.  The first stage uses a series of semi-joins to remove all the dangling tuples in $O(|D|)$ time, and the second stage performs pairwise joins to compute $Q(D)$ in $O(|Q(D)|)$ time.  The Dynamic Yannakakis algorithm \cite{idris17:_dynam} extends the algorithm to the dynamic case, but it deviates from the change propagation framework, making it harder to integrate into existing database systems.  \qichen{Our algorithm can also be viewed as a dynamic version of the Yannakakis algorithm, but it strictly follows the standard change propagation framework while achieving a better runtime.  The Dynamic Yannakakis algorithm has an update cost of $O(|D|)$ for free-connex queries, while our algorithm  achieves $O(\lambda)$ update time, where $\lambda$ is the  \textit{enclosureness} of the update sequence.  We have $\lambda \le |D|$ for all update sequences, while the former is usually much smaller on real-world update sequences.  Furthermore, Dynamic Yannakakis achieves $O(1)$ update time only for q-hierarchical queries, while our algorithm also achieves $O(1)$ update time for non-q-hierarchical queries if the update sequences enjoy some special properties, such as first-in-first-out or insertion-only (formally defined in Section~\ref{sec:enc}). The gap between Dynamic Yannakakis and our algorithm can be as large as $\Theta(|D|)$ on some non-q-hierarchical queries (see Example~\ref{ex:negative}).}%Furthermore, Dynamic Yannakakis can achieve $O(1)$ update time only for q-hierarchical queries, while our algorithm, thanks to the introduction of enclosureness $\lambda$, has $O(1)$ update time also for certain non-q-hierarchical if the update sequences possess special properties such as FIFO or insertion-only. 

\citet{bagan2007acyclic} observe that, in the static case, the second stage of the Yannakakis algorithm can be enhanced to support constant-delay enumeration.  We adapt their ideas to support enumeration in the dynamic case for our query plan.  However, as there is no notion of delta in the static case, we need some new ideas to support delta enumeration with constant delay, which non-trivially relies on some nice features of our query plan. 

\citet{kara2020trade} show that it is possible to increase the enumeration delay in exchange for faster update time, on hierarchical (but non-q-hierarchical) queries. We have not considered this trade-off, as we believe the constant delay is important, and our update cost $\lambda$ is low enough for most queries and update sequences already.  Furthermore, their trade-off only applies to full enumeration, not delta enumeration.  Nevertheless, for cases where $\lambda$ is high, it would be an interesting direction to explore such a trade-off.

In the standard change propagation framework, a single update to a base relation may incur many changes in the intermediate views.  Higher-Order Incremental View Maintenance (HIVM) \cite{ahmad2012dbtoaster} has been proposed to remedy this problem.  It takes the changes to a view as another query (delta query) and maintains this delta query recursively.  HIVM improves upon IVM for many complex queries in practice, and it can also extend to accelerate several machine learning tasks \cite{nikolic2018incremental,nikolic2020f}, but there is no theoretical guarantee on its update time.  Furthermore, HIVM still uses super-linear space.

The problem is also related to stream joins.  In particular, a cash-register stream corresponds to an insertion-only update sequence, while a turnstile stream is an update sequence with arbitrary insertions and deletions.  The sliding-window stream model is a special case of a FIFO update sequence.  Most stream processing systems like Flink \cite{carbone15:_apach_flink} and Trill \cite{chandramouli2014trill} use standard change propagation for multi-way stream joins, which we will compare against in Section \ref{sec:experiment}.  Some specialized systems are designed for two-way stream joins \cite{roy2014low,lin2015scalable,gedik2009celljoin,elseidy2014scalable,kang2003evaluating}, but they do not extend to multi-way joins.

%\begin{figure}
%    \centering
%    \includegraphics[scale=0.3]{img/illustration.pdf}
%    \caption{An illustration of our results.}
%    \label{fig:result}
%\end{figure}

%-------------------------------------------------------------------------------------------

\section{Preliminaries}
\label{sec:preliminaries}

\subsection{Problem Definition}
\noindent {\bf Conjunctive queries.} 
We focus on \emph{conjunctive queries (CQ)} of the following form:
\begin{equation}
  \label{q1}
  Q := \pi_{\y} \left( R_1(e_1) \Join R_2(e_2) \Join \cdots \Join  R_n(e_n) \right),
\end{equation}
where each $R_i$ is a relation with a set of attributes/variables $e_i$, 
$i=1,\dots,n$.  Each tuple $t \in R_i$ assigns a value to each attribute in $\x_i$.  For any $x\in e_i$, $t[x]=\pi_{x} t$ denotes the value of $t$ on  attribute $x$. Similarly, for a subset of attributes $\x \subseteq \x_i$, $t[\x] = \pi_\x t$ denotes the tuple formed by the values of $t$ on the attributes in $\x$.

Let $\V=e_1 \cup \cdots \cup e_n$ be the set of all attributes in the query.  We call $\y \subseteq \V$ the \textit{output attributes}, while $\bar{\y} = \V - \y$ are the {\em non-output attributes}, also known as the \emph{existential variables}. If $\y = \V$, such a query is known as a \qichen{{\em full join query}}; otherwise, it is said to be \qichen{{\em join-project query}}. %\textit{non-full}. 
For simplicity, we assume that each $R_i$ in $Q$ is distinct, i.e., the query does not have self-joins.  Nevertheless, self-joins can be taken care of easily: Suppose a relation $R$ appears twice in the query (with different attribute renamings).  Then we consider them as two identical copies of $R$, and for any update to $R$, we apply the update to both copies of $R$.

%$\sigma$ is the selection operator, $\phi_i$ is the predicate defined over relation $R_i$ and $\sigma_{\phi_i} R_i(e_i)$ selects out tuples from $R_i$ passing the predicate $\phi_i$. 

%As a simplification, when considering updates to a query $Q$ in the form of (\ref{q1}), we ignore all the selection operators, since it just takes $O(1)$ time to decide if $t$ passes the predicate $\phi_i$ and we just discard the update if it does not pass $\phi_i$.

Given a database $D$, we write $Q(D)$ for the query results of $Q$ on $D$. We use $Q(D\ltimes t)$ to denote the query results that depend on a given tuple $t$, and call $Q(D\ltimes t)$ the query results \textit{witnessed} by $t$.  Such a {\em witness query} will be frequently used in this paper. Given a query $Q$ in the form of (\ref{q1}) and a tuple $t \in R_i$, it is clear that 
\[    Q(D \ltimes t) = \pi_{\bm{y}} \left(R_1 \Join \cdots \Join R_{i-1} \Join  \{t\} \Join R_{i+1}\Join \cdots \Join R_n\right).
\]
Note that for a \qichen{full join} CQ, we have $Q(D\ltimes t) = Q(D+t) \ltimes t$; for \qichen{join-project queries}, $t$ itself may not appear in $Q(D\ltimes t)$ due to the projection on $\y$.  When analyzing the costs of algorithms, we adopt the notion of \textit{data complexity}, i.e., the size of the query $Q$ is taken as a constant while $|D|$ is an asymptotic parameter. 

\smallskip
\qichen{
\noindent{\bf Semi-joins.}  The semi-join $R_i(x_i) \ltimes R_j(x_j)$ is defined as
\[
    R_i(x_i) \ltimes R_j(x_j) = \{t|t \in \pi_{x_i} R_i \Join R_j\}.
\]
\smallskip
\noindent {\bf Updates and Deltas.}
An {\em update} to a database $D$ is either the insertion or deletion of a tuple $t$ in some relation $R_i$ of $D$.  In this paper, we adopt set semantics.  We denote $D+t$ as the database after inserting $t$ and $D-t$ as the database after deleting $t$.  In particular, this means that if $R_i$ already contains $t$, then inserting $t$ into $R_i$ will not change $R_i$; if $R_i$ does not contain $t$, deleting $t$ from $R_i$ has no effect, either. We ignore these non-effective updates.  

The {\em delta} of an update to $Q$ is defined as 
\[
    \Delta Q(D, t) = Q(D+t) - Q(D)
\]
in case of the insertion of $t$ and
\[
    \Delta Q(D, t) = Q(D) - Q(D-t)
\]
in the case of deletion.  For a \qichen{full join query},
$
    \Delta Q(D, t) = Q(D \ltimes t).
$
For \qichen{join-project queries}, $\Delta Q(D, t) \subseteq Q(D \ltimes t)$.  In particular, it is possible to have $\Delta Q(D, t) = \emptyset$ even if $Q(D \ltimes t) \ne \emptyset$.

We target \textit{constant delay}~\cite{bagan2007acyclic} for both full and delta enumeration, i.e., the time between the start of the enumeration process to the first tuple in $Q(D)$ (or $\Delta Q(D,t)$), the time between any consecutive pair of tuples, and the time between the last tuple and the termination of the enumeration process should all be bounded by a constant. 
}

\subsection{Classification of CQs}
\noindent {\bf Acyclic CQs.} 
There are several equivalent definitions of acyclic queries~\cite{beeri1983desirability, fagin1983degrees}, and here we adopt one based on generalized join tree~\cite{idris17:_dynam}. A {\em generalized relation} $R_e$ is defined on a subset of attributes $e \subseteq \V$ and is distinguished from the original relations.
\begin{definition}[Acyclic queries]
\label{def:acyclic}
    A CQ $Q = (\V,\E,\y)$ is acyclic if there exists a tree $\T$ in which each node corresponds to a distinct input relation or a generalized relation, while satisfying the following properties:
    \begin{itemize}[leftmargin=*]
    \item \textbf{(cover property)} each input relation in $\E$ corresponds to a distinct node in $\T$; moreover, each leaf node of $\T$ corresponds to an input relation in $\E$;
    \item \textbf{(connect property)} 
    for each attribute $x \in \V$, all nodes of $\T$ containing $x$ form a connected component of $\T$; 
\end{itemize}
\end{definition}

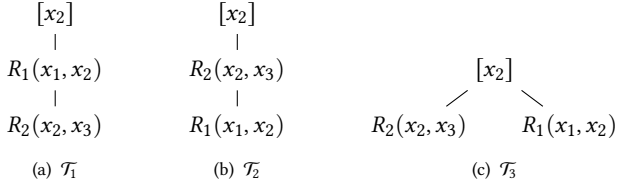
\begin{figure}[t]
\subfigure[$\T_1$]{
\centering
\begin{tikzpicture}
  \node (p1) at ( -2, -4) {$R_2(x_2, x_3)$ }; 
  \node (p2) at ( -2, -3.25) {$R_1(x_1,x_2)$};
  \node (p3) at ( -2, -2.5) {$[x_2]$};
  \begin{scope}[every path/.style={-}]
    \draw (p1) -- (p2);
    \draw (p2) -- (p3);
  \end{scope} 
\end{tikzpicture}
\label{fig:q_T1}
}
\hfill
\subfigure[$\T_2$]{
\centering
\begin{tikzpicture} 
  \node (p1) at ( -2, -4) {$R_1(x_1,x_2)$ }; 
  \node (p2) at ( -2, -3.25) {$R_2(x_2,x_3)$};
  \node (p3) at ( -2, -2.5) {$[x_2]$};
  \begin{scope}[every path/.style={-}]
    \draw (p1) -- (p2);
    \draw (p2) -- (p3);
  \end{scope} 
\end{tikzpicture}\label{fig:q_T2}
}
\hfill
\subfigure[$\T_3$]{
\centering
\begin{tikzpicture} 
  \node (p1) at ( -2, -4) {$R_2(x_2, x_3)$ }; 
  \node (p2) at ( -0, -4) {$R_1(x_1,x_2)$};
  \node (p3) at ( -1, -3.25) {$[x_2]$};
  
  \begin{scope}[every path/.style={-}]
    \draw (p1) -- (p3);
    \draw (p2) -- (p3);
  \end{scope} 
\end{tikzpicture}\label{fig:q_T3}
}
%\vspace{-1em}
\caption{Three free-connex join trees for $Q_1= R_1(x_1, x_2) \Join R_2(x_2, x_3)$. In \ref{fig:q_T3}, node $[x_2]$ is a generalized relation with one attribute $x_2$. The height of $\T_1, \T_2$ is $2$ and that of $\T_3$ is $1$. 
}
\label{fig:trees}
\end{figure}

$\T$ is called a generalized join tree for $\Q$. If all nodes in $\T$ correspond to relations in $\E$, $\T$ is called a traditional join tree. An example is given in Figure \ref{fig:q_T3}. In $\T$, we use $r$ to denote the root, and $\T_e$ for the subtree rooted at node $e$,  $\mathcal{C}_e$ for the set of children of node $e$ and $p_e$ for the parent of node $e$. If $e$ is a leaf, $\mathcal{C}_e = \emptyset$; for the root $r$, $p_r = \emptyset$. 
\qichen{Let $\key(e) = e \cap p_e$ be the \textit{join key} between node $e$ and $p_e$}.

\paragraph{Free-connex CQs} A CQ $Q = (\V,\E,\y)$ is {\em free-connex} if $Q$ and $(\V, \{e_1,\dots, e_n,\y\}, \y)$ are both acyclic~\cite{bagan2007acyclic}. By definition, any free-connex query must be acyclic, and an acyclic \qichen{full join query} must be free-connex. 
For our development, we need an equivalent definition based on {\em free-connex tree} (the equivalence is proved in Appendix~\ref{appendix:preliminaries}):
\begin{definition}[Free-connex CQs]
\label{def:free-connex-tree}
     A CQ $Q = (\V,\E,\y)$ is free-connex if there exists a tree $\T$ in which each node corresponds to a distinct input relation or a generalized relation, while satisfying the following properties:
    \begin{itemize}[leftmargin=*]
    \item \textbf{(cover property)} each input relation in $\E$ corresponds to a distinct node in $\T$; moreover, each leaf node of $\T$ corresponds to an input relation in $\E$;
    \item \textbf{(connect property)} 
    for each attribute $x$, all nodes of $\T$ containing $x$ form a connected component of $\T$; 
    \item \textbf{(guard property)} if node $e$ corresponds to a generalized relation, $e \subseteq e'$ holds for every child node $e'$ of $e$; 
    %\item \update{\textbf{(above property)} if node $e$ corresponds to a generalized relation, $e$ is either the root node, or its parent node $p_e$ is also a generalized relation.}  
    \item \textbf{(above property)} any node corresponding to a generalized relation appears above any node corresponding to an input relation; 
    \item \textbf{(connex property)} there exists a connected subtree $\con$ of $\T$ such that (\romannumeral 1) $\con$ contains the root of $\T$; (\romannumeral 2) for any node $e \in \con$, $\key(e) \subseteq \y$; (\romannumeral 3) $\y \subseteq \bigcup_{e \in \con} e$.
    \end{itemize}
\end{definition}
$\T$ is called a free-connex join tree of $\Q$, and $\con$ is called the connex subtree.  The height of $\T$ is defined as the maximum number of relations on any leaf-to-root path, without counting generalized relations. For example, for $Q'_1:=\pi_{x_2}R_1(x_1, x_2) \Join R_2(x_2, x_3)$, all three free-connex join trees in Figure \ref{fig:trees} are valid free-connex join trees, with the connex substree $\{[x_2]\}$. 

\smallskip 
\noindent {\bf Q-hierarchical CQs~\cite{berkholz17:_answer}.} 
A CQ $Q$ is {\em q-hierarchical} if (1) for every pair of attributes $x_1, x_2$, either $\E_{x_1} \subseteq \E_{x_2}$ or $\E_{x_2} \subseteq \E_{x_1}$ or $\E_{x_1} \cap \E_{x_2} = \emptyset$; and (2)	for every pair of attributes $x_1, x_2$, if $x_1 \in \y$ and $\E_{x_1} \subsetneq \E_{x_2}$, then $x_2 \in \y$, where $\E_x$ denote the set of relations containing attribute $x$. Interesting, a CQ is q-hierarchical if and only if there is a height-1 free-connex join tree (see Appendix~\ref{appendix:preliminaries}).

\medskip Firstly, a full join query can be evaluated in linear time in terms of input and output size if and only if it is acyclic; for join-project CQs, this complete class extends to free-connex queries. Furthermore, free-connex and q-hierarchical CQs have played important roles in query enumeration.
\cite{bagan2007acyclic} showed that in static settings, constant-delay enumeration after a linear-time preprocessing step is possible for a CQ if and only if it is free-connex. \citet{berkholz17:_answer} showed that in dynamic settings, constant-delay enumeration is possible for a CQ from a data structure that can be updated in constant time if and only if it is q-hierarchical.

\section{Change Propagation Without Joins}
\label{sec:propagation}

\subsection{A New Query Plan}
\label{sec:queryplan}
\qichen{Given a free-connex query $Q$, our new query plan is guided by a free-connex join tree $\T$ of $Q$.}  We illustrate the construction using the query in Figure \ref{fig:1} with the join tree highlighted in red (note that the join tree is not unique). A normal query plan following this join tree would compute a series of joins $(R_1 \Join R_2) \Join (R_3 \Join \pi_{x_4} R_4)$.  In our new query plan, we replace each join with a semi-join followed by a projection.  More precisely, we maintain two views for each node $e \in \T$, a semi-join view $V_s(R_e)$ and a projection view $V_p(R_e)$, defined recursively as follows. 

Every non-root node $e \in \T$ has a \textit{projection view} 
\begin{equation}
\label{eq:vp}    
    V_p(R_e) := \pi_{\key(e)} V_s(R_e).
\end{equation}
\qichen{Noted that the root node does not have a projection view.}

To define the \textit{semi-join view} $V_s(R_e)$, we distinguish three cases. 

\begin{enumerate}
    \item[(i)] If $e$ is a leaf, $R_e$ is an input relation, and $V_s(R_e) := R(e)$.
    \item[(ii)] If $e$ is an internal node and $R_e$ is an input relation, then
\begin{equation}
\label{eq:vs1}
    V_s(R_e) := R_e \ltimes V_{p}(R_{e_1}) \ltimes \cdots \ltimes V_p(R_{e_k}),
\end{equation}

where $\C_e =\{e_1,e_2, \dots, e_k\}$ are the children of $e$.
\item[(iii)] If $e$ is an internal node that corresponds to a generalized relation $R_e$, since all the $V_p(R_{e_i})$'s have the same attributes $\key(e_i) = e_i \cap e = e$ for every $i \in [k]$ (by the guard property in Definition \ref{def:acyclic}), \eqref{eq:vs1} simplifies to an intersection:
\begin{equation}
    \label{eq:vs2}
    V_s(R_e) := V_p(R_{e_1}) \cap \cdots \cap V_p(R_{e_k}).
\end{equation}
\end{enumerate}

\smallskip
Our query plan simply connects these views together using the \qichen{formulae} above. \qichen{Algorithm~\ref{alg:plan} takes as input a free-connex join tree, and outputs a new query plan under our framework. }
%More precisely, (\romannumeral 1) the root $r$ has only the semi-join view $V_s(R_r)$; (\romannumeral 2) each leaf node $e$ is mapped to a super-node consisting of $V_p(R_e)$ and $R_e$ (since $R_e = V_s(R_e)$ in this case), where $V_p(R_e)$ is the parent of $e$; (\romannumeral 3) each remaining internal node $e$ is mapped to a super-node consisting of $V_p(R_e), V_s(R_e), R_e$, where $V_p(R_e)$ is the parent of $V_s(R_e)$ and $V_s(R_e)$ is the parent of $R_e$ and (\romannumeral 4) for each node $e$ with its parent $p_e$, add $V_p(R_e)$ as a child of $V_s(R_{p_e})$. An example given in Figure~\ref{fig:np}. 
%
%\begin{example}
Figure~\ref{fig:np} shows the new query plan for the example query. %induced by the join tree in Figure~\ref{fig:gt}.  
Note that $R_2$ and $R_4$ fall into case (ii), while the root node $[x_3]$ is under case (iii).  
\begin{algorithm}[t]
\SetAlgoVlined
\caption{{\sc PlanGeneration}$(Q, T)$}
\label{alg:plan}
\SetKwInOut{Input}{Input}
\SetKwInOut{Output}{Output}

\qichen{
\Input{A free-connex join tree $T$ for query $Q$;}
\Output{A new query plan $T$ for $Q$;}
\ForEach{node $e$ in a postorder traversal of $T$}{
    Replace node $e$ with $V_s(R_e)$ in $T$\;
    \If{$e$ is not the root of $T$}{
        Add $V_p(R_e)$ between $V_s(R_e)$ and the parent of $e$\;
    }
} 
Return $T$\;}
\end{algorithm} 

As neither projection nor semi-join (including intersection as a special case) enlarges the input relations, the following is straightforward:

\begin{lemma}
  \label{lem:linearsize}
 All views in our query plan have size $O(|D|)$.
\end{lemma}

\begin{example}
Figure~\ref{fig:before} shows the initial index built for the query in Figure \ref{fig:1}.  For $R_1$ and $R_4$, both semi-join and projection views are defined as themselves. $V_s(R_2)$ contains tuples in $R_2$ that can join with $V_p(R_1)$, which include $(2, 2)$ and $(2, 4)$.
$V_s(R_3)$ is defined similarly including $4$ tuples from $R_3$.  For the generalized node $[x_3]$, we define the generalized relation $R([x_3]) = V_p(R_2) \cup V_p(R_3)$. Only tuple $(4)$ belongs to $R([x_3])$, since every other tuple in $R([x_3])$ fails to join with $V_p(R_2)$ and $V_p(R_3)$: their counters need to be $2$.  
\end{example}

\subsection{Change propagation}

Change propagation using our new query plan can be done using standard (actually, even simpler for certain operators) propagation \qichen{formulae} \cite{chirkova2012materialized}.  For completeness, we briefly describe them below, which are also needed to understand the algorithms in Section \ref{sec:enumerate}. 

\smallskip
\noindent {\bf \textsc{ S-Update}} When there is an update to $V_s(R_e)$ for some $e$, we use an {\sc S-Update} to update $V_p(R_e)$ by formula \eqref{eq:vp}.  This can be done in $O(1)$ time by {\em derivation counting} \cite{chirkova2012materialized}, a standard technique to propagate changes through a projection.  Specifically, we associate a counter $\textsf{count}[t']$ for each tuple $t'\in V_p(R_e)$ that stores the number of tuples $t \in V_s(R_e)$ such that $t[\key(e)] = t'$.  The detailed process, which needs to distinguish between an insertion and a deletion, is given in Algorithm~\ref{alg:S-update}.  Note that for the algorithm to run in $O(1)$ time, we need a hash index on $V_p(R_e)$.

\smallskip
\noindent {\bf \textsc{P-Update}}
Let $e_i$ be a child of $e$.  When there is an update to some $V_p(R_{e_i})$, we use a {\sc P-Update} to update $V_s(R_e)$ by formula \eqref{eq:vs1} in the case where $e$ is an input relation or \eqref{eq:vs2} in case $e$ is a generalized relation. We consider the former case first; the latter case is similar. 

The standard change propagation formula for a semi-join \cite{algebraic1998griffin} rewrites it as a join followed by a projection, e.g., $ R_e\ltimes R_{e_i} := \pi_e (R_e \Join R_{e_i})$. This defeats the whole purpose of avoiding joins.  However, observe that in our query plan, $R_{e_i}$ has already been projected onto $\key(e_i) = e_i \cap e \subseteq e$ before the semi-join, thus this allows a very simple and efficient way to maintain the whole multi-way semi-join \eqref{eq:vs1} as one operator, which can also be considered as a ``horizontal'' version of derivation counting.  
% \begin{lemma}
% \label{lem:semijoin}
%   For an internal node $e$, $t  \in V_s(R_e)$ holds if and only if $t \in R_e$ and $\pi_{\key(e')} t \in V_p(R_{e'})$ holds for every $e' \in C_e$.
% \end{lemma}
More precisely, we maintain a counter $\textsf{count}[t']$ for every tuple $t'$ in  $R_e$, storing the number of child nodes $e_i\in \C_e$ such that $t'[\key(e_i)] \in V_p(R_{e_i})$. A tuple $t'$ appears in $V_s(R_e)$ if and only if $\textsf{count}[t'] = |\C_e|$.
% With counters, we can simplify Lemma~\ref{lem:semijoin} by the following corollary:
% \begin{corollary}
% \label{pre:semijoin}
%     For an internal node, $t \in V_s(R_e)$ holds if and only if $t \in R_e$ and $\textsf{count}[t] = |C_e|$.
% \end{corollary}
The algorithm is then immediate, as shown in Algorithm \ref{alg:P-update}.  We also need a hash index (that needs to %\footnote{This hash index needs to 
support $e\cap e_i$ as the key for each $e_i\in \C_e$) on $R_e$ so that each counter change can be done in $O(1)$ time.  However, unlike the {\sc S-Update}, a {\sc P-Update} may take more than constant time since multiple tuples may change their counters.  In fact, this is the only place where the update time blows up during change propagation in our query plan.

\begin{algorithm}[t]
\SetAlgoVlined
\caption{{\sc S-Update}$(e,t)$}
\label{alg:S-update}

\SetKwInOut{Input}{Input}
\SetKwInOut{Output}{Output}
{
\Input{An update $t$ from $V_s(R_e)$;}
\Output{Updated $V_p(R_e)$;}
$t' \leftarrow t[\key(e)]$\;
\eIf{$t$ is an insertion into $V_s(R_e)$}{
    \lIf{$t' \in V_p(R_e)$}{
        $\textsf{count}[t'] \leftarrow \textsf{count}[t'] + 1$}
    \qichen{\uElse{
       %Add $t'$ into $V_p(R_e)$\;
       $V_p(R_e) \gets V_p(R_e) \cup \{t'\}$,
       $\textsf{count}[t'] \leftarrow 1$, \ 
       {\sc P-Update}($p_e, t'$)\;}}
}{
    \If{$\textsf{count}[t'] = 1$}{
    	%Delete $t'$ from $V_p(R_e)$\;
    	 $V_p(R_e) \gets V_p(R_e) - \{t'\}$, \ {\sc P-Update}($p_e, t'$)\;
    }
    \lElse{$\textsf{count}[t'] \leftarrow \textsf{count}[t'] - 1$}
}
}
\end{algorithm}

\begin{algorithm}[t]
\SetAlgoVlined
\caption{{\sc P-Update}($e,t$)}
\label{alg:P-update}
{
\SetKwInOut{Input}{Input}
\SetKwInOut{Output}{Output}

\Input{An update $t$ from $V_p(R_{e_i})$ for some $e_i \in \C_e$;}
\Output{Updated $V_s(R_e)$;}
\eIf{$t$ is an insertion into $V_p(R_{e_i})$}{
    \ForEach{$t' \in R_e$ with $t'[\key(e_i)] = t$}{
        $\textsf{count}[t'] \leftarrow \textsf{count}[t']+1$\;
        \If {$\textsf{count}[t'] = |\C_e|$} {
            %Add $t'$ into $V_s(R_e)$\;
    	    $V_s(R_e) \gets V_s(R_e) \cup \{t'\}$, \ {\sc S-Update}$(e, t')$\;
        }
    }
}{
    \ForEach{$t' \in R_e$ with $t'[\key(e_i)] = t$}{
        $\textsf{count}[t'] \leftarrow \textsf{count}[t']-1$\;
        \If {$\textsf{count}[t'] = |C_e|-1$} {
            %Delete $t'$ from $V_s(R_e)$\;
            $V_s(R_e) \gets V_s(R_e) - \{t'\}$, \ {\sc S-Update}$(e, t')$\;
        }
    }
}
}
\end{algorithm} 

\begin{algorithm}[t]
\SetAlgoVlined
\caption{{\sc R-Update}($e,t$)}
\label{alg:R-update}

\SetKwInOut{Input}{Input}
\SetKwInOut{Output}{Output}

{
\Input{An update $t$ from an input relation $R_e$;}
\Output{Updated $V_s(R_e)$;}

\eIf{$t$ is an insertion into $R_e$}{
    $\textsf{count}[t] \leftarrow 0$\;
    \ForEach {$e_i \in \C_e$ } {
        \lIf{$t[\key(e_i)] \in V_p(e_i)$}{$\textsf{count}[t] \leftarrow \textsf{count}[t] + 1$}
    }
    %Assign counter $\textsf{count}[t]$ to $t$\;
    \If{$\textsf{count}[t] = |\C_e|$} {
    	%Add $t$ into $V_s(R_e)$\;
    	$V_s(R_e) \gets V_s(R_e) \cup \{t\}$, \ {\sc S-Update}$(e, t)$\;
    }
}{
    \If{$\textsf{count}[t] = |\C_e|$} {
    	%Delete $t$ from $V_s(R_e)$\;
    	$V_s(R_e)\gets V_s(R_e) - \{t\}$, \ {\sc S-Update}$(e, t)$\;
    }
}
}
\end{algorithm}

\medskip 
\noindent {\bf {\textsc R-Update.}} The last case is when there is an update in an input relation $R_e$, we also need to update $V_s(R_e)$ by formula \eqref{eq:vs1}.  We call this an {\sc R-Update}. The detailed procedure, given in Algorithm \ref{alg:R-update}, simply maintains the counters in $R_e$, and then $V_s(R_e)$, in a straightforward manner. It is obvious that an {\sc R-Update} takes $O(1)$ time (also using the hash index on $V_p(R_{e_i})$).

\begin{example}
Figure~\ref{fig:after} shows the index after inserting $(1, 1)$ into $R_1$.  This new tuple first triggers an insertion to $V_p(R_1)$, which further increments counters of the three tuples in $V_s(R_2)$ with $x_2=1$, which are then brought into $V_s(R_2)$.  From here, the propagation diverges into three paths. Tuple $(1, 2)\in R_2$ increments the counter of $(1) \in V_p(R_2)$ but this propagation path stops here. Tuple $(1,1)\in V_s(R_2)$ first inserts a new tuple $(1)$ to $V_p(R_2)$, which then further increments the counter of tuple $(1)$ in the root, bringing it to $V_s([x_3])$. Tuple $(1, 4)\in R_2$ increments the counter of $(4) \in V_p(R_2)$ and the propagation stops.  

Figure~\ref{fig:deletion} shows the index after deleting  $(1,1)$ from $R_4$.  This deletion first decrements the counter of tuple $(1)\in V_p(R_4)$, removing it from $V_p(R_4)$, and further decrements the counter of $(1,1)\in V_s(R_3)$, removing it from $V_s(R_3)$ as well.  Finally, the counter of $(1) \in V_p(R_3)$ decreases from $2$ to $1$, and the propagation stops here.  
\end{example}

\begin{lemma}
  \label{lem:lineartime}
 All projection and semi-join views in our query plan can be updated in $O(|D|)$ time.
\end{lemma}

\section{Enumeration}
\label{sec:enumerate}

\subsection{Full Result Enumeration}
\label{sec:full-enumeration}

We first consider how to perform constant-delay enumeration of $Q(D)$ from our query plan. We need the following lemma:

\begin{lemma}
\label{lem:semi-join-view}
    For any node $e$, $V_s(R_e) = \pi_{e} (\Join_{e' \in \T_e} R_{e'})$.
\end{lemma}

\begin{proof}[Proof of Lemma~\ref{lem:semi-join-view}]
  We prove it by the induction on the height of free-connex join tree $\T$.  First, it holds for any leaf node $e$, since $V_s(R_e) = R_e$. %it is clear that for the leaf node $e$, the lemma holds as  %    \[V_s(R_e) := R_e.\]
 We next consider an arbitrary internal node $e$. Let $C_e = \{e_1,e_2,\cdots,e_k\}$ be the set of children of node $e$. By hypothesis, we assume this lemma holds for every $e_i \in C_e$, i.e.
 \[V_s(R_{e_i}) = \pi_{e_i}(\Join_{e' \in \T_{e_i}} R_{e'})\]
 Then, we can rewrite $V_s(R_e)$ as follows:
    \begin{align*}
        \Leftrightarrow \ & R_e \ltimes V_p(R_{e_1}) \ltimes \cdots \ltimes V_p(R_{e_k}) \\
        \Leftrightarrow \ & R_e \Join V_p(R_{e_1}) \Join \cdots \Join V_p(R_{e_k}) \\
        \Leftrightarrow \ & R_e \Join \left(\pi_{\key(e_1)} V_s(R_{e_1})\right) \Join \cdots \Join \left(\pi_{\key(e_k)} V_s(R_{e_k})\right) \\
        \Leftrightarrow \ & R_e \Join \left(\pi_{\key(e_1)}(\Join_{e' \in \T_{e_1}} R_{e'})\right) \Join \cdots \Join \left(\pi_{\key(e_k)} (\Join_{e' \in \T_{e_n}} R_{e'})\right) \\
        \Leftrightarrow \ & \pi_{e}(R\Join_{e' \in \T_{e}} R_{e'}))
    \end{align*}
    where the first equation follows the definition of semi-join views, the second equation follows the fact that $\key(e_i) = e_i \cap e \subseteq e$, the third equation follows the definition of projection views, the fourth equation follows the hypothesis, and the last equation follows the facts that $\T_e = \{e\} \cup \T_{e_1} \cup \T_{e_2} \cup \cdots \cup \T_{e_k}$ and $\key(e_i)$ is exactly the set of join attributes shared by $R_e$ and any relation in $\T_{e_i}$. 
\end{proof}

In plain language, the semi-join view of node $e$ is essentially the projection of the join results of relations in the subtree rooted at $e$, to attributes in $e$. An immediate corollary is 
\begin{corollary}
\label{cor:full}
$V_s(R_r) = \pi_{r} Q(D)$.
\end{corollary}
This means that the semi-join view at the root $r$ (recall that $r$ does not have a projection view) contains precisely all the query results projected onto $r$.  Using the notion of a witness query, this leads to the following useful fact for full enumeration, where $\biguplus$ denotes disjoint union:

\begin{lemma}
    \label{lem:full-enumerate-disjoint}
      $Q(D) = \biguplus_{t \in V_s(R_r)} Q(D \ltimes t)$.
\end{lemma}

\begin{algorithm}[t]
\SetAlgoVlined
\caption{{\sc FullEnum}$(\T,e,t)$}
\label{alg:full-enumerate}
{
\KwIn{A free-connex join tree $\T$ \update{with connex subtree $\con$}, a node $e \in \T$ and a key $t \in \pi_{\key(e)} R_e$;}
\KwOut{Query results over $\T$ that can be joined with $t$;}

\eIf{$e \notin \con$}{
    {\sc Yield} $\langle \rangle$
}{
    Let $\C_e = \{e_1, e_2, \cdots, e_k\}$\;
    \ForEach {$t_0 \in \pi_{\y \cap e} V_s(R_e)$ such that $\pi_{\key(e)} t_0 = t$} {
        \ForEach{$t_1 \in $ {\sc FullEnum}$(\T_{e_1}, e_1, t_0[\key(e_1)])$} {
            \ForEach{$t_2 \in $ {\sc FullEnum}$(\T_{e_2}, {e_2}, t_0[\key(e_2)]$} {
                $\cdots$\\
                \ForEach{$t_k \in $ {\sc FullEnum}$(\T_{e_k}, e_k, t_0[\key(e_k)])$} {
                    {\sc Yield} $t_0 \Join t_1 \Join t_2 \Join \cdots \Join t_k$\;
                }
            }
        }
    }
}
}
\end{algorithm}

\begin{algorithm}[t]
\caption{{\sc DeltaEnum}($\T, t$)}
\label{alg:delta-enumerate}

{
\KwIn{A free-connex join tree $\T$; an updated tuple $t$.}
\KwOut{Delta results induced by $t$.}

Let $e_0, e_1, \cdots, e_k=r$ be the nodes on $t$'s propagation path\;
\ForEach{witness tuple $t'$ of $t$}{
    Let $e_i$ be the node such that $t' \in \pi_{\y \cap e_i}\Delta V_s(R_{e_i},t)$\;
    $S\gets t' \Join V_l(R_{e_{i+1}}) \Join \cdots \Join V_l(R_{e_k})$\;
    \ForEach{$q \in S$}{
        $S_i \gets${\sc FullEnum}$(\T_{e_i}, e_i, q[\key(e_i)])$\;
        $S_j \gets${\sc FullEnum}$(\T_{e_j} - \T_{e_{j-1}}, e_j, q[\key(e_j)]), j\in [i+1,k]$\;
         {\sc Yield} $S_i \Join S_{i+1} \Join \cdots \Join S_k$\;
    } 
}
}
\end{algorithm}

Lemma \ref{lem:semi-join-view}, Corollary \ref{cor:full}, and Lemma \ref{lem:full-enumerate-disjoint} allow us to use essentially the same algorithm from \citet{bagan2007acyclic} to achieve constant-delay enumeration of $Q(D)$; see  Algorithm~\ref{alg:full-enumerate}, which takes as input a node $e \in \T$ and a key $t \in \pi_{\key(e)} R_e$, and yields the query results over $\T_e$ that can be joined with $t$.  To enumerate $Q(D)$, we simply invoke {\sc FullEnum}$(\T, r, \cdot)$. 
\begin{lemma}
\label{lem:full_runningtime}
    Algorithm~\ref{alg:full-enumerate} enumerates $Q(D)$ with $O(1)$ delay.
\end{lemma}

\begin{proof}
        We prove it by induction on the height of $\T$. Algorithm~\ref{alg:full-enumerate} stops if $r \notin \con$, which only happens when $\y=\emptyset$. For ease of expression, assume any node $e$ with \update{$e \notin \con$} is removed from $\T$.  We first establish a based case, in which $\T$ contains only one node. The algorithm returns $\pi_{\y \cap e} V_s(R_e)$ in $O(|\pi_{\y \cap e} V_s(R_e)|)$ time, since
        all tuples in $\pi_{\y \cap e} V_s(R_e)$ can be enumerated in $O(1)$ delay. 
        Hence, this base case can be handled with $O(1)$ delay.
        
        In general, we have the hypothesis holds on all child nodes $e_i$ of $e$ (line 6-9): Algorithm~\ref{alg:full-enumerate} can enumerate all join results that agree with values $\pi_{\key(e_i)} t$ over attributes $\key(e_i)$ in the subtree $\T_{e_i}$ \update{if $\key(e_i) \subseteq \y$}, \update{as long as $e_i$ belongs to the connex subtree, making $\key(e_i) \subseteq e \cap \y$}. Let $t_i$ be a join result returned from $\T_{e_i}$. From the properties of the \update{connex subtree}, line 10 will return a valid join result. Emitting every combination of join results over all subtrees of $e$ just takes $O(1)$ time.  
\end{proof}

\subsection{Delta Enumeration}

Delta enumeration is straightforward in a standard query plan, as the root node corresponds to $Q(D)$, so all changes propagated to the root are precisely $\Delta Q(D, t)$.  However, it becomes tricky in our new query plan, as no node corresponds to $Q(D)$, which is necessarily the case if a linear-size representation of $Q(D)$ is desired.  In our query plan, one cannot just inspect the root, because not every change propagates to the root, and many propagations stop mid-way, which is actually the main reason why our query plan is not only space-efficient but also time-efficient. Recall that the full enumeration algorithm relies on Lemma~\ref{lem:full-enumerate-disjoint}.  Then the key question is, can we have an analogy of Lemma~\ref{lem:full-enumerate-disjoint} for the delta $\Delta Q(D, t)$?  In other words, can we identify a set of witness tuples $t'$ for $t$ such that the delta $\Delta Q(D,t)$ is the disjoint union of $Q(D \ltimes t')$?  Fortunately, the answer is yes, but the answer is not as simple as Lemma~\ref{lem:full-enumerate-disjoint}.

Let's first consider the insertion case.  When we insert $t$ into some $R_e$, the propagation follows the path from $e$ to $r$, by (possibly) applying an {\sc R-Update} first, then an {\sc S-Update}, {\sc P-Update}, {\sc S-Update}, {\sc P-Update}, \dots. Recall that both S-update and R-update only propagate a single change upward (see line 8, 12 in Algorithm~\ref{alg:S-update} and Algorithm~\ref{alg:R-update}), but P-update may propagate multiple changes upward (see line 6, 12 in Algorithm~\ref{alg:S-update}. Hence, there could be multiple propagation paths starting from $t$.  To be more precise, we denote the nodes lying on the path from $e$ to $r$ as $e_0 = e, e_1, e_2, \cdots, e_k = r$.  Every propagation path inserts a tuple into each of the views on the path, and we denote the inserted tuples on such a path as $(t, t^s_{0}, t^p_{0}, t^s_{1}, t^p_{1}, \cdots, )$, where $t^s_{i} \in V_s(R_i)$ and $t^p_{i} \in V_p(R_i)$ for $i \in \{0,1,2,\cdots,k\}$.  

Now, we distinguish three cases of a propagation path with respect to its ending tuple: (1) $t$; (2) $t^p_{j}$ for some $j \in \{0,1,2,\cdots,k\}$; (3) $t^s_{i}$ for some $i \in \{0,1,2,\cdots,k\}$.

Case (1) happens when the first update is an {\sc R-Update} and does not propagate any further change. This means that in  Algorithm~\ref{alg:R-update}, there exists some child node $e'$ of $e$ such that $t[\key(e')] \notin V_p(R_{e'})$, i.e., $t$ cannot join with $\T_{e'}$. In this case, $t$ will not produce any change to $Q(D)$, thus can be ignored.

Case (2) happens when {\sc P-Update}($e_j, t^p_j$) does not propagate any further change. Putting it into Algorithm~\ref{alg:P-update}, this means that either there exists no tuple $t'\in R_{p_{e_j}}$ that can join with $t^p_j$, or if such a tuple exists, but it cannot join with \update{(or intersect with, if $R_{p_{e_j}}$ is a generalized relation)} any query result over $\T_e'$ for some child node $e'$ of $p_{e_j}$, since its counter is smaller than $|\C_{p_{e_j}}|$. In either case, this propagation path will not cause any change to $Q(D)$, thus can also be ignored.

Case (3) happens when {\sc S-update}($e_i, t^s_i$) does not propagate any further change. Putting it into Algorithm~\ref{alg:S-update}, this means that either we have reached the root, or there exists some other tuple $t' \in V_p(R_{i})$ such that $t' \neq t^s_i$ and $t^s_i[\key(e_i)] = t'[\key(e_i)]$.  \update{Note that $R_{i}$ can be either an input relation or a generalized relation.} This is the only case where changes to $Q(D)$ can \textit{possibly} happen. We will give a more detailed characterization of this case later. 

\paragraph{Live views} To support constant-delay delta enumeration, we maintain a {\em live view} for each node $e \in \con$:
\[
    V_l(R_e) := \pi_{\y \cap e} Q(D),
\]
which are the ``live'' tuples (i.e., appearing in the query results) projected onto $e$.  Note that $V_l(R_e) \subseteq \pi_{\y \cap e}V_s(R_e)$, which means for $e \subseteq \y$, it can be implemented by simply adding an extra bit in $V_s(R_e)$, indicating if the corresponding tuple is in $V_l(R_e)$.  

For the root $r$, there is no need to maintain $V_l(R_r)$ separately since $V_l(R_r) = V_s(R_r)$ by Corollary \ref{cor:full}. For the leaf nodes, their live views need not be maintained, either, since they will not be needed by delta enumeration.  The other live views can be maintained by the following observation:

\begin{lemma}
For any non-root node $e$ such that $e \in \con$ and any tuple $t \in \pi_{\y \cap e} V_s(R_e)$, $t\in V_l(R_e)$ if and only if $t \Join V_l(R_{p_e})\neq \emptyset$.  
\label{lem:live_con}
\end{lemma}

Based on the Lemma~\ref{lem:live_con}, the maintenance of $V_l(R_e)$ can piggyback on the delta enumeration: After enumerating a result $t' \in \Delta Q(D, t)$, we update the live views in a top-down fashion.  For every non-root $e$ such that $e \cap \y \neq \emptyset$, if the update is insertion, then we always add $t'[e]$ to $V_l(R_e)$; if the update is deletion, then we delete $t'[e]$ from $V_l(R_e)$ if $t'[e]$ cannot join with $V_l(R_{p_e})$, which can be done in $O(1)$ time with a hash index on $V_l(R_{p_e})$ (which is physically the same hash index on $V_s(R_{p_e})$ for $e\subseteq \y$).  This only adds another constant to the delay of delta enumeration.  
\begin{figure*}
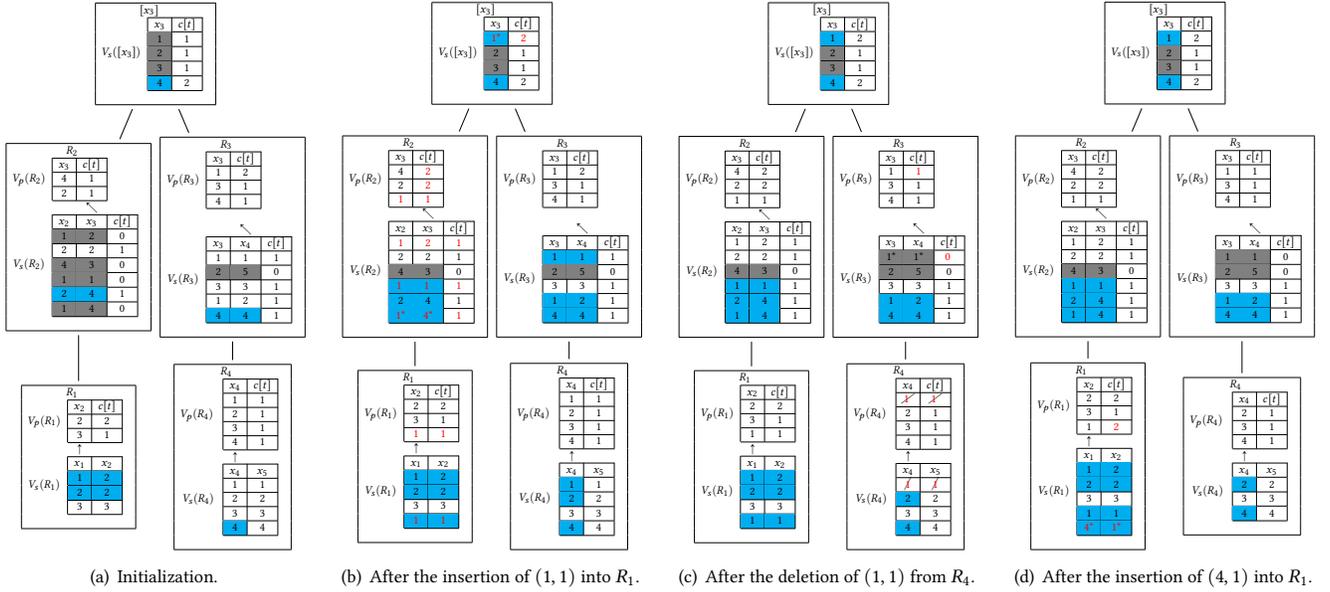

    \centering
    \subfigure[Initialization. 
    ]{
        \resizebox{0.237\linewidth}{!}{
            \input{img/example_static}
        }
        \label{fig:before}
    } \hfill
    \subfigure[After the insertion of $(1, 1)$ into $R_1$. 
    ]{
        \resizebox{0.237\linewidth}{!} {
            \input{img/example1.tex}
        }
        \label{fig:after}
    } \hfill
    \subfigure[
    After the deletion of $(1,1)$ from $R_4$. 
    ]{
        \resizebox{0.237\linewidth}{!} {
            \input{img/example_deletion.tex}
        }
        \label{fig:deletion}
    } \hfill
    \subfigure[After the insertion of $(4, 1)$ into $R_1$. 
    ]{
        \resizebox{0.237\linewidth}{!} {
            \input{img/example_multiple.tex}
        }
        \label{fig:multiple}
    }

    \caption{A running instance for query in Figure \ref{fig:1} using the plan in Figure \ref{fig:np}. Tuples in white are in $V_s(R)$, in grey are in $R\setminus V_s(R)$, in cyan are in $V_l(R)$ (live views for leaf nodes are not needed, but we still show them for clarity), with star symbols are the witness tuples.  Changes in each step are marked in red. }
    \label{fig:running_instance}
\end{figure*}

\paragraph{Witness tuples}
We now are ready to give a more precise characterization of the ending tuples falling into Case (3) that actually cause changes to $Q(D)$, called {\em witness tuples}:

\begin{definition}[Witness tuple]
\label{def:witness}
Suppose $t$ is inserted into or deleted from $D$. A tuple $t'$ is a witness of $t$ if 
\begin{equation}
\label{eq:witness1}
    t' \in  \Delta V_s(R_r,t), \textrm{ or }
\end{equation} 
\begin{equation}
\label{eq:witness2}
    t' \in \pi_{\y \cap e} \Delta V_s(R_e,t) \ltimes V_l(R_{p_e})
\end{equation} 
for some non-root $e$ such that $e \in \con$.
\end{definition}

Here $\Delta V_s(R_e,t)$ denotes the tuples to be inserted into (or deleted from) $V_s(R_e)$ due to $t$ and $V_l(R_{p_e})$ is the live view before the update.  We give some intuition behind Definition~\ref{def:witness}. First, \eqref{eq:witness1} is the counterpart of Corollary \ref{cor:full} for delta enumeration and such a $t'$ is guaranteed to generate changes to $Q(D)$.  \eqref{eq:witness2} is specific for delta enumeration, addressing the situation mentioned earlier, where the propagation stops mid-way yet still causes changes to $Q(D)$. Note that in this case, the attributes of $t'$ are $e \cap \y$.  Then \eqref{eq:witness2} implies that $t' \in \pi_{\y \cap e}\Delta V_s(R_e,t)$ and $t'[\key(e)] \in \pi_{\key(e)} V_l(R_{p_e})$. %Since $V_l(R_e) \subseteq V_p(R_e)$, 
Since $t'[\key(e)] \in \pi_{\key(e)} V_l(R_{p_e})$, it must have $t'[\key(e)] \in V_p(R_e)$, i.e. $t'[\key(e)] \notin \Delta V_p(R_e, t)$, which means that the propagation stops at node $e$ under case (3).  In addition, each witness tuple $t'$ should (\romannumeral 1) contribute to the delta over $\T_e$ induced by $t$, and (\romannumeral 2) join with tuples from the remaining relations in $\T - \T_e$. 
For (\romannumeral 1), it suffices to require $t' \in \Delta \left(\pi_{\y \cap e} V_s(R_e)\right)= \pi_{\y \cap e}\Delta V_s(R_e,t)$, since $\Delta\left(\pi_{e \cap \y} (\Join_{e' \in \T_e} R_{e'})\right) = \Delta \left(\pi_{\y \cap e} V_s(R_e)\right)$.
For (\romannumeral 2), it suffices to require $t' \ltimes V_l(R_{p_e}) \neq \emptyset$, and this is exactly the reason we introduced $V_l(R_e)$ in the first place.

\begin{lemma}
	\label{lem:witness}
	$\Delta Q(D, t)= \biguplus_{t':  \textit{a witness of $t$}} Q(D \ltimes t')$. 
\end{lemma}

We are now ready to state the counterpart of Lemma \ref{lem:full-enumerate-disjoint} for delta enumeration, in Lemma~\ref{lem:witness}. Unlike Lemma \ref{lem:full-enumerate-disjoint}, the proof of Lemma \ref{lem:witness} is nontrivial, and the details are given in Appendix~\ref{appendix:algorithm}.

\smallskip 
\noindent {\bf The algorithm.}
To perform delta enumeration using Lemma \ref{lem:witness}, we still need to address two issues: (1) how to find all witness tuples $t'$, and (2) how to enumerate $Q(D\ltimes t')$ with constant delay.  

To find all the witness tuples, we consider the two cases in Definition \ref{def:witness}: \eqref{eq:witness1} can be computed easily after updating $V_s(R_r)$; for \eqref{eq:witness2}, just an extra check with $V_l(R_{p_e})$ is needed, which can be done in $O(1)$ time using the hash index on $V_l(R_{p_e})$.  These steps only increase the update cost by a constant factor. 

It remains to describe how to enumerate $Q(D\ltimes t')$ for each witness $t'$.  As before, let $e_0,e_1,\dots, e_k=r$ be the nodes on the propagation path, and suppose we are given a witness tuple $t' \in \pi_{\y \cap e_i}\Delta V_s(R_{e_i},t)$ for some $i$. We first enumerate the query results participated by $t'$ together with relations on the path from $e_{i+1}$ to the root $r$, denoted as $S$. This can be done by joining $t$ with the live views associated with these nodes. For each such result $q \in S$, we enumerate the query results that participated by $q$. This enumeration is done by partitioning the whole free-connex join tree into disjoint subtrees $\T_{e_i}, \T_{e_{i+1}} - \T_{e_i}, \cdots, \T_{e_{k}} - \T_{e_{k-1}}$, and invoking {\sc FullEnum} for each subtree separately. Finally, we join these subtrees together. The detailed process is given in Algorithm~\ref{alg:delta-enumerate}.  Note that, as written, the algorithm does not achieve constant-delay enumeration.  However, this can be easily fixed. First, the join in line 4 can be enumerated with constant delay using (a variant of) {\sc FullEnum} starting from $t'$.  Then we interleave the two enumeration processes: After enumerating each $q\in S$, we immediately call line 6--8. Finally, line 6--8 can be rewritten into nested loops so as to enumerate the join $S_i \Join\cdots \Join S_k$ with constant delay.  In fact, this join is more like a cross product (common attributes must have the same value, the same as those in $q$), and a total of $\Pi_{j=i}^k|S_j|$ results will be yielded.

\begin{example}
    In figure~\ref{fig:before}, there are two query results $(1, 2, 4, 4)$ and $(2, 2, 4, 4)$. In figure~\ref{fig:after}, when the propagation stops, we have
    \begin{itemize}[leftmargin=*]
    \item Tuple $(1, 2)\in R_2$ is not a witness as it cannot join with any tuple in $V_l([x_3])$, thus no delta is produced;
    \item Tuple $(1) \in [x_3]$ is a witness, which triggers delta enumeration. For a witness in the root, {\sc DeltaEnum} simply degenerates to {\sc FullEnum}$(\T_r, r, (1))$, which outputs $\{(1, 1, 1, 1), (1, 1, 1, 2)\}$. 
    \item Tuple $(1,4)\in R_2$ is a witness, which triggers delta enumeration.  {\sc DeltaEnum} finds $S = (1,4) \Join V_l([x_3]) = \{(1,4)\}$.  For $(1,4) \in S$, it invokes {\sc FullEnum}$(\T_r-\T_{R_2}, r, (4))$ with $\{(4, 4)\}$ returned and {\sc FullEnum}$(\T_{R_2}, R_2, (1, 4))$ with $\{(1, 1, 4)\}$ returned. Joining them yields the delta $\{(1, 1, 4, 4)\}$. 
    \end{itemize}
    Finally, as each new result is enumerated, we update the live views.
    
    In figure~\ref{fig:deletion}, tuple $(1,1)\in \Delta V_s(R_3)$ is a witness. {\sc DeltaEnum} first finds $S = \{(1,1)\}$. For $(1,1) \in S$, it  invokes {\sc FullEnum}$(\T_r-\T_{R_3}, r, (1))$ with $\{(1, 1, 1)\}$ returned, and {\sc FullEnum}$(\T_{R_3}, R_3, (1, 1))$ with $\{(1, 1, 4)\}$ returned (delta enumeration upon a deletion is done before the tuple deletion so as to find the delta).  Joining them yields the delta $\{(1, 1, 1, 1)\}$.  Finally, we update live views with the delta.
\end{example}

\begin{lemma}
\label{lem:delta_runningtime}
    Algorithm \ref{alg:delta-enumerate} enumerates $\Delta Q(D, t)$ with constant delay.
\end{lemma}

We have now closed the loop: while enumerating $\Delta Q(D, t)$, we update the live views as described earlier, which are needed for enumerating the next delta.

\section{Update Cost Analysis}
\label{sec:Analysis}

We have shown that the enumeration delay of both full query results and deltas is a constant, and this holds for the query plan defined by any free-connex join tree as in Section \ref{sec:queryplan}.  On the other hand, the update cost differs for different query plans and can be as large as $O(|D|)$ in the worst case.  This is caused by {\sc P-Update}, which may trigger an {\sc S-Update} to every tuple in its parent node.  However, such a worst-case behavior only happens on contrived update sequences, and the actual update cost can be much better.  Characterizing the update cost will be important for constructing a good query plan, as there can be many free-connex join trees for a given free-connex query.  As we will see, the height of the join tree is an important parameter, and this is precisely the reason why we make our framework applicable to any free-connex join tree, as the height of a free-connex join tree can be lower than that of any standard join tree.  For example, the query in Figure \ref{fig:trees} has a free-connex join tree of height 1 while the two standard join trees have height 2; the query in Figure \ref{fig:1} has a free-connex join tree of height 2 while any standard join tree has height as least 3.

\subsection{Enclosureness}
\label{sec:enc}
\qichen{
\noindent {\bf Update sequences and lifespans.}
Given an update sequence $S_D$, the {\em lifespan} of tuple $t$ is an interval $I(t) = [t^+, t^-]$, where $t^+$ denotes the timestamp when $t$ is inserted into $D$ and $t^-$ denotes the timestamp when $t$ is deleted from $D$.  We set $t^+ = -\infty$ to indicate that $t$ exists in the initial $D$ and $t^- = +\infty$ indicates that $t$ still exists in $D$ after the update sequence.  Note that if a tuple is repeatedly inserted and deleted, it will be treated as multiple tuples, which have the same values but disjoint lifespans. 

Although our algorithms will be able to handle arbitrary update sequences, their performance can be better if the update sequences possess some nice properties.  In particular, the following two restrictive classes of update sequences are of practical importance:
\begin{itemize}[leftmargin=*]
    \item \textbf{First-in-first-out (FIFO)}. A update sequence $S_D$ is FIFO if for any two tuples $t_1, t_2 \in S_D$, $t^+_1 < t^+_2$ implies $t^-_1 < t^-_2$.  FIFO sequences are commonly used in practice, such as sliding-window or tumbling-window models over streaming data.
    \item \textbf{{Insertion-only} or { deletion-only}.} A update sequence $S_D$ is insertion-only (w.r.t. deletion-only) if for any tuple $t \in S_D$, $t^- = +\infty$ (w.r.t. $t^+=-\infty$). The two cases are symmetric, so we will only discuss the insertion-only case in this paper. 
\end{itemize}
}

The notion of \textit{enclosureness} was first introduced in \cite{wang2020maintaining} to give an instance-specific characterization of the hardness of the update sequence, which we briefly review next. %Recall that each tuple $t$ in the update sequence has a lifespan $I(t) = [t^+, t^-]$.

\begin{definition}[Enclosureness]
    Given an update sequence $S_D$, the {\it enclosureness} of a tuple $t\in  S_D$ is
    \begin{equation}
    \lambda(t) := \max_{\substack{\mathcal{J} \subseteq S_D \\ \forall t_1 \in\mathcal{J}, I(t_1) \subset I(t) \\ \forall t_2, t_3 \in \mathcal{J}, I(t_2) \cap I(t_3) = \emptyset}} |\mathcal{J}|,
    \label{equ:old_enc}
    \end{equation}
i.e., the largest number of disjoint lifespans in $S_D$ contained in $I(t)$. 
Then the enclosureness of the update sequence is the average enclosureness of all the tuples (but at least $1$), i.e.,
 \begin{equation}
    \lambda(S_D) := \max\left(\frac{\sum_{t \in S_D} \lambda(t)}{|S_D|}, 1\right).
\label{equ:fullquery_enc}
\end{equation}
We often omit $S_D$ and simply write $\lambda:=\lambda(S_D)$ for the enclosureness of an update sequence. 
\end{definition}
Then, they give an algorithm that can update any foreign-key acyclic query in $O(\lambda)$ time for any $S_D$ while supporting $O(1)$-delay enumeration.  This is appealing, since while $\lambda$ can be as large as $O(|S_D|)$ in the worst case, it is often a small constant for many common update sequences, including FIFO, FILO (first-in-last-out), and insertion-/deletion-only sequences.  The worst-case situation only happens when there are many tuples with long lifespans joining with many tuples with short lifespans, something that is uncommon in practice (i.e., many big but ephemeral changes to the query).

However, their analysis crucially relies on the nice property of foreign-key acyclic queries, that their result size is at most linear, which is not the case for non-key joins.  In fact, we show below that the $O(\lambda)$ update time is unachievable for free-connex queries, which follows from the negative result that we prove below:

\begin{theorem}
\label{the:lb-non-weak-hierarchical-full}
Consider the query $Q= R_1(x_1)\Join R_2(x_1,x_2) \Join R_3(x_2,x_3) \Join R_4(x_3,x_4) \Join R_5(x_4)$ over a FIFO update sequence.  If there is an algorithm for $Q$ with update time $O(|D|^{1/2-\epsilon})$ while supporting $O(|D|^{1-\epsilon})$-delay enumeration of full results for any constant $\epsilon > 0$, then the OuMv conjecture\footnote{The OuMv conjecture~\cite{henzinger2015unifying} is that the following problem cannot be solved in $O(n^{3-\epsilon})$ time for any constant $\epsilon > 0$: Given an $n\times n$ matrix $M$ and a sequence of $n$-dimensional vectors $u_1, v_1, u_2, v_2, \cdots, u_n, v_n$, compute $u_i M v_i$ for each $i$ over the Boolean semiring. The algorithm must return $u_iMv_i$ before $u_{i+1}, v_{i+1}$ are revealed.}  fails.
\end{theorem}

Note that this theorem separates the difficulty of (at least one of) free-connex queries from foreign-key acyclic queries, for which $O(1)$ update time is possible for FIFO sequences \cite{wang2020maintaining}.

\subsection{Join-tree-specific Enclosureness}

Hope is not all lost despite the negative result above.  First, Theorem \ref{the:lb-non-weak-hierarchical-full} only holds for a particular free-connex query; other queries may still be updated in $O(1)$ time. Secondly, the definition of enclosureness in \cite{wang2020maintaining} only considers the time dimension while ignoring the structure dimension, i.e., which relation each update is applied to.  These observations motivate a more refined definition of enclosureness that also depends on the join tree (which nodes the updates are applied to).  As we will see, a hard query like the one in Theorem \ref{the:lb-non-weak-hierarchical-full} can still be solved efficiently, when information from both the structural dimension and the time dimension is taken into account.

    \newcommand{\LI}{\widehat{I}}
    \newcommand{\RI}{\widecheck{I}}
    \begin{definition}[Effective lifespan of Tuples from Input Relations]
    \label{def:minimal-input}
    Given a free-connex query $Q$, a free-connex join tree $\T$ of $Q$, a database $D$, and an update sequence $S_D$, the two {\it effective lifespans} of an input tuple $t_1 \in R_e$ with lifespan $I(t_1) = [t_1^+, t_1^-]$ are 
	\begin{align*}
	   	\LI(t_1) =& \left[t_1^+,  \min\left(t_1^-, \min_{t_2 \in R_{e'}: e' \in \T_e - \{e\}, t_2^- > t_1^+} t_2^-\right)\right]; \\
	    	\RI(t_1) =& \left[\max\left(t_1^+, \max_{t_2 \in R_{e'}: e' \in \T_e - \{e\}, t_2^+ < t_1^-} t_2^+\right), t_1^-\right].
	\end{align*}
        where the minimum and maximum choices are taken over all tuples $t_2$ from any input relation $R_{e'}$ residing in the subtree $\T_e$. 
    \end{definition}
    In plain language, $\LI(t_1)$ is obtained from $I(t_1)$ by moving forward its ending time to the first deletion of an input tuple from any descendent of $e$, while to obtain $\RI(t_1)$, we move its starting to the last insertion of an input tuple from any descendent of $e$.

   %\update{
   %However, a generalized relation does not contain any input tuple. Here, we separately define the {\em virtual lifespans} as follows:
   %
   %\begin{definition}[Virtual lifespans for Generalized Relations]
   % \label{def:minimal-generalized}
   % Given a free-connex query $Q$, a free-connex join tree $\T$ of $Q$, a database $D$, an update sequence $S_D$, and a generalized relation $e$, every tuple $t_1$ from an input relation in $\T_e$ defines a virtual lifespan for $e$:
   % \begin{align*}
%     & \left[t_1^+,  \min\left(t_1^-, \min_{t_2 \in R_{e'}: e' \in \T_e - \{e\}, t_2^- > t_1^+} t_2^-\right)\right]
%     \end{align*}
%     where the minimum choice is taken over all tuples $t_2$ from any input relation $R_{e'}$ residing in the subtree $\T_e$.
% \end{definition}

%The virtual lifespans defined for a generalized relation also capture the updates we may apply to its semi-join view. %: some tuple with a virtual lifespan may be inserted into the generalized relation when there is an insertion in its subtree, and it may be deleted when there is any deletion after that insertion in its subtree.

% \begin{lemma}
%     Given an update sequence $S_D$, let $n$ be the total number of lifespans on the subtree $\T_e$, where $e$ is a generalized relation, then $e$ has $n$ effective lifespans.
% \end{lemma}
%}

We can now define the join-tree-specific enclosureness of a tuple:
\begin{definition}
\label{def:Tenclosure}
    Given a free-connex query $Q$, a free-connex join tree $\T$ of $Q$, a database $D$, and an update sequence $S_D$, for a node $e\in \T$ and a tuple $t\in R_e$, its {\em enclosureness} is 
    \begin{equation}
    \lambda_\T(t) = \max_{\substack{\qichen{\forall t' \in \mathcal{J}, \exists e' \in \mathcal{T}_e - \{e\}, t' \in R_{e'}} \\ \forall t_1 \in\mathcal{J}, \dot{I}(t_1) \subseteq I(t) \\ \forall t_2, t_3 \in \mathcal{J}, \dot{I}(t_2) \cap \dot{I}(t_3) = \emptyset}} |\mathcal{J}|,
    \label{equ:new_enc}
    \end{equation}
where each $\dot{I}$ is either $\LI$ or $\RI$, 
i.e., the largest number of disjoint effective lifespans of tuples in the descendants of $e$, which are contained in the lifespan of $t$.  Then the enclosureness of the update sequence is still the average:
\[
     \lambda_{\T}(S_D) := \max\left(\frac{\sum_{t \in S_D} \lambda_{\T}(t)}{|S_D|}, 1\right).
\]
We often write $\lambda_{\T}:=\lambda_{\T}(S_D)$ for the enclosureness of an update sequence with respect to $\T$. 
\end{definition}

\begin{figure}[t]
\centering
\includegraphics[width=\linewidth]{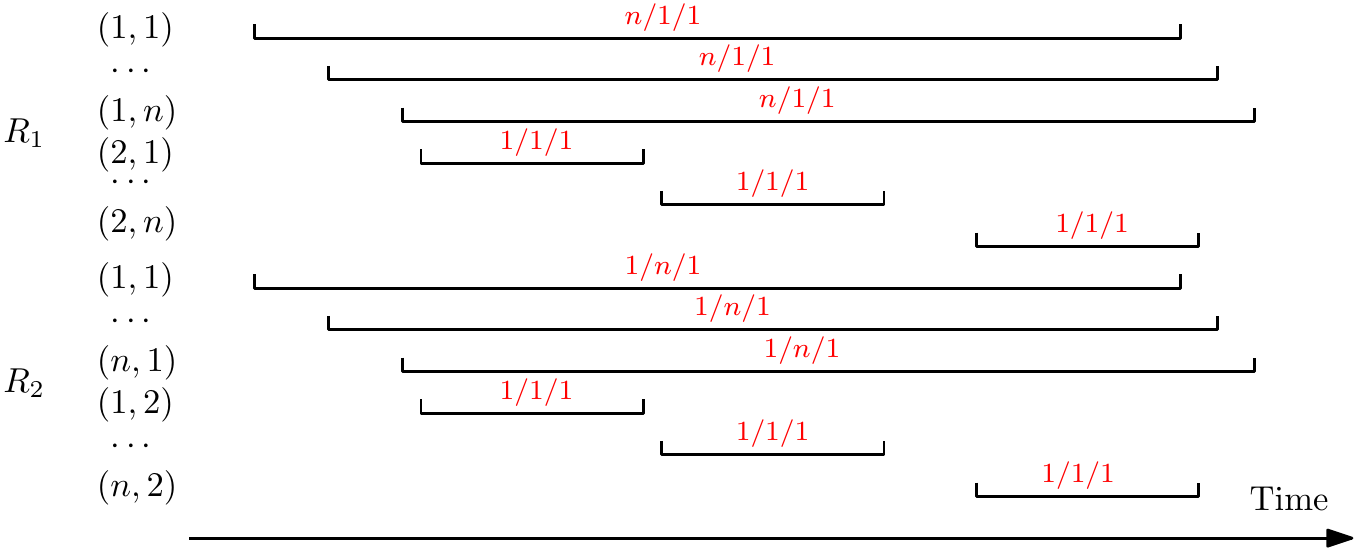}
\caption{As update sequence in Example~\ref{ex:enc}. Each interval is the lifespan of a tuple, and three numbers above each interval are its enclosureness over $\T_1$, $\T_2$ and $\T_3$ in Figure \ref{fig:trees}. 
}
\label{fig:example_seq}
%\end{minipage}
\end{figure}

\begin{example}
\label{ex:enc}
Consider $Q:= R_1(x_1, x_2) \Join R_2(x_2, x_3)$ in Figure \ref{fig:trees} with $\T_1, \T_2, \T_3$.  For the update sequence in Figure~\ref{fig:example_seq}, $\lambda_{\T_1} = \lambda_{\T_2} = n$ and $\lambda_{\T_3}=1$.  In fact, $\lambda_{\T_3}=1$ for any update sequence.
\end{example}

The main analytical result of this paper is the following theorem, whose proof is quite technical given in Appendix~\ref{appendix:analysis}:

\begin{theorem}
   \label{thm:main}
   For any free-connex query $Q$, the update cost of the query plan in Section \ref{sec:propagation} induced by any given free-connex join tree $\T$ of $Q$ is $O(\lambda_{\T})$ under any update sequence with enclosureness $\lambda_{\T}$.
\end{theorem}

This result is complemented with a matching lower bound, for at least one particular query: $Q = \pi_{x_1} (R_1(x_1, x_2) \Join R_2(x_2))$, which has one join tree as shown in Figure \ref{fig:q_T1} (one could add a generalized relation $[x_1]$ at the top, but it does not change the enclosureness).  Thus, for this query, $\lambda_{\T}$ does not really depend on $\T$.  

\begin{theorem}{\cite{wang2020maintaining}}
 \label{the:lb-non-q-hierarchical-projection}
   Suppose there is an algorithm for the query $Q = \pi_{x_1} (R_1(x_1, x_2)\Join R_2(x_2))$ with update time $O(\lambda^{1-\epsilon})$ while supporting $O(\lambda^{1-\epsilon})$-delay enumeration of full results for any constant $\epsilon > 0$,  
   then the OMv conjecture\footnote{The OMv conjecture is similar to the OuMv conjecture, except that the algorithm needs to compute $Mv_i$ for every $v_i$.} fails.
  \label{hardstream1}
\end{theorem}

\subsection{Implications of Enclosureness}
\label{sec:implications}

We present some implications of our join-tree-specific enclosureness and Theorem \ref{thm:main}, exhibiting an interesting trade-off between the hardness of update sequences and the complexity of queries.

\smallskip
\noindent {\bf Arbitrary update sequences.} For arbitrary update sequences, prior work \cite{berkholz17:_answer,idris17:_dynam} has shown how to achieve $O(1)$ update time while supporting $O(1)$-delay enumeration for any q-hierarchical query. \qichen{It turns out that this is an easy consequence of Theorem \ref{thm:main}, plus the following structural property of q-hierarchical queries, as well as the simple fact that $\lambda_{\T}=1$ if the height of $\T$ is $1$:}

\begin{lemma}
	Every q-hierarchical CQ has a height-1 free-connex join tree.
	\label{lem:q-hierarchical}
\end{lemma}

For arbitrary update sequences, q-hierarchical queries are precisely the class of queries for which $O(1)$ update time is possible \cite{berkholz17:_answer}.  Thus, for queries outside this class, we must restrict the update sequence in order to achieve $O(1)$ update time.  We consider the following two classes of update sequences.
 
\smallskip
\noindent {\bf FIFO sequences.}
The update time is shown to be $O(1)$ for foreign-key acyclic joins over FIFO sequences~\cite{wang2020maintaining}, but nothing is known for non-key joins (except for q-hierarchical queries which do not rely on FIFO).  We present the first extension in this direction:

\begin{lemma}
\label{lem:weak-hierarchical}
    For any free-connex query $Q$ with a free-connex join tree $\T$ of height at most $2$, $\lambda_\T = 1$ for any FIFO sequence.
\end{lemma}

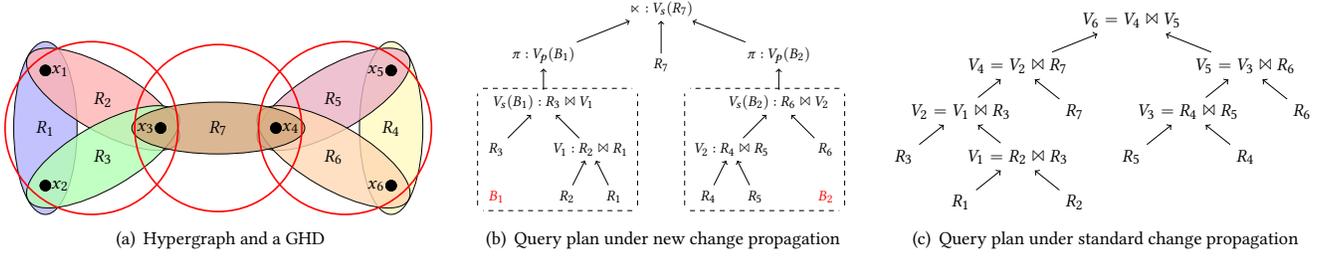
\begin{figure*}
    \subfigure[Hypergraph and a GHD]{
    \resizebox{0.33\linewidth}{!}{
        \centering
        \begin{tikzpicture}
    \node (v1) at (-3,1) {};
    \node (v2) at (-3,-1) {};
    \node (v3) at (-1,0) {};
    \node (v4) at (1,0) {};
    \node (v5) at (3,1) {};
    \node (v6) at (3,-1) {};
    \node (w1) at (-3, 0) {};
    \node (w2) at (-2, 0.5) {};
    \node (w3) at (-2, -0.5) {};
    \node (w4) at (3, 0) {};
    \node (w5) at (2, 0.5) {};
    \node (w6) at (2, -0.5) {};
    \node (l1) at (-1, 1) {};
    \node (r1) at (1, 1) {};

    \begin{scope}[fill opacity=0.8]
    \filldraw[fill=yellow!30, rotate=90] ($(w4)$) ellipse (1.5 and 0.55);
    \filldraw[fill=purple!30, rotate=30] ($(w5)$) ellipse (1.5 and 0.55);
    \filldraw[fill=orange!30, rotate=150] ($(w6)$) ellipse (1.5 and 0.55);
    \filldraw[fill=blue!30, rotate=90] ($(w1)$) ellipse (1.5 and 0.55);
    \filldraw[fill=red!30, rotate=150] ($(w2)$) ellipse (1.5 and 0.55);
    \filldraw[fill=green!30, rotate=30] ($(w3)$) ellipse (1.5 and 0.55);
    \filldraw[fill=brown!70] ($(0, 0)$) ellipse (1.5 and 0.45);
    \draw[color=red, thick] ($(-2.2, 0)$) circle (1.5);
    \draw[color=red, thick] ($(2.2, 0)$) circle (1.5);
    \draw[color=red, thick] ($(0, 0)$) circle (1.45);
    \end{scope}

    \foreach \v in {1,2,...,6} {
        \fill (v\v) circle (0.1);
    }

    \fill (v1) circle (0.1) node [right] {$x_1$};
    \fill (v2) circle (0.1) node [right] {$x_2$};
    \fill (v3) circle (0.1) node [left] {$x_3$};
    \fill (v4) circle (0.1) node [right] {$x_4$};
    \fill (v5) circle (0.1) node [left] {$x_5$};
    \fill (v6) circle (0.1) node [left] {$x_6$};

    \node at (w1) {$R_1$};
    \node at (w2) {$R_2$};
    \node at (w3) {$R_3$};
    \node at (w4) {$R_4$};
    \node at (w5) {$R_5$};
    \node at (w6) {$R_6$};
    \node at (0, 0) {$R_7$};

\end{tikzpicture}
        \label{fig:dumbbell}
        }
    } 
    \hfill
    \subfigure[Query plan under new change propagation]{
        \resizebox{0.29\linewidth}{!}{
            \centering
            \begin{tikzpicture}
  \node (p1) at (0, 0) {$\ltimes : V_s(R_7)$};
  \node (p2) at (-2.5, -1) {$\pi : V_p(B_1)$};
  \node (p3) at (0, -1.2) {$R_7$};
  \node (p4) at (2.5, -1) {$\pi : V_p(B_2)$};
  \node (p5) at (-2.5, -2) {$V_s(B_1) : R_3 \Join V_1$};
  \node (p6) at (2.5, -2) {$V_s(B_2) : R_6 \Join V_2$};
  \node (p7) at (-3.5, -3) {$R_3$};
  \node (p8) at (-1.5, -3) {$V_1 : R_2 \Join R_1$};
  \node (p9) at (1.5, -3) {$V_2 : R_4 \Join R_5$};
  \node (p10) at (3.5, -3) {$R_6$};
  \node (p11) at (-2, -4) {$R_2$};
  \node (p12) at (-1, -4) {$R_1$};
  \node (p13) at (1, -4) {$R_4$};
  \node (p14) at (2, -4) {$R_5$};
  \node (a1) at (-0.5, -4.3) {};
  \node (a2) at (-3.9, -4.3) {};
  \node (a3) at (-3.9, -1.7) {};
  \node (a4) at (-0.5, -1.7) {};
  \node[text=red] (a5) at (-3.5, -4) {$B_1$};
  \node (b1) at (0.5, -4.3) {};
  \node (b2) at (3.9, -4.3) {};
  \node (b3) at (3.9, -1.7) {};
  \node (b4) at (0.5, -1.7) {};
  \node[text=red] (b5) at (3.5, -4) {$B_2$};
  \begin{scope}[every path/.style={<-}]
    \draw (p1) -- (p2);
    \draw (p1) -- (p3);
    \draw (p1) -- (p4);
    \draw (p2) -- (p5);
    \draw (p4) -- (p6);
    \draw (p5) -- (p7);
    \draw (p5) -- (p8);
    \draw (p6) -- (p9);
    \draw (p6) -- (p10);
    \draw (p8) -- (p11);
    \draw (p8) -- (p12);
    \draw (p9) -- (p13);
    \draw (p9) -- (p14);
  \end{scope}
  \begin{scope}
    \draw[dashed] (a1) -- (a2);
    \draw[dashed] (a2) -- (a3);
    \draw[dashed] (a3) -- (a4);
    \draw[dashed] (a1) -- (a4);
  \end{scope}
  \begin{scope}
    \draw[dashed] (b1) -- (b2);
    \draw[dashed] (b2) -- (b3);
    \draw[dashed] (b3) -- (b4);
    \draw[dashed] (b1) -- (b4);
  \end{scope}
\end{tikzpicture}
            \label{fig:dumbbell_newplan}
        }
    }
    \hfill
    \subfigure[Query plan under standard change propagation]{
        \resizebox{0.33\linewidth}{!}{
            \centering
            \begin{tikzpicture}
  \node (p1) at (0, 0) {$V_6 = V_4 \Join V_5$};
  \node (p2) at (-2, -0.8) {$V_4 = V_2 \Join R_7$};
  \node (p3) at (2, -0.8) {$V_5 = V_3 \Join R_6$};
  \node (p4) at (-3, -1.6) {$V_2 = V_1 \Join R_3$};
  \node (p5) at (-1, -1.6) {$R_7$};
  \node (p6) at (1, -1.6) {$V_3 = R_4 \Join R_5$};
  \node (p7) at (3, -1.6) {$R_6$};
  \node (p8) at (-4, -2.4) {$R_3$};
  \node (p9) at (-2, -2.4) {$V_1 = R_2 \Join R_3$};
  \node (p10) at (0, -2.4) {$R_5$};
  \node (p11) at (2, -2.4) {$R_4$};
  \node (p12) at (-3, -3.2) {$R_1$};
  \node (p13) at (-1, -3.2) {$R_2$};
  \begin{scope}[every path/.style={<-}]
    \draw (p1) -- (p2);
    \draw (p1) -- (p3);
    \draw (p2) -- (p4);
    \draw (p2) -- (p5);
    \draw (p3) -- (p6);
    \draw (p3) -- (p7);
      \draw (p4) -- (p8);
    \draw (p4) -- (p9);
    \draw (p6) -- (p10);
    \draw (p6) -- (p11);
    \draw (p9) -- (p12);
    \draw (p9) -- (p13);
    
  \end{scope}
\end{tikzpicture}
            \label{fig:dp_dumbbell}
        }
    }
    \caption{\ref{fig:dumbbell} is the hypergraph of the ``dumbbell'' query $Q = R_1(x_1,x_2)\Join R_2(x_1,x_3) \Join R_3(x_2, x_3) \Join R_4(x_5,x_6) \Join R_5(x_4,x_5) \Join R_6(x_4,x_6) \Join R_7(x_3,x_4)$, with GHD illustrated in red circle.  \ref{fig:dp_dumbbell},\ref{fig:dumbbell_newplan} are query plans under the standard, new change propagation framework respectively. In \ref{fig:dp_dumbbell}, $B_1, B_2$ are treated as two basic relations, on which projection and semi-join views are constructed. }
    \label{fig:my_label}
\end{figure*}

Note that the height limit of $2$ is the best one can hope for, since the query in Theorem \ref{the:lb-non-weak-hierarchical-full} has a join tree of height $3$ and the theorem shows that it cannot be updated in $O(1)$ time over FIFO sequences.  Although the height-$2$ limitation restricts the class of queries, this already includes some fairly complex queries, such as the one in Figure~\ref{fig:1}; more examples can be found in Section~\ref{sec:experiment}.  

\smallskip
\noindent {\bf Insertion-only sequences.}  As we restrict the update sequence further, we can handle more queries in $O(1)$ time.  For simplicity, the following result only considers insertion-only sequences, but the same result holds for deletion-only or FILO sequences as well.

\begin{lemma}
	For any free-connex CQ $Q$ and free-connex join tree $\T$,  $\lambda_\T =1$ for any insertion-only update sequence.
	\label{lem:insertion-only}
\end{lemma}

\qichen{Combining Theorem~\ref{thm:main} and Lemma~\ref{lem:insertion-only}, the following theorem is straightforward. 

\begin{theorem}
For a free-connex query $Q$, there is an index that can be updated in $O(1)$ amortized time under any insertion-only update sequence, while supporting $O(1)$-delay enumeration.
	\label{thm:insertion-only}
\end{theorem}
}

Note that Lemma \ref{lem:insertion-only} incorporates the static result~\cite{bagan2007acyclic} as a special case.  Given a static database $D$, we can simply insert every tuple from $D$ into our query plan.  By Lemma \ref{lem:insertion-only}, this builds a data structure in $O(|D|)$ time that supports $O(1)$-delay enumeration of $Q(D)$.  Also, the dichotomy result of \cite{bagan2007acyclic} states that $O(|D|)$-time preprocessing and $O(1)$-delay enumeration are possible only for free-connex queries, thus Lemma \ref{lem:insertion-only} cannot be extended to beyond free-connex queries, either, even over insertion-only sequences.

\qichen{
\begin{example}
\label{ex:negative}
    Consider an insertion-only update sequence for the query in Figure~\ref{fig:1}: (1) tuples $(i, j) \in [n] \times [n]$ are inserted into $R_2$, $R_3$ and $R_4$ initially;  (2) tuples $(i, j) \in [n] \times [n]$ are inserted into $R_1$ later.  Standard change propagation or HIVM needs to materialize $\Delta(R_1 \Join R_2 \Join R_3)$, hence incurs $O(n^3)$ cost; the Dynamic Yannakakis algorithm~\cite{idris17:_dynam} needs to scan all tuples $(i', j') \in R_2$ for $i' = j$, once $(i, j)$ is inserted into $R_1$, hence incurs $\Theta(n)$ cost; and our framework only incurs $O(1)$ cost.
\end{example}
}

\smallskip
\noindent {\bf Query plan optimization.}
If the given query and/or the update sequence do not fall into any of the three cases above where $O(1)$ update time can be guaranteed, our enclosureness analysis still yields an effective heuristic for choosing a good $\T$, which in turn determines the query plan.  First, it is clear that $\T$ with a smaller height is always preferred.  Furthermore, Definition \ref{def:Tenclosure} suggests that we should put nodes with more updates higher in $\T$, as a tuple in a node might increase the enclosureness of tuples in its ancestors.  Thus, in our implementation, we construct all join trees and use the one that minimizes
$\displaystyle{\sum_{e \in \T} d(e) N(e)}$,
where $d(e)$ is the depth of $e$ in $\T$ (not counting generalized relations and itself) and $N(e)$ is the number of updates to $e$.  If $N(e)$ is unavailable, we can estimate it by observing (and buffering) the first few updates.

\section{Extensions to General Queries}
\label{sec:extend}

\subsection{General CQs}
\noindent {\bf Acyclic but non-free-connex queries.} Consider such a query $\pi_{x_1,x_3} R_1(x_1, x_2) \Join R_2(x_2, x_3)$.  We simply add $x_2$ as an output attribute to turn it into a free-connex query, and then do a projection over $x_1,x_3$ during enumeration.  Note that enumeration may contain duplicates. Thus, if a {\tt DISTINCT} keyword is declared explicitly, duplicates need to be removed, hence making the delay more than constant, but this is inevitable due to the lower bound \cite{bagan2007acyclic}.

\smallskip
\noindent {\bf Cyclic queries.}  Cyclic queries can also be easily handled in our framework by resorting to
\emph{Generalized Hypertree Decomposition} (GHD) \cite{gottlob2002hypertree}. More specifically, by grouping several relations into a {\em bag}, an arbitrary CQ can be converted into a free-connex one.  For example, Figure~\ref{fig:dumbbell} shows a GHD for the ``dumbbell'' query with 3 bags.

We can use standard change propagation within each bag, and apply our framework across the bags.  
This results in the query plan in Figure~\ref{fig:dumbbell_newplan}, which has $O(N^2)$ space and $O(N)$ update time while supporting constant-delay enumeration.
On the other hand, the standard change propagation framework would use a query plan like the one in Figure~\ref{fig:dp_dumbbell}, which has $O(N^3)$ space and update time. Of course, all these are worst-case bounds; on realistic inputs, the costs are lower, but our new query plan is still order-of-magnitude better than the old plan, as shown in Section \ref{sec:experiment}.

If one is interested in further improving the theoretical bounds, the algorithm for maintaining the query results inside each bag can be replaced by a better algorithm.  For example, Kara et al.~\cite{kara2019icdt} present an algorithm for maintaining the triangle join.  Replacing with the new algorithm can improve the space usage from $O(N^2)$ to $O(N\sqrt{N})$.   On the other side, although the algorithm~\cite{kara2019icdt} can improve the update cost for each bag to $O(\sqrt{N})$, the ``dumbbell'' query still suffers from $O(N)$ update cost. This is indeed unavoidable as a single tuple update can change as large as $O(N)$ results materialized for one bag, which further propagates to the overall framework. Hence, the update cost for this GHD-based change propagation framework is determined by updates not only inside each bag but also across bags.  

Beyond the triangle join, not many results are known.  This is still an actively researched problem; any improvement here will also improve general CQs when plugged into our framework.

\begin{theorem}
    Given a CQ $Q$ with a free-connex GHD of {\em width}\footnote{The definition of width depends on the algorithm used for maintaining query results inside each bag. If adopting the standard change propagation framework, the width is defined as the maximum width over all bags, where the width of a bag is the optimal integral edge covering number of the corresponding subquery derived for this bag.} $w$, there is an index of $O(N^{w})$ size that can be updated in $\Omega(N^{w})$ time while supporting $O(1)$-delay enumeration. 
\end{theorem}

\qichen{

\begin{proof}
   Maintaining any bag of relations requires $\Omega(N^{w})$ time, and it needs $O(N^{w})$ space to store all query results in the bag.  After maintaining each bag of relations, the algorithms proposed in Section~\ref{sec:propagation} can use to maintain between each bag, which takes $O(|D|)$ time for maintenance.  Noted that current database size $|D|$ is bounded by the largest bag size, which will be $N^{w}$, makes the total maintenance time to $\Omega(N^{w})$ and space cost to $O(N^{w})$.    
\end{proof}

The following lemma can be easily derived from the above theorem.

}

\begin{lemma}
\label{lem:dubbell}
    For the ``dumbbell'' query, there is an index of $O(N^{1.5})$ size that can be updated in $O(N)$ time per tuple update, while supporting $O(1)$-delay enumeration.
\end{lemma}

\subsection{Selection, union, and set difference}
The query plan in Section \ref{sec:queryplan} works for CQs with joins and projections, but it can be equipped with other operators easily.
\begin{itemize}[leftmargin=*]
    \item If there is a selection $\sigma_\phi$ on an input relation $R_e$ where $\phi$ is a predicate on $e$, then for an update with tuple $t\in R_e$, we simply check if $\phi(t)$ is true, and discard this update if not.  This only adds $O(1)$ time to the update cost. 
    \item For the union of CQs $Q = Q_1 \cup \cdots \cup Q_k$, we just maintain each $Q_i$ separately.  Full enumeration can be supported with $O(1)$ delay using the technique in  \cite{carmeli2019enumeration}.  We note that \cite{carmeli2019enumeration} assumes that the data structure on each $Q_i(D)$ can check if $t\in Q(D_i)$ in $O(1)$ time for any given $t$, which is indeed supported by our query plan.  For delta enumeration, we can use the same technique to enumerate $\Delta Q_1(D) \cup \cdots \cup \Delta Q_k(D)$. However, this is not the same as $\Delta (Q_1 \cup \cdots \cup Q_k)$, and we need to check, say, if some new result $t\in \Delta(Q_1)$ already exists in $Q_2(D)$.  Thus, while  the technique of \cite{carmeli2019enumeration} is still correct for delta enumeration, the delay is not bounded by a constant. How to support $O(1)$-delay delta enumeration for UCQs remains an interesting open problem.
    \item For a query like $Q=Q_1 - Q_2$, we can as above maintain $Q_1$ and $Q_2$ separately.  For  enumeration, we enumerate every $t\in Q_1(D)$ and check if $t\in Q_2(D)$, although this does not guarantee constant delay. In fact, even in the static case, it is an open question whether $Q_1(D) - Q_2(D)$ can be enumerated in $O(1)$ delay after linear-time preprocessing where $Q_1$ and $Q_2$ are both free-connex.
\end{itemize}

\subsection{Aggregations}

Standard relational algebra can be extended to support aggregations, and we adopt the following formalism~\cite{abo2016faq,joglekar16:_ajar}.  Let $(S, \oplus, \otimes)$ be a commutative ring\footnote{In the static case,  $(S, \oplus, \otimes)$ is only required to be a semi-ring, but we need additive inverses to support deletions. }. Every tuple $t \in R_e$ has an annotation $w(t) \in S$. For a full CQ $Q$ in the form of $(\ref{q1})$, the annotation for any join result  $t \in Q(D)$ is defined as 
$w(t) := \mathop{\otimes}\limits_{e \in Q} w(t[e])$.
For a non-full query $\pi_\y Q$, the projection becomes {\tt GROUP BY} $\y$, and the annotation for each result $t \in \pi_\y Q(D)$ (i.e., the aggregate of each group) is 
$w(t) := \mathop{\oplus}\limits_{t' \in Q(D): \pi_{\y} t'= t} w(t')$.  

Our new change propagation framework can be extended to support aggregations easily. For any relation $e$, let $w(t)$ be the annotation for tuple $t \in R_e$, $w_s(t)$ be the annotation for $t \in V_s(R_e)$ and $w_p(t)$ be the annotation for $t \in V_p(R_e)$. Following the definitions of semi-join view $V_s$ and projection view $V_p$ (see Section~\ref{sec:propagation}), the annotation of $t \in V_s(R_e)$ is defined as
\begin{equation}
    w_s(t) := \left(\mathop{\otimes}\limits_{e' \in \C_e: e' \notin \con} w_p(\pi_{\key(e')} t) \right) \otimes w(t).
    \label{eq:agg1}
\end{equation}
The annotation of $t \in V_p(R_e)$ is defined as  
\begin{equation}
    w_p(t) := \mathop{\oplus}\limits_{t' \in V_s(R_e) : \pi_{\key(e)} t' = t} w_s(t').
    \label{eq:agg2}
\end{equation}

To simplify the enumeration process, we define an additional view $V_a(R_e) = \pi_{e \cap \y} V_s(R_e)$ for each node $e \in \con$, where the annotation for any tuple $t \in V_a(R_e)$ is defined as 
\begin{equation}
   w_a(t) := \mathop{\oplus} \limits_{t' \in V_s(R_e) : \pi_{e \cap \y} t' = t} w_s(t').
    \label{eq:agg3}  
\end{equation}
Moreover, if $\y = \emptyset$, we maintain a special annotation $w_a(r)$ on top of the root $r$:
\begin{equation}
    w_a(r) := \mathop{\oplus}\limits_{t \in V_s(R_r)} w_s(t).
    \label{eq:agg5}
\end{equation}

\paragraph{Update} We store these annotations alongside their counters in the query plan.  Then, the change propagation and enumeration procedures should be modified according to the formulas above.  More precisely, whenever the counter of a tuple is updated, we also update its annotation according to \eqref{eq:agg1}, \eqref{eq:agg2}, \eqref{eq:agg3} and \eqref{eq:agg5}. 

\paragraph{Enumeration} We distinguish two cases for enumeration. If $\y \neq \emptyset$, we compute the annotation $w(t)$ for each query result $t \in Q$ as:
\begin{equation}
    w(t) := \mathop{\otimes}\limits_{e \in \con} w_a(\pi_{e \cap \y} t).
    \label{eq:agg4}
\end{equation}
within $O(1)$ time, as long as $V_a(\cdot)$ is well maintained for each node $e \in \con$.
If $\y=\emptyset$, we simply return $w_a(r)$.

\begin{lemma}
\label{lem:empty}
    If $\y = \emptyset$, $w_a(r)$ is the final aggregation. 
\end{lemma}

\begin{lemma}
\label{lem:non-empty}
  If $\y \neq \emptyset$, \eqref{eq:agg4} is the annotation of query result $t \in Q$.
\end{lemma}

\medskip
Below, we prove the correctness via Lemma~\ref{lem:empty} and Lemma~\ref{lem:non-empty}. Given a free-connex join tree $\T$, we define {\em non-connex subtree} for node $e$ as $\X_e$, which is the largest subtree of $\T$ rooted at $e$ such that $(\X_e - \{e\}) \cap \con = \emptyset$. If $e \notin \con$, $\C_e \subseteq \X_e$. 
    It is straightforward to see, for any pair of distinct nodes $e,e' \in \con$, $\X_e$ and $\X_{e'}$ must be node-disjoint.  We further define $\Q_f(e, t)$ be the set of join results $\Join_{e' \in \X_e} e'$, such that for any $t_i \in \Q_f(e, t)$, $\pi_{e} t_i = t$.

    \begin{lemma}
    \label{lem:support1}
        Given a free-connex query $Q$ with free-connex join tree $\T$, and an arbitrary node $e \notin \con$, then for any tuple $t \in V_s(R_e)$ %with $t_1, t_2, \cdots, t_\ell$ as the full join results of $Q_f(e) := \Join_{e' \in \X_e} e'$ such that $\pi_{e} t_i = t$,
        \[
            %w_s(t) := \oplus_{i \in [\ell]} w(t_i).
            w_s(t) := \oplus_{t_i \in \Q_f(e,t)} w(t_i).
        \]
    \end{lemma}

     \begin{proof}[Proof of Lemma~\ref{lem:support1}]
        We can prove it by induction on the size of $\X_e$.  It trivially holds when $\X_e$ contains only $e$ since there is only one join result $t$ with $w_s(t) = w(t)$. In general case, we assume that it holds for every child node of $e$. Then, 
       \[
        \begin{aligned}
        w_s(t)  &= w(t) \otimes \left(\bigotimes_{e' \in \C_e} \left(w_p(\pi_{\key(e')} t\right)\right) & \text{by }\eqref{eq:agg1}\\
                &= w(t) \otimes  \left(\bigotimes_{e' \in \C_e} \left(\bigoplus_{t_j \in \Q_f(e', \pi_{\key(e')} t)} w_s(t_j)\right)\right) & \text{by }\eqref{eq:agg2},
                %&= w(t) \otimes \left(\bigoplus_{i \in [\ell]} \left(\bigotimes_{e' \in \C_e} w_s(\pi_{e'} t_i)\right)\right) & \\
        %&= w(t) \otimes \left(\bigoplus_{i \in [\ell]} \left(\bigotimes_{e' \in \X_e -\{e\}} w(\pi_{e'} t_i) \right)\right) &\\ 
             %&= w(t) \otimes \left(\bigoplus_{i \in [\ell]} \left(\bigotimes_{e' \in \C_e} w_s(\pi_{e'} t_i)\right)\right) &\\
             %&= w(t) \otimes  \left(\bigotimes_{e' \in \C_e} \left(\bigoplus_{i \in [\ell]} w_s(\pi_{e'} t_i)\right)\right) & \\
             %&= w(t) \otimes \left(\bigotimes_{e' \in \C_e} \left(w_p(\pi_{\key(e')} t\right)\right) & 
        \end{aligned}
    \]
     The join between any pair of child nodes in $\C_e$ can be degenerated to Cartesian product, because for any $e_1, e_2 \in \C_e$, with any $t_1 \in \pi_{e_1} \Q_f(e, t)$ and $t_2 \in \pi_{e_2} \Q_f(e, t)$, either $e_1 \cap e_2 = \emptyset$, or $e_1 \cap e_2 \subseteq e$ and $\pi_{e_1 \cap e_2} t_1 = \pi_{e_1 \cap e_2} t_2 = \pi_{e_1 \cap e_2} t$.  In either case, 
    \[
    % \begin{aligned}
    %     & \pi_{e_1} \Q_f(e, t) \Join \pi_{e_2} \Q_f(e, t) \\
    %     =& \pi_{e_1 \cup e_2} \Q_f(e, t) \\
    %     =& \pi_{e_1} \Q_f(e_1, \pi_{\key(e_1)} t) \times \pi_{e_2} \Q_f(e_2, \pi_{\key(e_2)} t).
    % \end{aligned}
    \begin{aligned}
       \Q_f(e, t)
        =& t \Join (\times_{e' \in \C_e} \Q_f(e', \pi_{\key(e')} t)).
    \end{aligned}
    \]
    We further decompose $t_i \in \Q_f(e,t)$ as $t_i := \Join_{e' \in \X_e} (\pi_{e'} t_i)$, then
    \[
        \oplus_{t_i \in \Q_f(e,t)} w(t_i) = \oplus_{t_i \in \Q_f(e,t)} \left( \otimes_{e' \in \X_{e}} w(\pi_{e'} t_{i}) \right)
    \]
    Together with $w_s(t') := \oplus_{t'_i \in \Q_f(e,t)} w(t'_i)$ for every $e' \in \C_e$, we can further rewrite $w_s(t)$ as
    \[\begin{aligned}
        &=w(t) \otimes  \left(\bigotimes_{e' \in \C_e} \left(\bigoplus_{t_j \in \Q_f(e', \pi_{\key(e')} t)} w_s(t_j)\right)\right) \\
        &=w(t) \otimes  \left(\bigotimes_{e' \in \C_e} \left(\bigoplus_{t_j \in \Q_f(e',\pi_{\key(e')} t)} \left( \bigotimes_{e'' \in \X_{e'}} w(\pi_{e''} t_{j}) \right)\right)\right) \\
        &=w(t) \otimes \left(\bigoplus_{t_i \in \Q_f(e, t)} \left(\bigotimes_{e' \in \X_e -\{e\}} w(\pi_{e'} t_i) \right)\right) \\  % Using the cartesian product property to switch sum and product, and extend each t_j \in Q_f(e') to separate input tuples. 
        &=\left(\bigoplus_{t_i \in \Q_f(e, t)} \left(w(t) \times \bigotimes_{e' \in \X_e -\{e\}} w(\pi_{e'} t_i) \right)\right) =\bigoplus_{t_i \in \Q_f(e, t)} w(t_i)\end{aligned}
    \]
    which completes the proof. 
    \end{proof}

    Hence, if $\y = \emptyset$, $\X_r = \T$. The correctness of Lemma~\ref{lem:empty} follows the correctness of Lemma~\ref{lem:support1} on $e = r$.

    \begin{lemma}
    \label{lem:support2}
        Given a free-connex query $Q$ with free-connex join tree $\T$, and an arbitrary node $e \in \con$, then for any $t \in V_s(R_e)$% with $t_1, t_2 \cdots, t_\ell$ as the full join results of $\Join_{e' \in \X_e} R_{e'}$ such that $\pi_{e} t_i = t$,
        \[
            w_s(t) := \oplus_{t_i \in \Q_f(e, t)} w(t_i).
        \] 
    \end{lemma} 
    
    \begin{proof}[Proof of Lemma~\ref{lem:support2}]
    Let $\{e_1,e_2,\cdots,e_k\} = \C_e \cap \X_e$. From Lemma~\ref{lem:support1}, we know that for every tuple $t_i \in V_s(R_{e_i})$,
    \[w_s(t_i) = \oplus_{t' \in \Q_f(e, t_i)} w(t').\]
    Given any $t \in V_s(R_e)$, we can rewrite $w_s(t)$ as follow: %and full join results $t_1, t_2, \cdots, t_\ell$ such that $\pi_{\y} t_i = t$ for any $i \in [\ell]$
    \[
        \begin{aligned} 
         &= w(t) \otimes \left(\bigotimes_{e' \in \C_e \cap \X_e} \left(w_p(\pi_{\key(e')} t\right)\right) \\ 
                &= w(t) \otimes  \left(\bigotimes_{e' \in \C_e \cap \X_e} \left(\bigoplus_{t_j \in \Q_f(e', \pi_{\key(e')} t)} w_s(t_j)\right)\right) \\ 
                &= w(t) \otimes \left(\bigotimes_{e' \in \C_e \cap \X_e} \left(\bigoplus_{t_j \in \Q_f(e', \pi_{\key(e')} t)} \left(\bigotimes_{e'' \in \X_{e'}} w(\pi_{e''} t_j)\right)\right)\right) \\
                &=w(t) \otimes \left(\bigoplus_{t_i \in \Q_f(e, t)} \left(\bigotimes_{e' \in \X_e -\{e\}} w(\pi_{e'} t_i) \right)\right) \\
                &= \left(\bigoplus_{t_i \in \Q_f(e, t)} \left(w(t) \times \bigotimes_{e' \in \X_e -\{e\}} w(\pi_{e'} t_i) \right)\right) = \bigoplus_{t_i \in \Q_f(e, t)} w(t_i)
        \end{aligned}
    \]

    \end{proof}

    \begin{proof}[Proof of Lemma~\ref{lem:non-empty}]
    Give any query result $t \in \Q$ and full join results $t_1, t_2, \cdots, t_\ell$ such that $\pi_{\y} t_i = t$ for any $i \in [\ell]$.  From the definition of annotation, 
    \begin{equation}
        w(t) := w(t_1) \oplus w(t_2) \oplus \cdots \oplus w(t_\ell).
        \label{eq:proof1}
    \end{equation}
    We start from the case where $\T = \con$. %Suppose $\con = \{e_1, e_2, \cdots, e_n\}$.  
    In this case, for any $e \in \con$, we have $w_s(t) = w(t)$ for every tuple $t \in V_s(R_{e})$. This way, we can rewrite \eqref{eq:proof1} as 
    \begin{equation}
        w(t) := \bigotimes_{e \in \E}\left(\bigoplus_{i \in [\ell]} w(\pi_{e} t_i)\right),
    \end{equation}
    and for any $i \in [\ell]$, 
    \[\bigoplus_{i \in [\ell]} w(\pi_{e} t_i) = w_a(\pi_{\y \cap e} t)\]
    based on the definition.  By replacing all terms, we can obtain equation \eqref{eq:agg4}.

    On the other hand, if $\con \subsetneq \T$, we note that $\T = \uplus_{e \in \con} \X_{e}$, where $\oplus$ denotes the disjoint union. Consider any query result $t$ of $Q$. This way, we have 
    \[
        \begin{array}{rl}
            w(t) &= \bigoplus_{i \in [\ell]} w(t_i)\\
                &= \bigoplus_{i \in [\ell]} \left(\bigotimes_{e \in \con} \left(\bigotimes_{e' \in \X_e} w(\pi_{e'} t_i) \right)\right)\\
                &= \bigotimes_{e \in \con} \left( \bigoplus_{t_j \in \Q_f(e, \pi_{\key(e)} t)} w(t_j) \right) \\
                &= \bigotimes_{e \in \con} \left( \bigoplus_{t_k \in V_s(R_e) : \pi_{e \cap y} t_k = \pi_{e \cap y} t} w_s(t_k) \right) \\
                &= \bigotimes_{e \in \con} \left(w_a(\pi_{e \cap y} t)\right) 
        \end{array}
    \]
    which completes the whole proof.
    \end{proof}

\section{Experiments}
\label{sec:experiment}

\subsection{Setup}
\noindent {\bf Prototype implementation.} We have implemented our algorithms and built a system prototype called CROWN (\textbf{C}hange p\textbf{RO}pagation \textbf{W}ithout joi\textbf{N}s) on top of Flink DataStream API.  All of our algorithms are implemented as DataStream functions, which take as input an update stream.  Each tuple in the update stream is associated with a flag indicating whether the update is an insertion or deletion, as well as the name of the updated relation.  After processing an update, the DataStream function outputs the deltas triggered by this update.  Enumeration of full query results can be invoked upon the user's request.
Implementing the prototype over Flink allows us to inherit all the benefits of Flink, such as fault-tolerance and the ability to work with a variety of data sources and sinks.  To dispatch tuples in a load-balanced fashion, we borrow a similar idea from massively parallel algorithms, such as HyperCube \cite{afrati11:_optim, beame13:_commun, wang2020maintaining}.  

We have evaluated our algorithms in both centralized and distributed settings.  The centralized version runs on a single machine with a single thread, where we disable certain Flink features such as false tolerance, serialization, and dispatching.  This is for a fair comparison with other centralized systems (DBToaster and Trill) that do not support these features.  The distributed version has all these features enabled.  It runs over two machines, each equipped with two Intel Xeon 2.1GHz processors with 48 cores and 416 GB memory.  The machine runs Linux, with Scala 2.11.12, dotnet 5.0.403, Flink 1.13.5, and Spark 2.2.3.  Each query is evaluated 10 times on each engine and we report the average runtime.  We set a 4-hour time limit for each run.

\smallskip
\noindent {\bf Query processing engines compared.} We compare CROWN with (1) DBToaster \cite{ahmad2012dbtoaster}, the best HIVM engine that supports multi-way joins over arbitrary update streams in centralized settings; (2) DBToaster Spark \cite{nikolic2016win}, which can support IVM with batch updates in a distributed/parallel setting; (3) Trill \cite{chandramouli2014trill}, a continuous query evaluation system over streaming data using the standard change propagation framework; and (4)  the native Flink SQL engine over streaming data.
\begin{table}
\begin{footnotesize}
\setlength\tabcolsep{3pt} 
\begin{tabular}{|c|c|c|c|c|c|} 
\hline
& \multirow{2}{*}{CROWN} & \multirow{2}{*}{Flink} & \multirow{2}{*}{DBToaster} & DBToaster& \multirow{2}{*}{Trill} \\
& & & & Spark & \\\hline
Distributed &  \checkmark  & \checkmark  & &  \checkmark & \\ \hline
Full & \multirow{2}{*}{\checkmark} & \multirow{2}{*}{\checkmark} &  \multirow{2}{*}{\checkmark} &  \multirow{2}{*}{\checkmark} &\\ 
enumeration & & & & &\\ \hline
Delta & \multirow{2}{*}{\checkmark}  &  & & & \multirow{2}{*}{\checkmark} \\
enumeration & & & & & \\\hline
Updates & Arbitrary  & FIFO & Arbitrary & Batch & Arbitrary  \\ \hline
\multirow{3}{*}{Internal} & \multirow{2}{*}{This} & Standard & \multirow{3}{*}{HIVM} & \multirow{3}{*}{HIVM} & Standard \\
& \multirow{2}{*}{paper} & change & & & change \\
& & propagation &  & & propagation \\
\hline
\end{tabular}
\end{footnotesize}
\caption{Comparison of different query processing engines. }
\label{tbl:system}
\end{table}

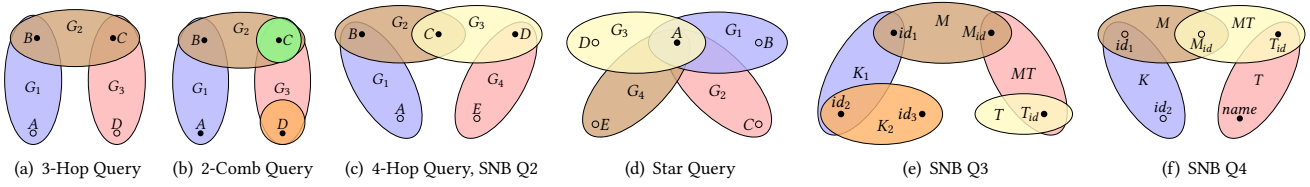
\begin{figure*}
\subfigure[3-Hop Query]{
    \resizebox{0.11\linewidth}{!}{%
    	\centering
    	\begin{tikzpicture}[font = \LARGE]
    \node (v1) at (-1.1,-2.5) {};
    \node (v2) at (-1.1,-1.3) {};
    \node (v3) at (-1,0) {};
    \node (v4) at (1,0) {};
    \node (v5) at (1.1,-1.3) {};
    \node (v6) at (1.1,-2.5) {};
    \node (v7) at (0, 0.4) {};

    \begin{scope}[fill opacity=0.8]

    \filldraw[fill=blue!30, shift={(-1.1, -1.1)}, rotate=90] ($(0, 0)$) ellipse (1.7 and 0.75);
    \filldraw[fill=red!30, shift={(1.1, -1.1)}, rotate=270] ($(0, 0)$) ellipse (1.7 and 0.75);
    \filldraw[fill=brown!70] ($(0, 0)$) ellipse (1.7 and 0.75);
    \end{scope}

    \draw (v1) circle (0.08) node [above] {$A$};
    \fill (v3) circle (0.08) node [left] {$B$};
    \fill (v4) circle (0.08) node [right] {$C$};
    \draw (v6) circle (0.08) node [above] {$D$};

    \node at (v2) {$G_1$};
    \node at (0,0.3) {$G_2$};
    \node at (v5) {$G_3$};
\end{tikzpicture}
    	\label{fig:L3}
	}
}
\subfigure[2-Comb Query]{
    \resizebox{0.11\linewidth}{!}{
        \centering
        \begin{tikzpicture}[font = \LARGE]
    \node (v1) at (-1.1,-2.5) {};
    \node (v2) at (-1.1,-1.3) {};
    \node (v3) at (-1,0) {};
    \node (v4) at (1,0) {};
    \node (v5) at (1.1,-1.3) {};
    \node (v6) at (1.1,-2.5) {};
    \node (v7) at (0, 0.4) {};

    \begin{scope}[fill opacity=0.8]

    \filldraw[fill=blue!30, shift={(-1.1, -1.1)}, rotate=90] ($(0, 0)$) ellipse (1.7 and 0.75);
    \filldraw[fill=red!30, shift={(1.1, -1.1)}, rotate=270] ($(0, 0)$) ellipse (1.7 and 0.75);
    \filldraw[fill=brown!70] ($(0, 0)$) ellipse (1.7 and 0.75);
    \filldraw[fill=green!45] (v4) circle (0.6);
    \filldraw[fill=orange!60] ($(1.1, -2.2)$) circle (0.6);
    \end{scope}

    \fill (v1) circle (0.08) node [above] {$A$};
    \fill (v3) circle (0.08) node [left] {$B$};
    \fill (v4) circle (0.08) node [right] {$C$};
    \fill (v6) circle (0.08) node [above] {$D$};

    \node at (v2) {$G_1$};
    \node at (0,0.3) {$G_2$};
    \node at (v5) {$G_3$};
\end{tikzpicture}
        \label{fig:2Comb}
    }
}
\subfigure[4-Hop Query, SNB Q2]{
    \resizebox{0.16\linewidth}{!}{%
        \centering
        \begin{tikzpicture}[font=\LARGE]
    \node (v1) at (0,-2.2) {};
    \node (v2) at (-0.5,-1.2) {};
    \node (v3) at (-1,0) {};
    \node (v4) at (1,0) {};
    \node (v5) at (2.5,-1.2) {};
    \node (v6) at (3,0) {};
    \node (v7) at (2, -2.2) {};
    \node (v8) at (2, -0.25) {};

    \begin{scope}[fill opacity=0.8]

    \filldraw[fill=blue!30, rotate=120] ($(v2)$) ellipse (1.7 and 0.75);
    \filldraw[fill=red!30, rotate=60] ($(v5)$) ellipse (1.7 and 0.75);
    \filldraw[fill=brown!70] ($(0, 0)$) ellipse (1.7 and 0.75);
    \filldraw[fill=yellow!30] ($(2, 0)$) ellipse (1.7 and 0.75);

    \end{scope}
    \draw (v1) circle (0.08) node [above] {$A$};
    \fill (v3) circle (0.08) node [left] {$B$};
    \fill (v4) circle (0.08) node [left] {$C$};
    \fill (v6) circle (0.08) node [right] {$D$};
    \draw (v7) circle (0.08) node [above] {$E$};

    \node at (v2) {$G_1$};
    \node at (0,0.3) {$G_2$};
    \node at (v5) {$G_4$};
    \node at (2, 0.3) {$G_3$};
\end{tikzpicture}
        \label{fig:L4}
    }
}
\subfigure[Star Query]{
    \resizebox{0.17\linewidth}{!}{%
    	\centering
        \begin{tikzpicture}[font=\LARGE]
    \node (v1) at (2,0) {};
    \node (v3) at (2,-2) {};
    \node (v4) at (0,0) {};
    \node (v5) at (2.5,-1) {};
    \node (v6) at (-2,0) {};
    \node (v7) at (-2, -2) {};

    \begin{scope}[fill opacity=0.8]

    \filldraw[fill=red!30, shift={(1, -1)}, rotate=135] ($(0, 0)$) ellipse (1.7 and 0.75);
    \filldraw[fill=brown!70, shift={(-1, -1)}, rotate=45] ($(0, 0)$) ellipse (1.7 and 0.75);
    \filldraw[fill=blue!30, shift={(1, 0)}] ($(0, 0)$) ellipse (1.7 and 0.75);
    \filldraw[fill=yellow!30, shift={(-1, 0)}] ($(0, 0)$) ellipse (1.7 and 0.75);

    \end{scope}

    % \foreach \v in {1,2,...,8} {
    %     \fill (v\v) circle (0.1);
    % }

    \draw (v1) circle (0.08) node [right] {$B$};
    \draw (v3) circle (0.08) node [left] {$C$};
    \fill (v4) circle (0.08) node [above] {$A$};
    \draw (v6) circle (0.08) node [left] {$D$};
    \draw (v7) circle (0.08) node [right] {$E$};

    \node at (1.4, 0.3) {$G_1$};
    \node at (1, -1.2) {$G_2$};
    \node at (-1, -1.2) {$G_4$};
    \node at (-1.4, 0.3) {$G_3$};
\end{tikzpicture}
    	\label{fig:Star}
	}
}
\subfigure[SNB Q3]{
    \resizebox{0.20\linewidth}{!}{%
    	\centering
        \begin{tikzpicture}[font=\LARGE]
    \node (v1) at (-2.5,-2) {};
    \node (v2) at (-2,-1) {};
    \node (v3) at (-1.2,0) {};
    \node (v4) at (1.2,0) {};
    \node (v5) at (2,-1) {};
    \node (v6) at (2.5,-2) {};
    \node (v7) at (-0.5, -2) {};
    \node (v8) at (2, -0.25) {};

    \begin{scope}[fill opacity=0.8]

    \filldraw[fill=blue!30, rotate=60] ($(v2)$) ellipse (1.7 and 0.75);
    \filldraw[fill=red!30, rotate=120] ($(v5)$) ellipse (1.7 and 0.75);
    \filldraw[fill=yellow!30, shift={(2,-2)}, rotate=180] ($(0, 0)$) ellipse (1.2 and 0.5); 
    \filldraw[fill=orange!60, shift={(-1.5,-2)}, rotate=180] ($(0, 0)$) ellipse (1.5 and 0.75);    
    \filldraw[fill=brown!70] ($(0, 0)$) ellipse (1.7 and 0.75);
    \end{scope}

    \fill (v1) circle (0.08) node [above] {$id_2$};
    \fill (v3) circle (0.08) node [right] {$id_1$};
    \fill (v7) circle (0.08) node [left] {$id_3$};
    \fill (v4) circle (0.08) node [left] {$M_{id}$};
    \fill (v6) circle (0.08) node [left] {$T_{id}$};

    \node at (v2) {$K_1$};
    \node at (-1.4, -2.3) {$K_2$};
    \node at (0,0.3) {$M$};
    \node at (v5) {$MT$};
    \node at (1.4, -2) {$T$};
\end{tikzpicture}
    	\label{fig:SNBQ3}
	}
}
\subfigure[SNB Q4]{
    \resizebox{0.16\linewidth}{!}{%
        \centering
        \begin{tikzpicture}[font=\LARGE]
    \node (v1) at (0,-2.2) {};
    \node (v2) at (-0.5,-1.2) {};
    \node (v3) at (-1,0) {};
    \node (v4) at (1,0) {};
    \node (v5) at (2.5,-1.2) {};
    \node (v6) at (3,0) {};
    \node (v7) at (2, -2.2) {};
    \node (v8) at (2, -0.25) {};

    \begin{scope}[fill opacity=0.8]

    \filldraw[fill=blue!30, rotate=120] ($(v2)$) ellipse (1.7 and 0.75);
    \filldraw[fill=red!30, rotate=60] ($(v5)$) ellipse (1.7 and 0.75);
    \filldraw[fill=brown!70] ($(0, 0)$) ellipse (1.7 and 0.75);
    \filldraw[fill=yellow!30] ($(2, 0)$) ellipse (1.7 and 0.75);

    \end{scope}

    % \foreach \v in {1,2,...,8} {
    %     \fill (v\v) circle (0.1);
    % }

    \draw (v1) circle (0.08) node [above] {$id_2$};
    \draw (v3) circle (0.08) node [below] {$id_1$};
    \draw (v4) circle (0.08) node [below] {$M_{id}$};
    \fill (v6) circle (0.08) node [below] {$T_{id}$};
    \fill (v7) circle (0.08) node [above] {$name$};

    \node at (v2) {$K$};
    \node at (0,0.3) {$M$};
    \node at (v5) {$T$};
    \node at (2.1,0.3) {$MT$};
\end{tikzpicture}
        \label{fig:SNBQ4}
    }
}
\caption{The relational hypergraphs of queries. The solid dots are output attributes for \qichen{join-project} and aggregation queries.}
\label{fig:expqueries}
\end{figure*}

 \begin{figure*}
    \includegraphics[width=\linewidth]{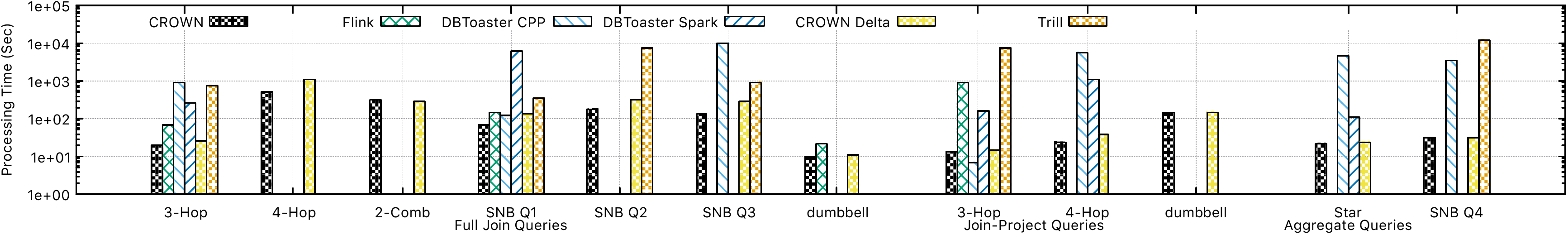}
    \caption{Processing times of CROWN, Flink, DBToaster, and Trill}
    \label{fig:runningtime}
 \end{figure*}
 
 \begin{figure*}[t]
\centering
\subfigure[SNB Q1]{
    \resizebox{0.32\linewidth}{!}{
        \centering
        \includegraphics[width=\linewidth]{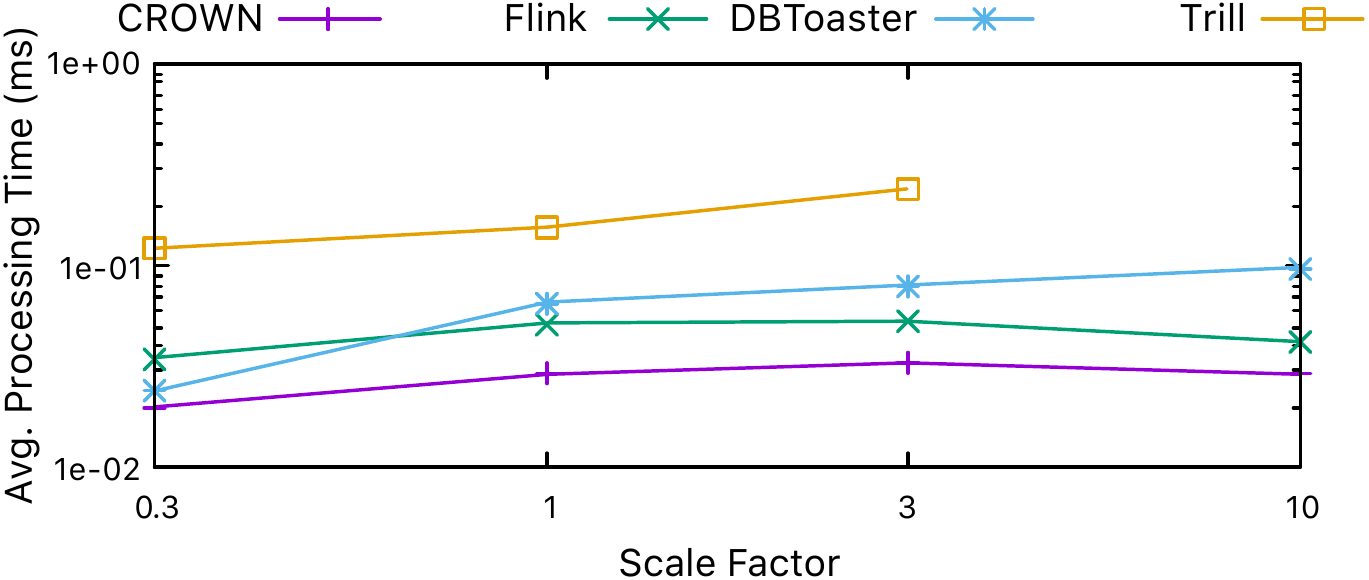}
        \label{fig:ScaleSNBQ1}
    }
} 
\hfill
\subfigure[SNB Q2]{
    \resizebox{0.32\linewidth}{!}{
        \centering
        \includegraphics[width=\linewidth]{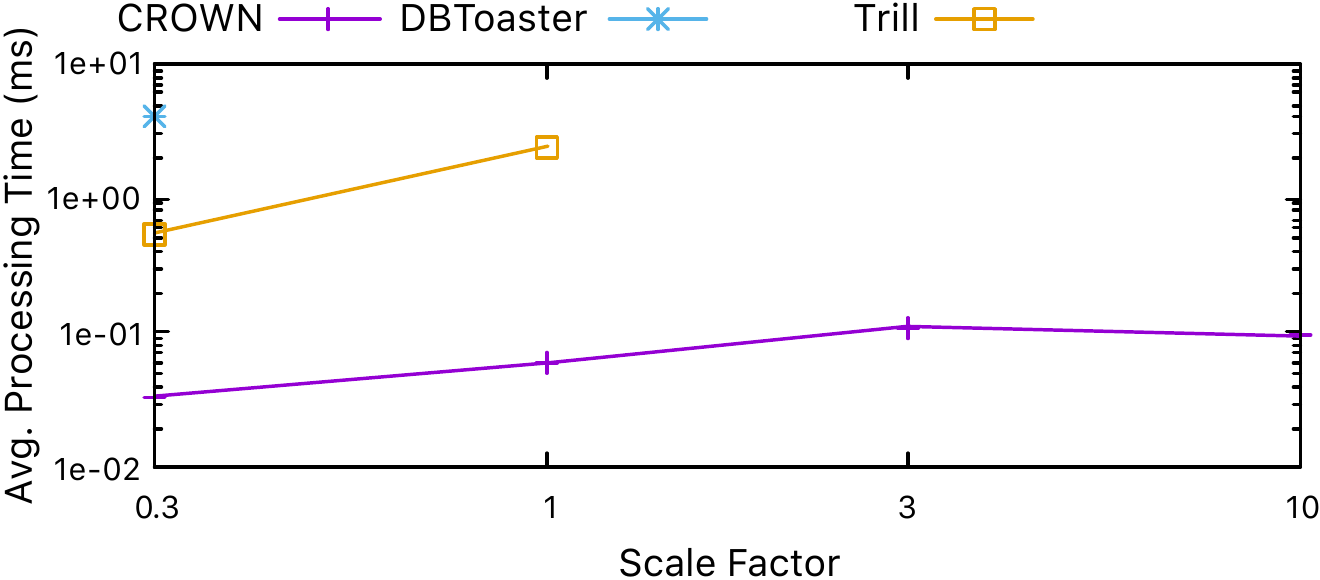}
        \label{fig:ScaleSNBQ2}
    }
}
\hfill
\subfigure[SNB Q4]{
    \resizebox{0.32\linewidth}{!}{
        \centering
        \includegraphics[width=\linewidth]{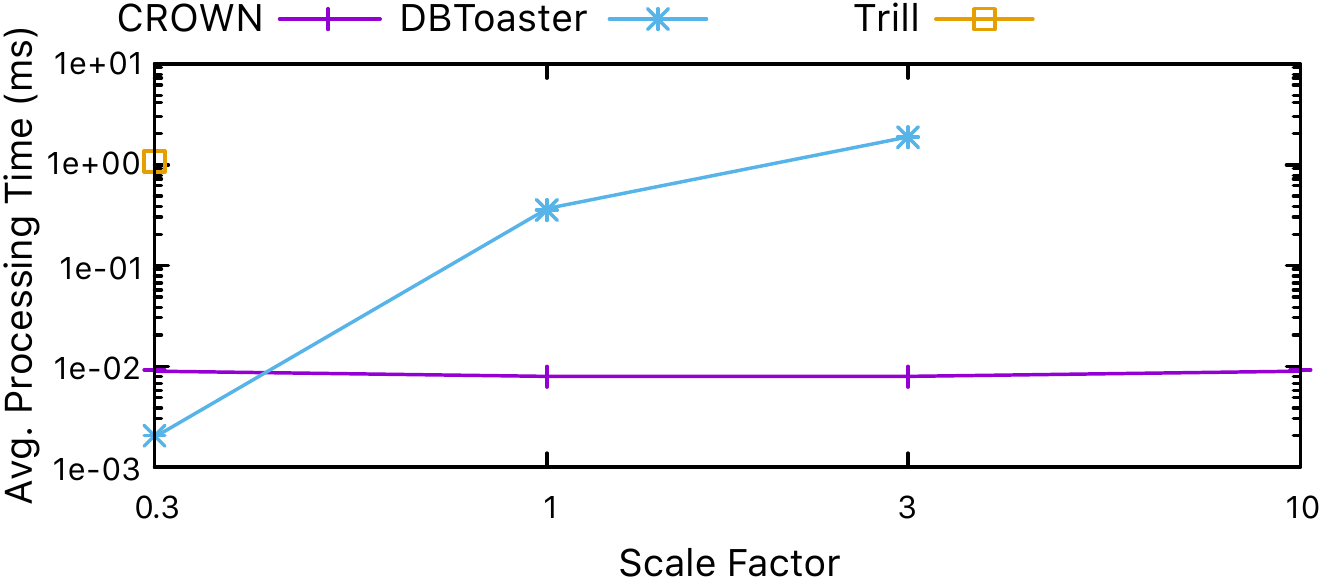}
        \label{fig:ScaleSNBQ4}
    }
}
\caption{Average Processing Time v.s. Scale Factor}
\label{fig:Scalability}
\end{figure*}

\begin{figure*}
\begin{minipage}{0.33\linewidth}
    \centering
    \includegraphics[width=\linewidth]{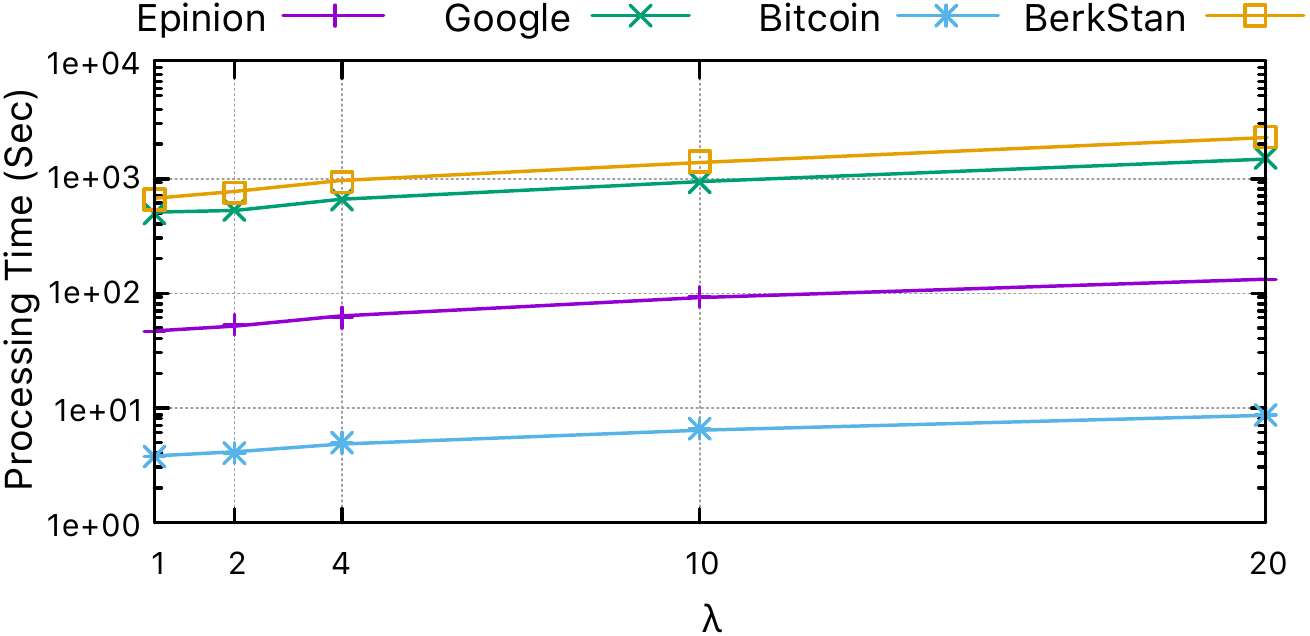}
    \caption{\small{Runtime v.s. enclosureness $\lambda$.}}
\label{fig:lambda}
\end{minipage}
\begin{minipage}{0.33\linewidth}
    \centering
    \includegraphics[width=\linewidth]{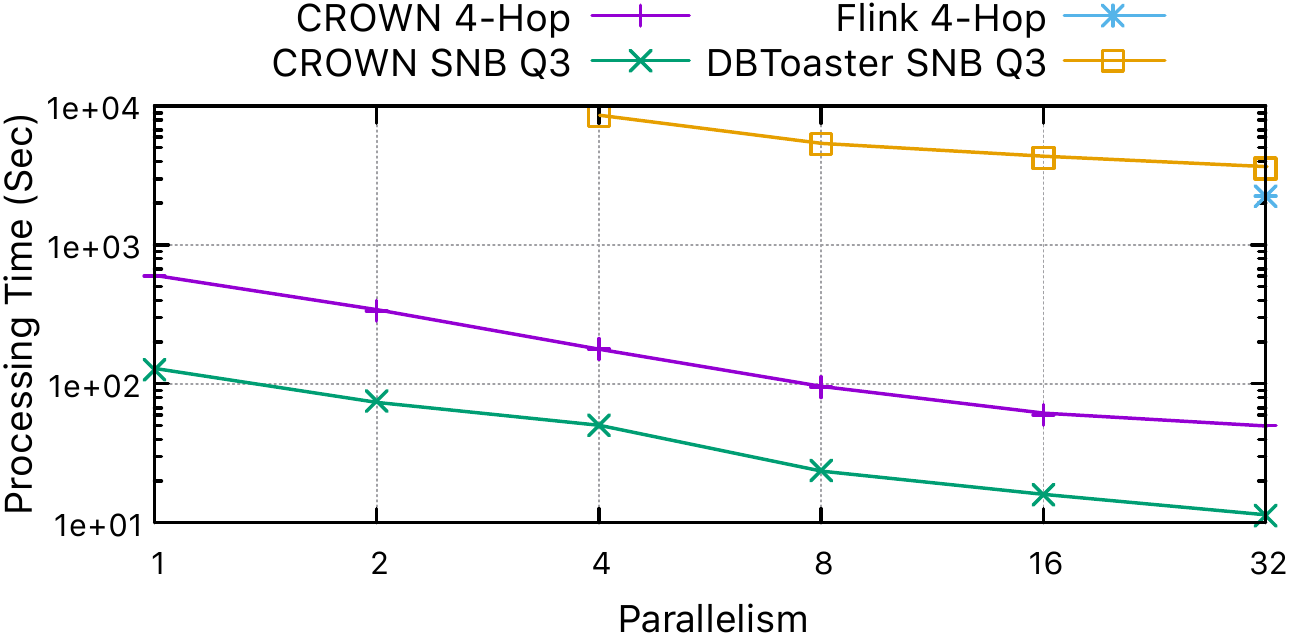}
    \caption{\small{Runtime v.s. parallelism $p$.}}
    \label{fig:scaleup}
\end{minipage}
\begin{minipage}{0.33\linewidth}
    \centering
    \includegraphics[width=\linewidth]{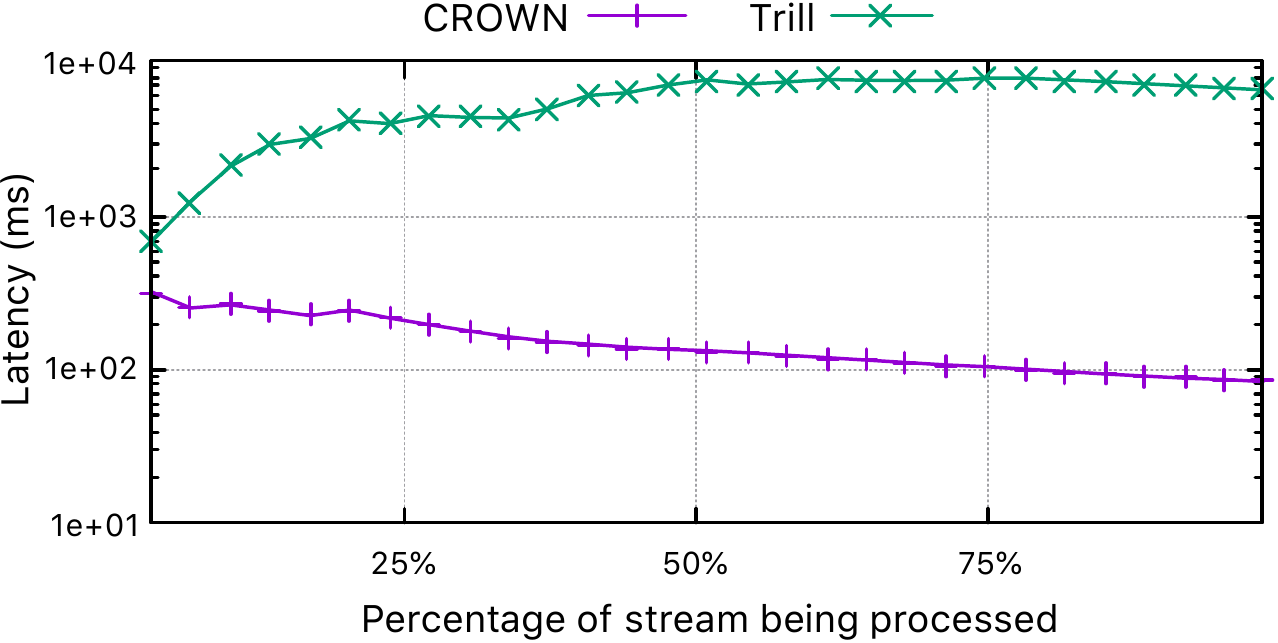}
    \caption{\small{Average latency.}}
    \label{fig:latency}
\end{minipage}
\end{figure*}

Table~\ref{tbl:system} summarizes various features of these systems. Note that only CROWN supports both full enumeration and delta enumeration.   Flink can support insertion-only update streams or window streams, but not arbitrary update streams.  We run every experiment twice: one for delta enumeration, and the other for full enumeration.  For full enumeration, we request the full query results after processing every 10\% of the update sequence.  As Trill does not support full enumeration, we ask Trill to report the entire delta stream for full enumeration.   

\smallskip \noindent {\bf Queries and updates.}  \qichen{We evaluate all systems over two classes of queries.  The first class contains graph pattern queries from the benchmark by Nguyen et al. \cite{nguyen2015join}, over the SNAP dataset (Stanford Network Analysis Project) \cite{snapnets}. Such a benchmark evaluates the performance of each system for join queries over static data, and we modify it to adapt to the dynamic scenario.   We test all acyclic queries from the benchmark, such as hop (path) queries, star queries and comb queries.  We also test the dumbbell query, which is a variant of the lollipop query.  The detailed query definition is given in the Appendix~\ref{appendix:queries} and }
%We evaluate all systems over two classes of queries. The first class contains graph pattern queries over the SNAP dataset (Stanford Network Analysis Project) \cite{snapnets}, such as 3-Hop, 4-Hop, and star.
%We select some real graphs from SNAP (Stanford Network Analysis Project) \cite{snapnets} and find the 3-Hop, 4-Hop, and star patterns from the graph.  
one example of the 3-Hop query is given below, where we use a filter over to control the output size.
	\lstset{upquote=true}
	\lstdefinestyle{mystyle}{
		commentstyle=\color{green},
		keywordstyle=\color{purple},
		stringstyle=\color{purple},
		basicstyle=\small\ttfamily,
		breaklines=true,
		columns=fullflexible,
		frame=single,
	}
	\lstset{style=mystyle}
\begin{lstlisting}[ language=SQL,
	deletekeywords={IDENTITY},
	deletekeywords={[2]INT},
	morekeywords={clustered},
	mathescape=true,
	xleftmargin=-1pt,
	framexleftmargin=-1pt,
	frame=tb,
	framerule=0pt ]
  SELECT G1.src as A, G2.src as B, G3.src as C, G3.dst as D
  FROM G G1, G G2, G G3
  WHERE G1.dst = G2.src AND G2.dst = G3.src
  AND FILTER OVER (G3.dst)
\end{lstlisting}

\qichen{The second class includes more complex analytical queries over the LDBC Social Network Benchmark (LDBC-SNB) \cite{LDBCSNB}}, which accesses the neighborhood of a given node in the graph with continuous updates. %sWe select four queries from the benchmark and extract the join structures.  
The following shows one example, which finds the number of distinct messages associated with a particular tag ID, while satisfying the filter conditions:
	\lstset{upquote=true}
	\lstdefinestyle{mystyle}{
		commentstyle=\color{green},
		keywordstyle=\color{purple},
		stringstyle=\color{purple},
		basicstyle=\small\ttfamily,
		breaklines=true,
		columns=fullflexible,
		frame=single,
	}
	\lstset{style=mystyle}
	\begin{lstlisting}[ language=SQL,
	deletekeywords={IDENTITY},
	deletekeywords={[2]INT},
	morekeywords={clustered},
	mathescape=true,
	xleftmargin=0pt,
	framexleftmargin=0pt,
	frame=tb,
	framerule=0pt]
  SELECT t_name, t_tagid, COUNT(DISTINCT m_messageid)
  FROM tag, message, message_tag, knows
  WHERE m_messageid = mt_messageid AND mt_tagid = t_tagid 
    AND m_creatorid = k_person2id AND m_c_replyof IS NULL 
    AND FILTER OVER (k_person1id) 
  GROUP BY t_name, t_tag_ids
\end{lstlisting}

%which tries to find the number of distinct messages associated with a particular tag ID.  Such messages should be posted by a person that has a friend who satisfies the given filter condition.
Figure~\ref{fig:expqueries} shows the join hypergraphs of all queries. %(that of the dumbbell query is in Figure~\ref{fig:dumbbell}).
Except for 2-Comb, SNB Q3 and Q4, they have a height-2 free-connex join tree. The star query (figure~\ref{fig:Star}) has a height-1 free-connex join tree, so it is q-hierarchical. The 4-Hop query (figure~\ref{fig:L4}) and SNB Q4 query (figure~\ref{fig:SNBQ4}) have the same hypergraph structure but different output attributes, and the 4-Hop query has a height-2 free-connex join trees while SNB Q4 query does not. 

We create FIFO streams with a parameter $w$.  For graph queries, we assign a distinct integer $t_e$ to each edge $e$ in the graph, where $e$ has its lifespan $[t_e, t_e+w]$. For LDBC-SNB queries, each tuple $t$ in the benchmark already has an insertion timestamp $t^+$, and we set its deletion time as $w$ days after its insertion, i.e., $t^- = t^+ + w$.
%in the meantime, we only query the records within $w$ days, i.e., each record will be removed $w$ days after the insertion.  
Note that the sliding window for graph queries is count-based, i.e., the window always contains the same number of tuples.  On the other hand, the window for LDBC-SNB queries is time-based, so the number of tuples in a window fluctuates over time.

% For each graph, we create multiple FIFO streams.  We assign a number $t$ for each edge in the graph as the timestamp.  The edge will be inserted in time $t$ and deleted in time $t+w$, where $w$ is a pre-defined window size.  During the experiments, we will test multiple $w$, from $0.01N$ to $N$, where $N$ represents the input size.  

% For each query, we create two update streams.  One is an arbitrary update stream.  Each tuple in the LDBC-SNB benchmark contains two timestamps, one is for the insertion time, and the other one is the deletion time.  Each record will be inserted at the insertion time and deleted at the deletion time.  As the deletion time of each record might not follow the first-in-first-out rule, the update stream is an arbitrary update stream.  

% Another type of update stream is the FIFO stream.  Each record will be inserted at the insertion time; however, we only query the records within $w$ days, i.e., each record will be removed $w$ days after the insertion.  Such update streams do not have a fixed window size.  However, it has a maximum window size of $w$ times the maximum number of updates allowed per day.   

\subsection{Experiment Results}

{\bf Runtime.} Figure~\ref{fig:runningtime} shows the total runtime of evaluating each graph query over a mid-sized graph {\em Epinions} and each SNB query in the centralized setting.  \qichen{The graph contains approximately 500K edges and 76K vertices, as well as 3.7B 3-Hop paths and 378B 4-Hop paths.  On the other hand, we use the default scale factor of $1$ for all SNB queries.  Under the scale factor, the total size of raw data is $1.5GB$, and the largest relation contains 15 attributes.} %(Note that the y-axis uses a logarithmic scale).  
We set a filter condition that only keeps $10\%$ of the designated endpoints for all queries.  A missing bar in the figure indicates that the corresponding system did not finish within the 4-hour limit or aborted with an error (mostly out-of-memory errors and garbage collection timeout).  Only CROWN can finish all queries successfully.  Trill only handles a few graph queries.  One possible explanation is that graph queries tend to generate a large number of deltas.  On the other hand, Flink ran out of memory when evaluating SNB Q2, Q3, and Q4.  For those queries where the systems can finish, we see that CROWN provides a speedup from 2x to 67x compared with Flink, 1.8x to 234x compared with DBToaster, and 2.7x to 523x compared with Trill.  Moreover, in handling \qichen{join-project queries}, CROWN requires much less time than handling \qichen{the corresponding full join queries}, while Flink requires more time.  In addition, CROWN performs well for both full and delta enumeration, and different modes of output do not affect the overall performance of CROWN.  

\smallskip 
\noindent {\bf Enclosureness.} To test the influences of enclosureness, we create multiple update sequences with different $\lambda$, over different graphs from the SNAP dataset.  We disable the output to see how the update cost would change with different $\lambda$.  The experiment results are shown in Figure~\ref{fig:lambda}.  From the results, we can see the maintenance cost of CROWN increases almost linear as $\lambda$ increases.   

\smallskip 
\noindent {\bf Distributed processing.}  To compare CROWN with DBToaster Spark and Flink in a distributed setting, we built a small cluster with 32 task slots, %and compared the performance amount DBToaster Spark, Flink, and CROWN. 
and tested 4-Hop as well as SNB Q3 query, on which DBToaster and Flink cannot finish in a centralized setting.  Figure~\ref{fig:scaleup} shows the results; missing data points or lines indicate the system cannot finish within the time limit.  Although we adopt the HyperCube algorithm to dispatch all tuples, CROWN can still obtain linear speedup with $p < 16$, where $p$ is the number of workers.  When more workers are available, the margin gain becomes smaller.  This is as expected, since (1) speedup becomes sublinear when adding more workers implied by HyperCube; (2) the processing time is already short, causing the system's overhead to dominate the entire runtime. For all finished data points, CROWN can provide a speedup from 45x to 324x.  

As Flink and DBToaster cannot finish all experiments with 128GB memory, so we increase the memory usage for these two systems to 500GB, where these two systems still only complete a tiny portion of the experiments. On the other hand, CROWN can finish all experiments with only 128GB of memory. If we further limit the memory usage of CROWN to 16GB, i.e., 500MB per worker, CROWN still works well without much change in its performance. %\xiao{Double check: no figure shown for memory usage?}

%\paragraph{Memory Efficiency and Latency} In the distributed setting, Flink and DBToaster require more than 500GB of memory but only can complete partial experiments; on the other hand, CROWN can finish all experiments with only 128GB memory. If we further limit the memory usage of CROWN to 16GB, i.e., 500MB per worker, CROWN still work well without much change in its performance. 

\smallskip 
\noindent {\bf Latency.}
Finally, we tested the latency of delta enumeration, i.e., the time between an update being received and its deltas being outputted. Figure~\ref{fig:latency} shows the result.  The average latency of CROWN is less than 90ms, while that of Trill is more than 6s.   In addition, the average latency is stable for CROWN, but it keeps growing for Trill, making it infeasible to process streams for long periods.

\begin{figure}[h]
\centering
\subfigure[3-Hop query]{
    \resizebox{\linewidth}{!}{
        \centering
        \includegraphics[width=\linewidth]{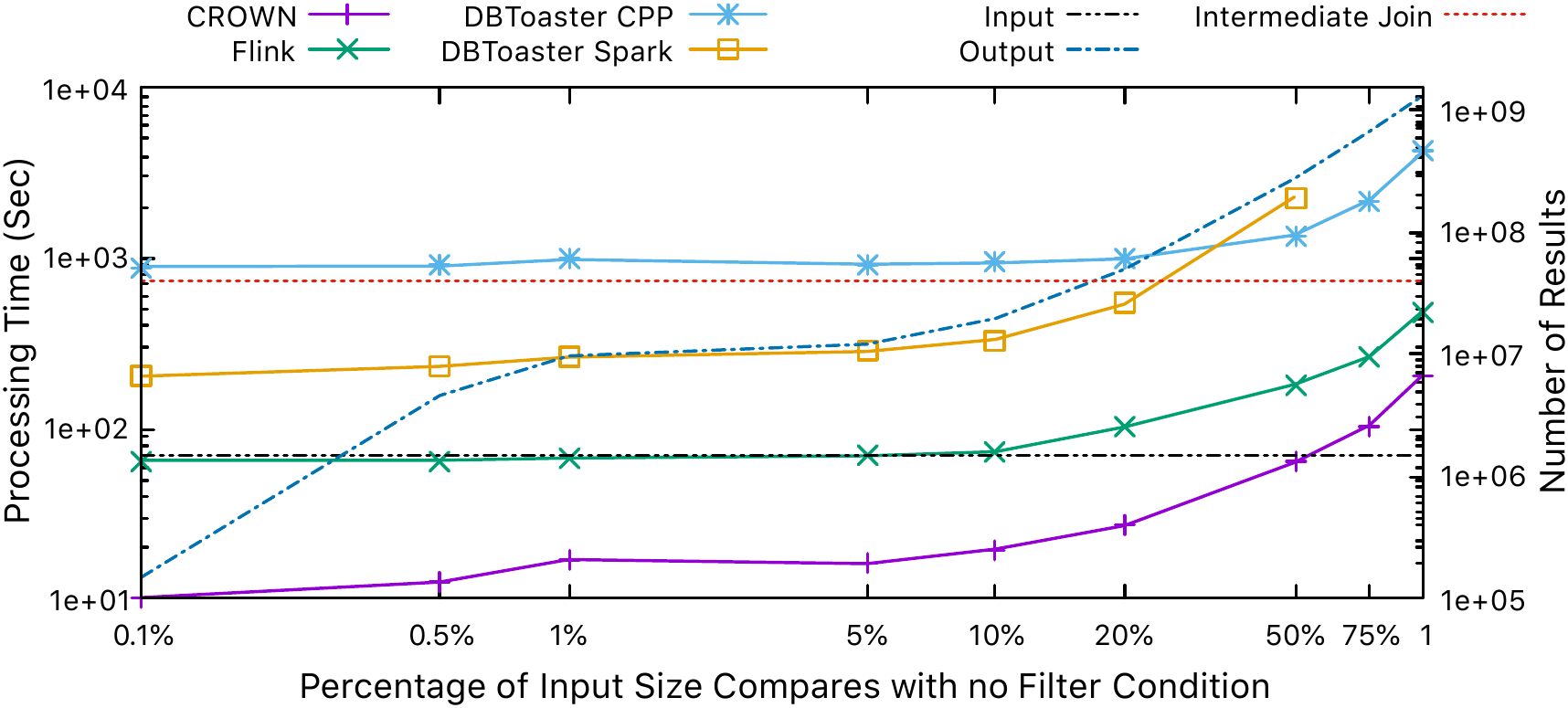}
        \label{fig:selectivityL3}
    }
} \\
\subfigure[4-Hop query]{
    \resizebox{\linewidth}{!}{
        \centering
        \includegraphics[width=\linewidth]{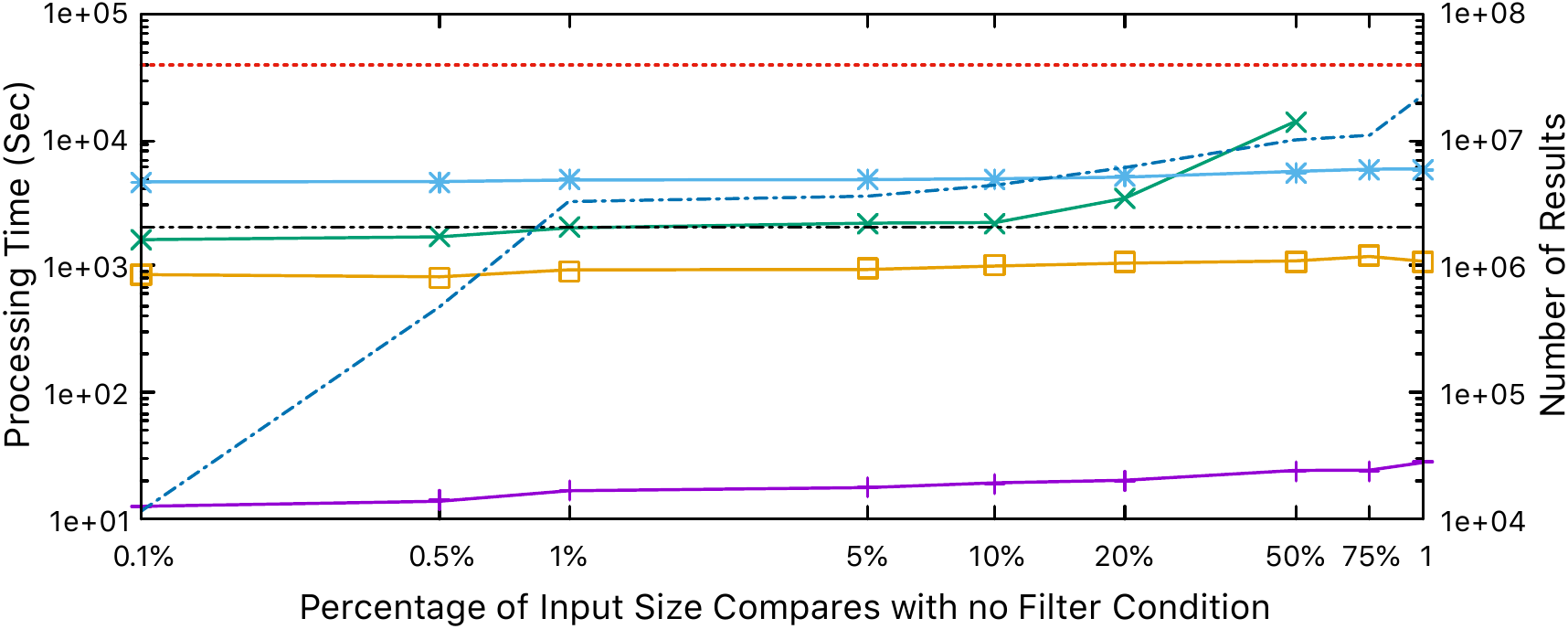}
        \label{fig:selectivityL4}
    }
}
\caption{Runtime v.s. selectivity}
\end{figure}

\qichen{
\smallskip 
\noindent {\bf Scalability.} To test the scalability of different platforms, we change the scale factor of the SNB benchmark and compare the average update cost between different platforms.  The experiment results are shown in Figure~\ref{fig:Scalability}.  The results show that the average processing time of CROWN is stable under different data sizes.  In contrast, the data size will affect the average processing time of other platforms, suggesting CROWN has better scalability than the competitors.  
}

\smallskip \noindent {\bf Selectivity.}  Figure~\ref{fig:selectivityL3} shows the runtime when varying selectivity of join conditions. For standard change propagation and HIVM, the maintenance cost depends not only on the input and output size, but also on the size of intermediate views. For the 3-Hop query 
$G_1(A, B) \Join G_2(B, C) \Join G_3(C, D)$ for $G_1 = G_2 = G$ and $G_3 = \textsc{Filter}(G)$, %\xiao{In the SQL example, three relations are logical copies of graph $G$, then why distinguish $R$ and $R'$?}
the maintenance cost will be bounded by the size of the view $G_1 \Join G_2$
%$|R(A, B) \Join R(B, C)|$, 
even when $G_3$ is empty.  In the meantime, the maintenance cost of CROWN only depends on the input and output size.  To better show such a property, we adjust the filter condition in the 3-Hop query, which only changes $|G_3|$ instead of $|G_1 \Join G_2|$.  Trill is omitted here as it exceeded the 4-hour limit for all data points except for the first one.  When $|G_3| \ge 0.5\% |G|$, the output size exceeds the input size; and when $|G_3| \ge 20\%|G|$, the output size exceeds the intermediate join size $|G_1 \Join G_2|$.  From the results, we can see the runtime of CROWN scales almost linearly as $|G| + |Q|$, which is as expected since the update sequence has $\lambda = 1$.   On the other hand, the runtime of the DBToaster and Flink scales proportionally to $|G_1 \Join G_2|+|Q|$, which leads to poor performance when $|G_3| \le 20\%|G|$.   A larger gap can be observed in Figure~\ref{fig:selectivityL4} when evaluating the 4-Hop query with projection, where the intermediate join size %$|G_1 \Join G_2 \Join G_4(C, D)|$ for $G_4 = G$ 
exceeds the size of the query results, even without any filter conditions.  The runtime of Flink and DBToaster on the 4-Hop query exceeds the 3-Hop query, even with a small output size.  Meanwhile, the runtime of CROWN is much smaller, which only depends on the input and output size.

%\clearpage 
\balance
\bibliographystyle{ACM-Reference-Format}
\bibliography{paper}


\providecommand{\noopsort}[1]{}
\begin{thebibliography}{37}

%%% the Web address.

\ifx \showCODEN    \undefined \def \showCODEN     #1{\unskip}     \fi
\ifx \showDOI      \undefined \def \showDOI       #1{#1}\fi
\ifx \showISBNx    \undefined \def \showISBNx     #1{\unskip}     \fi
\ifx \showISBNxiii \undefined \def \showISBNxiii  #1{\unskip}     \fi
\ifx \showISSN     \undefined \def \showISSN      #1{\unskip}     \fi
\ifx \showLCCN     \undefined \def \showLCCN      #1{\unskip}     \fi
\ifx \shownote     \undefined \def \shownote      #1{#1}          \fi
\ifx \showarticletitle \undefined \def \showarticletitle #1{#1}   \fi
\ifx \showURL      \undefined \def \showURL       {\relax}        \fi
% invisible to TeX
\providecommand\bibfield[2]{#2}
\providecommand\bibinfo[2]{#2}
\providecommand\natexlab[1]{#1}
\providecommand\showeprint[2][]{arXiv:#2}

\bibitem[\protect\citeauthoryear{Abo~Khamis, Ngo, and Rudra}{Abo~Khamis
  et~al\mbox{.}}{2016}]%
        {abo2016faq}
\bibfield{author}{\bibinfo{person}{Mahmoud Abo~Khamis}, \bibinfo{person}{Hung~Q
  Ngo}, {and} \bibinfo{person}{Atri Rudra}.} \bibinfo{year}{2016}\natexlab{}.
\newblock \showarticletitle{FAQ: questions asked frequently}. In
  \bibinfo{booktitle}{\emph{Proceedings of the 35th ACM SIGMOD-SIGACT-SIGAI
  Symposium on Principles of Database Systems}}. \bibinfo{pages}{13--28}.
\newblock


\bibitem[\protect\citeauthoryear{Afrati and Ullman}{Afrati and Ullman}{2011}]%
        {afrati11:_optim}
\bibfield{author}{\bibinfo{person}{Foto~N. Afrati} {and}
  \bibinfo{person}{Jeffrey~D. Ullman}.} \bibinfo{year}{2011}\natexlab{}.
\newblock \showarticletitle{Optimizing Multiway Joins in a Map-Reduce
  Environment}.
\newblock \bibinfo{journal}{\emph{IEEE Transactions on Knowledge and Data
  Engineering}} \bibinfo{volume}{23}, \bibinfo{number}{9}
  (\bibinfo{year}{2011}), \bibinfo{pages}{1282--1298}.
\newblock


\bibitem[\protect\citeauthoryear{Ahmad, Kennedy, Koch, and Nikolic}{Ahmad
  et~al\mbox{.}}{2012}]%
        {ahmad2012dbtoaster}
\bibfield{author}{\bibinfo{person}{Yanif Ahmad}, \bibinfo{person}{Oliver
  Kennedy}, \bibinfo{person}{Christoph Koch}, {and} \bibinfo{person}{Milos
  Nikolic}.} \bibinfo{year}{2012}\natexlab{}.
\newblock \showarticletitle{{DBToaster}: Higher-order delta processing for
  dynamic, frequently fresh views}.
\newblock \bibinfo{journal}{\emph{Proceedings of the VLDB Endowment}}
  \bibinfo{volume}{5}, \bibinfo{number}{10} (\bibinfo{year}{2012}),
  \bibinfo{pages}{968--979}.
\newblock


\bibitem[\protect\citeauthoryear{Atserias, Grohe, and Marx}{Atserias
  et~al\mbox{.}}{2013}]%
        {atserias2013size}
\bibfield{author}{\bibinfo{person}{Albert Atserias}, \bibinfo{person}{Martin
  Grohe}, {and} \bibinfo{person}{D{\'a}niel Marx}.}
  \bibinfo{year}{2013}\natexlab{}.
\newblock \showarticletitle{Size bounds and query plans for relational joins}.
\newblock \bibinfo{journal}{\emph{SIAM J. Comput.}} \bibinfo{volume}{42},
  \bibinfo{number}{4} (\bibinfo{year}{2013}), \bibinfo{pages}{1737--1767}.
\newblock


\bibitem[\protect\citeauthoryear{Bagan, Durand, and Grandjean}{Bagan
  et~al\mbox{.}}{2007}]%
        {bagan2007acyclic}
\bibfield{author}{\bibinfo{person}{Guillaume Bagan}, \bibinfo{person}{Arnaud
  Durand}, {and} \bibinfo{person}{Etienne Grandjean}.}
  \bibinfo{year}{2007}\natexlab{}.
\newblock \showarticletitle{On Acyclic Conjunctive Queries and Constant Delay
  Enumeration}. In \bibinfo{booktitle}{\emph{Computer Science Logic}}.
  \bibinfo{publisher}{Springer Berlin Heidelberg}, \bibinfo{address}{Berlin,
  Heidelberg}, \bibinfo{pages}{208--222}.
\newblock


\bibitem[\protect\citeauthoryear{Beame, Koutris, and Suciu}{Beame
  et~al\mbox{.}}{2017}]%
        {beame13:_commun}
\bibfield{author}{\bibinfo{person}{Paul Beame}, \bibinfo{person}{Paraschos
  Koutris}, {and} \bibinfo{person}{Dan Suciu}.}
  \bibinfo{year}{2017}\natexlab{}.
\newblock \showarticletitle{Communication Steps for Parallel Query Processing}.
\newblock \bibinfo{journal}{\emph{J. ACM}} \bibinfo{volume}{64},
  \bibinfo{number}{6}, Article \bibinfo{articleno}{40} (\bibinfo{date}{oct}
  \bibinfo{year}{2017}), \bibinfo{numpages}{58}~pages.
\newblock
\showISSN{0004-5411}
\urldef\tempurl%
\url{https://doi.org/10.1145/3125644}
\showDOI{\tempurl}


\bibitem[\protect\citeauthoryear{Beeri, Fagin, Maier, and Yannakakis}{Beeri
  et~al\mbox{.}}{1983}]%
        {beeri1983desirability}
\bibfield{author}{\bibinfo{person}{C. Beeri}, \bibinfo{person}{R. Fagin},
  \bibinfo{person}{D. Maier}, {and} \bibinfo{person}{M. Yannakakis}.}
  \bibinfo{year}{1983}\natexlab{}.
\newblock \showarticletitle{On the desirability of acyclic database schemes}.
\newblock \bibinfo{journal}{\emph{JACM}} \bibinfo{volume}{30},
  \bibinfo{number}{3} (\bibinfo{year}{1983}), \bibinfo{pages}{479--513}.
\newblock


\bibitem[\protect\citeauthoryear{Berkholz, Keppeler, and Schweikardt}{Berkholz
  et~al\mbox{.}}{2017}]%
        {berkholz17:_answer}
\bibfield{author}{\bibinfo{person}{Christoph Berkholz}, \bibinfo{person}{Jens
  Keppeler}, {and} \bibinfo{person}{Nicole Schweikardt}.}
  \bibinfo{year}{2017}\natexlab{}.
\newblock \showarticletitle{Answering Conjunctive Queries under Updates}. In
  \bibinfo{booktitle}{\emph{Proceedings of the 36th ACM SIGMOD-SIGACT-SIGAI
  Symposium on Principles of Database Systems}} (Chicago, Illinois, USA)
  \emph{(\bibinfo{series}{PODS '17})}. \bibinfo{publisher}{Association for
  Computing Machinery}, \bibinfo{address}{New York, NY, USA},
  \bibinfo{pages}{303–318}.
\newblock
\showISBNx{9781450341981}
\urldef\tempurl%
\url{https://doi.org/10.1145/3034786.3034789}
\showDOI{\tempurl}


\bibitem[\protect\citeauthoryear{Carbone, Katsifodimos, Ewen, Markl, Haridi,
  and Tzoumas}{Carbone et~al\mbox{.}}{2015}]%
        {carbone15:_apach_flink}
\bibfield{author}{\bibinfo{person}{Paris Carbone}, \bibinfo{person}{Asterios
  Katsifodimos}, \bibinfo{person}{Stephan Ewen}, \bibinfo{person}{Volker
  Markl}, \bibinfo{person}{Seif Haridi}, {and} \bibinfo{person}{Kostas
  Tzoumas}.} \bibinfo{year}{2015}\natexlab{}.
\newblock \showarticletitle{Apache {Flink}: Stream and Batch Processing in a
  Single Engine}.
\newblock \bibinfo{journal}{\emph{IEEE Data Engineering Bulletin}}
  \bibinfo{volume}{38}, \bibinfo{number}{4} (\bibinfo{year}{2015}),
  \bibinfo{pages}{28--38}.
\newblock


\bibitem[\protect\citeauthoryear{Carmeli and Kr{\"o}ll}{Carmeli and
  Kr{\"o}ll}{2019}]%
        {carmeli2019enumeration}
\bibfield{author}{\bibinfo{person}{Nofar Carmeli} {and} \bibinfo{person}{Markus
  Kr{\"o}ll}.} \bibinfo{year}{2019}\natexlab{}.
\newblock \showarticletitle{On the Enumeration Complexity of Unions of
  Conjunctive Queries}. In \bibinfo{booktitle}{\emph{Proceedings of the 38th
  ACM SIGMOD-SIGACT-SIGAI Symposium on Principles of Database Systems}}. ACM,
  \bibinfo{pages}{134--148}.
\newblock


\bibitem[\protect\citeauthoryear{Chandramouli, Goldstein, Barnett, DeLine,
  Fisher, Platt, Terwilliger, and Wernsing}{Chandramouli et~al\mbox{.}}{2014}]%
        {chandramouli2014trill}
\bibfield{author}{\bibinfo{person}{Badrish Chandramouli},
  \bibinfo{person}{Jonathan Goldstein}, \bibinfo{person}{Mike Barnett},
  \bibinfo{person}{Robert DeLine}, \bibinfo{person}{Danyel Fisher},
  \bibinfo{person}{John~C Platt}, \bibinfo{person}{James~F Terwilliger}, {and}
  \bibinfo{person}{John Wernsing}.} \bibinfo{year}{2014}\natexlab{}.
\newblock \showarticletitle{Trill: A high-performance incremental query
  processor for diverse analytics}.
\newblock \bibinfo{journal}{\emph{Proceedings of the VLDB Endowment}}
  \bibinfo{volume}{8}, \bibinfo{number}{4} (\bibinfo{year}{2014}),
  \bibinfo{pages}{401--412}.
\newblock


\bibitem[\protect\citeauthoryear{Chirkova and Yang}{Chirkova and Yang}{2012}]%
        {chirkova2012materialized}
\bibfield{author}{\bibinfo{person}{Rada Chirkova} {and} \bibinfo{person}{Jun
  Yang}.} \bibinfo{year}{2012}\natexlab{}.
\newblock \showarticletitle{Materialized views}.
\newblock \bibinfo{journal}{\emph{Foundations and Trends{\textregistered} in
  Databases}} \bibinfo{volume}{4}, \bibinfo{number}{4} (\bibinfo{year}{2012}),
  \bibinfo{pages}{295--405}.
\newblock


\bibitem[\protect\citeauthoryear{Elseidy, Elguindy, Vitorovic, and
  Koch}{Elseidy et~al\mbox{.}}{2014}]%
        {elseidy2014scalable}
\bibfield{author}{\bibinfo{person}{Mohammed Elseidy}, \bibinfo{person}{Abdallah
  Elguindy}, \bibinfo{person}{Aleksandar Vitorovic}, {and}
  \bibinfo{person}{Christoph Koch}.} \bibinfo{year}{2014}\natexlab{}.
\newblock \showarticletitle{Scalable and Adaptive Online Joins}.
\newblock \bibinfo{journal}{\emph{Proc. VLDB Endow.}} \bibinfo{volume}{7},
  \bibinfo{number}{6} (\bibinfo{date}{feb} \bibinfo{year}{2014}),
  \bibinfo{pages}{441–452}.
\newblock
\showISSN{2150-8097}
\urldef\tempurl%
\url{https://doi.org/10.14778/2732279.2732281}
\showDOI{\tempurl}


\bibitem[\protect\citeauthoryear{Erling, Averbuch, Larriba-Pey, Chafi,
  Gubichev, Prat, Pham, and Boncz}{Erling et~al\mbox{.}}{2015}]%
        {LDBCSNB}
\bibfield{author}{\bibinfo{person}{Orri Erling}, \bibinfo{person}{Alex
  Averbuch}, \bibinfo{person}{Josep Larriba-Pey}, \bibinfo{person}{Hassan
  Chafi}, \bibinfo{person}{Andrey Gubichev}, \bibinfo{person}{Arnau Prat},
  \bibinfo{person}{Minh-Duc Pham}, {and} \bibinfo{person}{Peter Boncz}.}
  \bibinfo{year}{2015}\natexlab{}.
\newblock \showarticletitle{The LDBC Social Network Benchmark: Interactive
  Workload}. In \bibinfo{booktitle}{\emph{Proceedings of the 2015 ACM SIGMOD
  International Conference on Management of Data}} (Melbourne, Victoria,
  Australia) \emph{(\bibinfo{series}{SIGMOD '15})}.
  \bibinfo{publisher}{Association for Computing Machinery},
  \bibinfo{address}{New York, NY, USA}, \bibinfo{pages}{619–630}.
\newblock
\showISBNx{9781450327589}
\urldef\tempurl%
\url{https://doi.org/10.1145/2723372.2742786}
\showDOI{\tempurl}


\bibitem[\protect\citeauthoryear{Fagin}{Fagin}{1983}]%
        {fagin1983degrees}
\bibfield{author}{\bibinfo{person}{R. Fagin}.} \bibinfo{year}{1983}\natexlab{}.
\newblock \showarticletitle{Degrees of acyclicity for hypergraphs and
  relational database schemes}.
\newblock \bibinfo{journal}{\emph{JACM}} \bibinfo{volume}{30},
  \bibinfo{number}{3} (\bibinfo{year}{1983}), \bibinfo{pages}{514--550}.
\newblock


\bibitem[\protect\citeauthoryear{Gedik, Bordawekar, and Yu}{Gedik
  et~al\mbox{.}}{2009}]%
        {gedik2009celljoin}
\bibfield{author}{\bibinfo{person}{Bu{\u{g}}ra Gedik},
  \bibinfo{person}{Rajesh~R Bordawekar}, {and} \bibinfo{person}{Philip~S Yu}.}
  \bibinfo{year}{2009}\natexlab{}.
\newblock \showarticletitle{CellJoin: a parallel stream join operator for the
  cell processor}.
\newblock \bibinfo{journal}{\emph{The VLDB journal}} \bibinfo{volume}{18},
  \bibinfo{number}{2} (\bibinfo{year}{2009}), \bibinfo{pages}{501--519}.
\newblock


\bibitem[\protect\citeauthoryear{Gottlob, Leone, and Scarcello}{Gottlob
  et~al\mbox{.}}{2002}]%
        {gottlob2002hypertree}
\bibfield{author}{\bibinfo{person}{Georg Gottlob}, \bibinfo{person}{Nicola
  Leone}, {and} \bibinfo{person}{Francesco Scarcello}.}
  \bibinfo{year}{2002}\natexlab{}.
\newblock \showarticletitle{Hypertree decompositions and tractable queries}.
\newblock \bibinfo{journal}{\emph{J. Comput. System Sci.}}
  \bibinfo{volume}{64}, \bibinfo{number}{3} (\bibinfo{year}{2002}),
  \bibinfo{pages}{579--627}.
\newblock


\bibitem[\protect\citeauthoryear{Griffin and Kumar}{Griffin and Kumar}{1998}]%
        {algebraic1998griffin}
\bibfield{author}{\bibinfo{person}{Timothy Griffin} {and}
  \bibinfo{person}{Bharat Kumar}.} \bibinfo{year}{1998}\natexlab{}.
\newblock \showarticletitle{Algebraic Change Propagation for Semijoin and
  Outerjoin Queries}.
\newblock \bibinfo{journal}{\emph{SIGMOD Rec.}} \bibinfo{volume}{27},
  \bibinfo{number}{3} (\bibinfo{date}{Sept.} \bibinfo{year}{1998}),
  \bibinfo{pages}{22–27}.
\newblock
\showISSN{0163-5808}
\urldef\tempurl%
\url{https://doi.org/10.1145/290593.290597}
\showDOI{\tempurl}


\bibitem[\protect\citeauthoryear{Henzinger, Krinninger, Nanongkai, and
  Saranurak}{Henzinger et~al\mbox{.}}{2015}]%
        {henzinger2015unifying}
\bibfield{author}{\bibinfo{person}{Monika Henzinger},
  \bibinfo{person}{Sebastian Krinninger}, \bibinfo{person}{Danupon Nanongkai},
  {and} \bibinfo{person}{Thatchaphol Saranurak}.}
  \bibinfo{year}{2015}\natexlab{}.
\newblock \showarticletitle{Unifying and Strengthening Hardness for Dynamic
  Problems via the Online Matrix-Vector Multiplication Conjecture}. In
  \bibinfo{booktitle}{\emph{Proceedings of the Forty-Seventh Annual ACM
  Symposium on Theory of Computing}} (Portland, Oregon, USA)
  \emph{(\bibinfo{series}{STOC '15})}. \bibinfo{publisher}{Association for
  Computing Machinery}, \bibinfo{address}{New York, NY, USA},
  \bibinfo{pages}{21–30}.
\newblock
\showISBNx{9781450335362}
\urldef\tempurl%
\url{https://doi.org/10.1145/2746539.2746609}
\showDOI{\tempurl}


\bibitem[\protect\citeauthoryear{Idris, Ugarte, and Vansummeren}{Idris
  et~al\mbox{.}}{2017}]%
        {idris17:_dynam}
\bibfield{author}{\bibinfo{person}{Muhammad Idris}, \bibinfo{person}{Martin
  Ugarte}, {and} \bibinfo{person}{Stijn Vansummeren}.}
  \bibinfo{year}{2017}\natexlab{}.
\newblock \showarticletitle{The Dynamic Yannakakis Algorithm: Compact and
  Efficient Query Processing Under Updates}. In
  \bibinfo{booktitle}{\emph{Proceedings of the 2017 ACM International
  Conference on Management of Data}} (Chicago, Illinois, USA)
  \emph{(\bibinfo{series}{SIGMOD '17})}. \bibinfo{publisher}{Association for
  Computing Machinery}, \bibinfo{address}{New York, NY, USA},
  \bibinfo{pages}{1259–1274}.
\newblock
\showISBNx{9781450341974}
\urldef\tempurl%
\url{https://doi.org/10.1145/3035918.3064027}
\showDOI{\tempurl}


\bibitem[\protect\citeauthoryear{Idris, Ugarte, Vansummeren, Voigt, and
  Lehner}{Idris et~al\mbox{.}}{2019}]%
        {idris2019efficient}
\bibfield{author}{\bibinfo{person}{Muhammad Idris},
  \bibinfo{person}{Mart{\'\i}n Ugarte}, \bibinfo{person}{Stijn Vansummeren},
  \bibinfo{person}{Hannes Voigt}, {and} \bibinfo{person}{Wolfgang Lehner}.}
  \bibinfo{year}{2019}\natexlab{}.
\newblock \showarticletitle{Efficient query processing for dynamically changing
  datasets}.
\newblock \bibinfo{journal}{\emph{ACM SIGMOD Record}} \bibinfo{volume}{48},
  \bibinfo{number}{1} (\bibinfo{year}{2019}), \bibinfo{pages}{33--40}.
\newblock


\bibitem[\protect\citeauthoryear{Idris, Ugarte, Vansummeren, Voigt, and
  Lehner}{Idris et~al\mbox{.}}{2020}]%
        {idris2020general}
\bibfield{author}{\bibinfo{person}{Muhammad Idris},
  \bibinfo{person}{Mart{\'\i}n Ugarte}, \bibinfo{person}{Stijn Vansummeren},
  \bibinfo{person}{Hannes Voigt}, {and} \bibinfo{person}{Wolfgang Lehner}.}
  \bibinfo{year}{2020}\natexlab{}.
\newblock \showarticletitle{General dynamic Yannakakis: conjunctive queries
  with theta joins under updates}.
\newblock \bibinfo{journal}{\emph{The VLDB Journal}} \bibinfo{volume}{29},
  \bibinfo{number}{2} (\bibinfo{year}{2020}), \bibinfo{pages}{619--653}.
\newblock


\bibitem[\protect\citeauthoryear{Joglekar, Puttagunta, and R\'{e}}{Joglekar
  et~al\mbox{.}}{2016}]%
        {joglekar16:_ajar}
\bibfield{author}{\bibinfo{person}{Manas~R. Joglekar}, \bibinfo{person}{Rohan
  Puttagunta}, {and} \bibinfo{person}{Christopher R\'{e}}.}
  \bibinfo{year}{2016}\natexlab{}.
\newblock \showarticletitle{AJAR: Aggregations and Joins over Annotated
  Relations}. In \bibinfo{booktitle}{\emph{Proceedings of the 35th ACM
  SIGMOD-SIGACT-SIGAI Symposium on Principles of Database Systems}} (San
  Francisco, California, USA) \emph{(\bibinfo{series}{PODS '16})}.
  \bibinfo{publisher}{Association for Computing Machinery},
  \bibinfo{address}{New York, NY, USA}, \bibinfo{pages}{91–106}.
\newblock
\showISBNx{9781450341912}
\urldef\tempurl%
\url{https://doi.org/10.1145/2902251.2902293}
\showDOI{\tempurl}


\bibitem[\protect\citeauthoryear{Kang, Naughton, and Viglas}{Kang
  et~al\mbox{.}}{2003}]%
        {kang2003evaluating}
\bibfield{author}{\bibinfo{person}{Jaewoo Kang}, \bibinfo{person}{Jeffrey~F
  Naughton}, {and} \bibinfo{person}{Stratis~D Viglas}.}
  \bibinfo{year}{2003}\natexlab{}.
\newblock \showarticletitle{Evaluating window joins over unbounded streams}. In
  \bibinfo{booktitle}{\emph{Proceedings 19th International Conference on Data
  Engineering (Cat. No. 03CH37405)}}. IEEE, \bibinfo{pages}{341--352}.
\newblock


\bibitem[\protect\citeauthoryear{Kara, Ngo, Nikolic, Olteanu, and Zhang}{Kara
  et~al\mbox{.}}{2020a}]%
        {kara2019icdt}
\bibfield{author}{\bibinfo{person}{Ahmet Kara}, \bibinfo{person}{Hung~Q. Ngo},
  \bibinfo{person}{Milos Nikolic}, \bibinfo{person}{Dan Olteanu}, {and}
  \bibinfo{person}{Haozhe Zhang}.} \bibinfo{year}{2020}\natexlab{a}.
\newblock \showarticletitle{Maintaining Triangle Queries under Updates}.
\newblock \bibinfo{journal}{\emph{ACM Trans. Database Syst.}}
  \bibinfo{volume}{45}, \bibinfo{number}{3}, Article \bibinfo{articleno}{11}
  (\bibinfo{date}{aug} \bibinfo{year}{2020}), \bibinfo{numpages}{46}~pages.
\newblock
\showISSN{0362-5915}
\urldef\tempurl%
\url{https://doi.org/10.1145/3396375}
\showDOI{\tempurl}


\bibitem[\protect\citeauthoryear{Kara, Nikolic, Olteanu, and Zhang}{Kara
  et~al\mbox{.}}{2020b}]%
        {kara2020trade}
\bibfield{author}{\bibinfo{person}{Ahmet Kara}, \bibinfo{person}{Milos
  Nikolic}, \bibinfo{person}{Dan Olteanu}, {and} \bibinfo{person}{Haozhe
  Zhang}.} \bibinfo{year}{2020}\natexlab{b}.
\newblock \showarticletitle{Trade-offs in static and dynamic evaluation of
  hierarchical queries}. In \bibinfo{booktitle}{\emph{Proceedings of the 39th
  ACM SIGMOD-SIGACT-SIGAI Symposium on Principles of Database Systems}}.
  \bibinfo{pages}{375--392}.
\newblock


\bibitem[\protect\citeauthoryear{Lee, Son, and Kim}{Lee et~al\mbox{.}}{2001}]%
        {cikm01:lee}
\bibfield{author}{\bibinfo{person}{Ki~Yong Lee}, \bibinfo{person}{Jin~Hyun
  Son}, {and} \bibinfo{person}{Myoung~Ho Kim}.}
  \bibinfo{year}{2001}\natexlab{}.
\newblock \showarticletitle{Efficient Incremental View Maintenance in Data
  Warehouses}. In \bibinfo{booktitle}{\emph{Proceedings of the Tenth
  International Conference on Information and Knowledge Management}} (Atlanta,
  Georgia, USA) \emph{(\bibinfo{series}{CIKM '01})}.
  \bibinfo{publisher}{Association for Computing Machinery},
  \bibinfo{address}{New York, NY, USA}, \bibinfo{pages}{349–356}.
\newblock
\showISBNx{1581134363}
\urldef\tempurl%
\url{https://doi.org/10.1145/502585.502644}
\showDOI{\tempurl}


\bibitem[\protect\citeauthoryear{Leskovec and Krevl}{Leskovec and
  Krevl}{2014}]%
        {snapnets}
\bibfield{author}{\bibinfo{person}{Jure Leskovec} {and} \bibinfo{person}{Andrej
  Krevl}.} \bibinfo{year}{2014}\natexlab{}.
\newblock \bibinfo{title}{{SNAP Datasets}: {Stanford} Large Network Dataset
  Collection}.
\newblock \bibinfo{howpublished}{\url{http://snap.stanford.edu/data}}.
\newblock


\bibitem[\protect\citeauthoryear{Lin, Ooi, Wang, and Yu}{Lin
  et~al\mbox{.}}{2015}]%
        {lin2015scalable}
\bibfield{author}{\bibinfo{person}{Qian Lin}, \bibinfo{person}{Beng~Chin Ooi},
  \bibinfo{person}{Zhengkui Wang}, {and} \bibinfo{person}{Cui Yu}.}
  \bibinfo{year}{2015}\natexlab{}.
\newblock \showarticletitle{Scalable distributed stream join processing}. In
  \bibinfo{booktitle}{\emph{Proceedings of the 2015 ACM SIGMOD International
  Conference on Management of Data}}. \bibinfo{pages}{811--825}.
\newblock


\bibitem[\protect\citeauthoryear{Nguyen, Aref, Bravenboer, Kollias, Ngo,
  R{\'e}, and Rudra}{Nguyen et~al\mbox{.}}{2015}]%
        {nguyen2015join}
\bibfield{author}{\bibinfo{person}{Dung Nguyen}, \bibinfo{person}{Molham Aref},
  \bibinfo{person}{Martin Bravenboer}, \bibinfo{person}{George Kollias},
  \bibinfo{person}{Hung~Q Ngo}, \bibinfo{person}{Christopher R{\'e}}, {and}
  \bibinfo{person}{Atri Rudra}.} \bibinfo{year}{2015}\natexlab{}.
\newblock \showarticletitle{Join processing for graph patterns: An old dog with
  new tricks}.
\newblock In \bibinfo{booktitle}{\emph{Proceedings of the GRADES'15}}.
  \bibinfo{pages}{1--8}.
\newblock


\bibitem[\protect\citeauthoryear{Nikolic, Dashti, and Koch}{Nikolic
  et~al\mbox{.}}{2016}]%
        {nikolic2016win}
\bibfield{author}{\bibinfo{person}{Milos Nikolic}, \bibinfo{person}{Mohammad
  Dashti}, {and} \bibinfo{person}{Christoph Koch}.}
  \bibinfo{year}{2016}\natexlab{}.
\newblock \showarticletitle{How to win a hot dog eating contest: Distributed
  incremental view maintenance with batch updates}. In
  \bibinfo{booktitle}{\emph{Proc. ACM SIGMOD International Conference on
  Management of Data}}. ACM, \bibinfo{pages}{511--526}.
\newblock


\bibitem[\protect\citeauthoryear{Nikolic and Olteanu}{Nikolic and
  Olteanu}{2018}]%
        {nikolic2018incremental}
\bibfield{author}{\bibinfo{person}{Milos Nikolic} {and} \bibinfo{person}{Dan
  Olteanu}.} \bibinfo{year}{2018}\natexlab{}.
\newblock \showarticletitle{Incremental view maintenance with triple lock
  factorization benefits}. In \bibinfo{booktitle}{\emph{Proc. ACM SIGMOD
  International Conference on Management of Data}}. ACM,
  \bibinfo{pages}{365--380}.
\newblock


\bibitem[\protect\citeauthoryear{Nikolic, Zhang, Kara, and Olteanu}{Nikolic
  et~al\mbox{.}}{2020}]%
        {nikolic2020f}
\bibfield{author}{\bibinfo{person}{Milos Nikolic}, \bibinfo{person}{Haozhe
  Zhang}, \bibinfo{person}{Ahmet Kara}, {and} \bibinfo{person}{Dan Olteanu}.}
  \bibinfo{year}{2020}\natexlab{}.
\newblock \showarticletitle{F-IVM: learning over fast-evolving relational
  data}. In \bibinfo{booktitle}{\emph{Proceedings of the 2020 ACM SIGMOD
  International Conference on Management of Data}}.
  \bibinfo{pages}{2773--2776}.
\newblock


\bibitem[\protect\citeauthoryear{Ross, Srivastava, and Sudarshan}{Ross
  et~al\mbox{.}}{1996}]%
        {ross1996materialized}
\bibfield{author}{\bibinfo{person}{Kenneth~A. Ross}, \bibinfo{person}{Divesh
  Srivastava}, {and} \bibinfo{person}{S. Sudarshan}.}
  \bibinfo{year}{1996}\natexlab{}.
\newblock \showarticletitle{Materialized View Maintenance and Integrity
  Constraint Checking: Trading Space for Time}. In
  \bibinfo{booktitle}{\emph{Proceedings of the 1996 ACM SIGMOD International
  Conference on Management of Data}} (Montreal, Quebec, Canada)
  \emph{(\bibinfo{series}{SIGMOD '96})}. \bibinfo{publisher}{Association for
  Computing Machinery}, \bibinfo{address}{New York, NY, USA},
  \bibinfo{pages}{447–458}.
\newblock
\showISBNx{0897917944}
\urldef\tempurl%
\url{https://doi.org/10.1145/233269.233361}
\showDOI{\tempurl}


\bibitem[\protect\citeauthoryear{Roy, Teubner, and Gemulla}{Roy
  et~al\mbox{.}}{2014}]%
        {roy2014low}
\bibfield{author}{\bibinfo{person}{Pratanu Roy}, \bibinfo{person}{Jens
  Teubner}, {and} \bibinfo{person}{Rainer Gemulla}.}
  \bibinfo{year}{2014}\natexlab{}.
\newblock \showarticletitle{Low-latency handshake join}.
\newblock \bibinfo{journal}{\emph{Proceedings of the VLDB Endowment}}
  \bibinfo{volume}{7}, \bibinfo{number}{9} (\bibinfo{year}{2014}),
  \bibinfo{pages}{709--720}.
\newblock


\bibitem[\protect\citeauthoryear{Wang and Yi}{Wang and Yi}{2020}]%
        {wang2020maintaining}
\bibfield{author}{\bibinfo{person}{Qichen Wang} {and} \bibinfo{person}{Ke Yi}.}
  \bibinfo{year}{2020}\natexlab{}.
\newblock \showarticletitle{Maintaining Acyclic Foreign-Key Joins under
  Updates}. In \bibinfo{booktitle}{\emph{Proceedings of the 2020 ACM SIGMOD
  International Conference on Management of Data}}.
  \bibinfo{pages}{1225--1239}.
\newblock


\bibitem[\protect\citeauthoryear{Yannakakis}{Yannakakis}{1981}]%
        {yannakakis1981algorithms}
\bibfield{author}{\bibinfo{person}{Mihalis Yannakakis}.}
  \bibinfo{year}{1981}\natexlab{}.
\newblock \showarticletitle{Algorithms for acyclic database schemes}. In
  \bibinfo{booktitle}{\emph{Proc. International Conference on Very Large Data
  Bases}}. \bibinfo{pages}{82--94}.
\newblock


\end{thebibliography}

\appendix
\newpage

\section{Missing Proofs in Section~\ref{sec:preliminaries}}
\label{appendix:preliminaries}

\begin{lemma}
\label{lem:generalized-join-tree}
    Given an acyclic CQ $\Q = (\V,\E,\y)$, it is free-connex if and only if it has a free-connex join tree $\T$ as defined in Section~\ref{sec:preliminaries}. 
\end{lemma}

\begin{proof}
    \underline{\textit{If Direction.}} Suppose there is a free-connex join tree $\T$ as defined in Section~\ref{sec:preliminaries}. We simply add a node $r$ containing exactly all output attributes in $\y$ as the root of $\T$, and start the following procedure:
    \begin{itemize}[leftmargin=*]
        \item We visit the root in a bottom-up way. For every node $e$ with $e \in \T_\textsf{con}$, we remove the edge between $e$ and its parent, and then move $e$ together with the current subtree rooted at $e$ as a child of $r$. Let $\T'$ be the resulted tree. 
    \end{itemize}
    It is obvious that every node in $\T'$ corresponds to a relation, or a generalized relation of $\E \cup \{\y\}$. Moreover, implied by the definition of $\T$, for every node $e \in \T_\textsf{con}$ and its parent node $p_e$, we have $e \cap p_e \subseteq \y$. This way, the connect condition of output attribute in $\y$ is preserved since $\y \subseteq r$. Implied by the definition of acyclic CQs in~\cite{idris17:_dynam}, there is a free-connex join tree for $(\V, \E \cup \{\y\}, \y)$, hence $(\V, \E \cup \{\y\}, \y)$ is also acyclic. Implied by the definition of free-connex CQs in~\cite{bagan2007acyclic}, $\Q$ is free-connex.

    \underline{\textit{Only-If Direction.}} Suppose we are given a free-connex CQ $\Q=(\V,\E,\y)$. We next show how to construct a free-connex join tree as defined in Section~\ref{sec:preliminaries}. Let $\E_\y = \{e \cap \y : e \in \E\}$. We start the following helper lemma: 

    \begin{lemma}[\cite{bagan2007acyclic}, Lemma 21]
    \label{lem:free-connex-helper}
    For any free-connex CQ $\Q = (\V, \E, \y)$, there exists a traditional join tree $\T'$ for $(\V, \E \cup \E_\y, \y)$ and a subset of relations $\E_{\textsf{con}} \subseteq \E \cup \E_\y$ such that the corresponding nodes of $\E_{\textsf{con}}$ form a connex subtree of $\T'$, i.e., $\E_{\textsf{con}}$ includes the root of $\T'$, $\y = \bigcup_{e \in \E_{\textsf{con}}} e$ and the subtree is connected.  
    \end{lemma}

    Let $\T'$ be such a traditional join tree for $(\V, \E \cup \E_\y, \y)$ rooted at node $r$. Note that $r \subseteq \y$. We will transform $\T'$ into a free-connex join tree for $\Q$ via the following steps: 
    
    \paragraph{Step 1: Remove all nodes $e$ with $e \cap \y = \emptyset$} We start with one observation. For any pair of $e, e' \in  \E \cup \E_\y$, such that $e \cap \y = \emptyset$ and $e' \cap \y \neq \emptyset$, $e$ cannot be an ancestor of $e'$ in $\T'$. Suppose not, $e' \notin \con$ since $e \notin \con$. Let $x \in e' \cap\y$ be an output attribute in $e'$. Implied by the property of $\con$, there must exist some node $e'' \in \con$ with $x \in e''$. Moreover, $e'' \notin \T_{e}$, implied by the facts that $e\notin \con$ and $\con$ forms a connect subtree. Hence, all nodes lying on the path between $e''$ and $e''$ (including $e$) must contain $x$, implied by the connect property of $x$. This contradicts the fact that $x \notin e$ since $e\cap\y=\emptyset$.  %implied by the connect property of output attribute(s) in $r \cap e'$ for root node $r$. 

    This way, we remove all relations $e \cap \y = \emptyset$ from $\T'$ as follows. For any relation $e \in \E$ with $e \cap \y = \emptyset$ and $p_e \cap \y \neq \emptyset$, we remove the subtree root at $e$ as a whole. From our observation above, all relations residing in the subtree rooted at $e$ do not contain any output attribute. At last, we will put this subtree back as a child of relation $p_e$ if $p_e - \y \neq \emptyset$, or as a child of arbitrary relation if $p_e -\y =\emptyset$. 

    \paragraph{Step 2: Remove all nodes $e' \in \E_\y$ such that $e' \subsetneq e$ for some $e \in \con$} If there exists a pair of nodes $e \in \E_\y, e' \in \con$ such that $e \subsetneq e'$, we can remove $e$ and add each of its children nodes (not including $e'$ if $e'$ is a child of $e$) as a new child node of $e'$. It can be easily checked that the connect property is preserved. The reduced $\con$ is still a valid connex subtree of the updated $\T'$. After this step, we assume no pair of nodes $e \in \E_\y, e' \in \con$ with $e \subsetneq e'$. 
    
    %Let $r$ be the root node of $\T'$. Let $\T'_e$ be the subtree rooted at node $e$.% A node $e \in \con$ is a {\em boundary node} if $e' \notin \con$ holds for some child node $e'$ of $e$.

    \begin{lemma}
    \label{lem:boundary}
        %For any boundary node $e \in \con$ and some child node $e' \notin \con$ , $e'' \cap \y \subseteq e$ holds for any node $e'' \in \T'_{e'}$.
        For any node $e \notin \con$ and its lowest ancestor $e' \in \con$, $e \cap \y \subseteq e'$.
    \end{lemma}

    \begin{proof}
        Suppose not, assume an output attribute $x \in e \cap \y - e'$. Implied by the connect property, no other node in $\T'- \T'_{e'}$ contains $x$. Let $e''$ be the child of $e'$ lying on the path from $e'$ to $e$. Moreover, no other node in $\T'_{e''}$ except belongs to $\con$, since $e'$ is the lower ancestor of $e$ in $\con$. Hence, no node in $\con$ contains $x$, violating the property of $\con$. 
    \end{proof}

    \begin{lemma}
    \label{lem:step-2}
        After step 2, $\E_\y \subseteq \con$.
    \end{lemma}

    \begin{proof}
      Suppose not, assume $e' \in \E_\y - \con$. Let $e \in \con$ be the lowest ancestor of $e'$. Implied by Lemma~\ref{lem:boundary}, $e' \subseteq e$, hence $e'$ will be removed in step 2.
    \end{proof}

    %\paragraph{Step 3: Move all relations $e' \in \E - \con$ with $e' \subseteq \y$ upward} Consider any node $e' \in \E - \con$ with $e' \subseteq \y$. Let $e \in \con$ be the boundary node that is also an ancestor of $e'$. Implied by Lemma~\ref{lem:boundary}, $e' \subseteq e$. We remove the subtree rooted at $e'$ and add it as a child node of $e$. We next show that the connect property is preserved. Due to the connect property of $\T'$ before transformation, $e' \cap p_{e'} = e' \cap p_{e'} \cap \y \subseteq e$, hence this step preserves the connect property.
    
    \paragraph{Step 3: Add a guard for every relation in $\E_\y$} Consider any node $e \in \E - \con$ and its parent node $p_{e}$ with $p_{e} \subseteq \y$ and $p_{e} \neq e \cap \y$. We remove the subtree rooted at $e$ and add it as a child node of $e' \in \con$ with $e' = e \cap \y$ if such a node $e'$ exists. We next show that the connect property is preserved. Due to the connect property of $\T'$ before transformation, $e' \cap p_{e'} = e' \cap p_{e'} \cap \y \subseteq e$, hence this step preserves the connect property. 
    
   \begin{lemma}
   \label{lem:guard}
       After step 3, for any node $e' \in \E_\y$, it has a child node $e \in\E$ such that $e' = e\cap \y$. 
   \end{lemma}
   
   \begin{proof}
       Consider an arbitrary node $e' \in \E_\y$. Note that $e' \in \con$, implied by Lemma~\ref{lem:step-2}. By contradiction, assume that any node $e \in \E$ such that $e' = e \cap \y$ is not the child of $e'$. 
       
       If $e'$ is an ancestor of $e$, let $e''$ be the child of $e'$ lying on the path from $e$ to $e'$. There must be $e' = e' \cap e \subseteq e''$. If $e'' \in \con$, $e' \subseteq e''$, coming to a contradiction of Step 2. If $e'' \notin \con$, then $e' = e' \cap e = e' \cap e'' = e'' \cap \y$ implied by Lemma~\ref{lem:boundary}, $e''$ is such a child node for $e'$, coming to a contradiction.

       Otherwise, $e' = e' \cap e \subseteq p_{e'}$. As $e \in \con$, $p_{e'} \in \con$. Together, $e' \subseteq p_{e'}$, coming to a contradiction of Step 2. 
   \end{proof}
 
   %\paragraph{Step 4: Replace nodes with generalized relations} At last, we start visiting nodes $e \in \E_\y$ in a top-down manner. Consider an arbitrary node $e'$. Note that $e' \subseteq \y$. If $e' \subseteq e$ holds for every child node $e$ of $e'$, we replace $e'$ as a generalized relation $[e']$. Otherwise, it must have some child node $e \in \E$ with $e \cap \y = e'$ after Step 3. We just replace $e'$ with $e$. If $e \notin \con$, we add it to $\con$. For any child node $e'' \in \con$ of $e'$, which is now a child node of $e$, we have $e'' \cap e' = e'' \cap e' \cap \y = e'' \cap e \subseteq \y$.  It can be easily checked that the connect property is preserved and the guard property is established for every generalized relation.

   \paragraph{Step 4: Replace nodes with generalized relations} At last, we start visiting nodes in $\E_\y$ in a top-down manner. Consider an arbitrary node $e'$. Note that $e' \subseteq \y$. From Lemma~\ref{lem:guard}, it must have some child node $e \in \E$ with $e \cap \y = e'$ after Step 3. If $e' \subseteq e$ holds for every child node $e$ of $e'$, and $p_{e'}$ is a generalized relation if $p_{e'}$ exists, we replace $e'$ as a generalized relation $[e']$. Otherwise, we replace $e'$ with $e$. If $e \notin \con$, we add it to $\con$. For any child node $e'' \in \con$ of $e'$, which is now a child node of $e$, we have $e'' \cap e' = e'' \cap e' \cap \y = e'' \cap e \subseteq \y$.  It can be easily checked that the connect property is preserved and the guard property is established for every generalized relation.  Moreover, the parent node of any generalized relation if exists is also a generalized relation. Hence, the above property is preserved. % \update{In addition, the parent node $p_{e'}$ of any non-root generalized relations $[e']$ is also a generalized relation.}

  \smallskip After these four steps, the resulted tree is a free-connex join tree with all properties satisfied.
\end{proof}

\section{Missing proofs in Section~\ref{sec:enumerate}}
\label{appendix:algorithm}

\qichen{

\begin{proof}[Proof of Lemma~\ref{lem:live_con}]
We first prove the ``only if" direction.  For any $t \in V_l(R_e)$, there exists a $t' \in Q(D)$, such that $\pi_{e} t' = t$.  Meanwhile, it indicates that $t'' = \pi_{p_{e}} t'$ must satisfy $t'' \in V_l(R_{p_{e}})$, because $t'' \in \pi_{p_{e}} Q(D)$.  Hence, $t''$ can join $t$, indicates $t \Join V_l(R_{p_{e}}) \neq \emptyset$. 

For the ``if" direction.  Let $t' \in V_l(R_{p_{e}})$ be a tuple that can join with $t$. We divide the join tree $\T$ into two subtrees $(\T_e, \T \setminus \T_e)$ and divide the output attributes $\y$ into two sets $(\y_e, \y \setminus \y_e)$ accordingly.   Because $t' \in V_l(R_{p_{e}})$, $ \pi_{\y \setminus \y_e} Q(D \ltimes t') \neq \emptyset$ and we let $t''$ be one tuple from $ \pi_{\y \setminus \y_e} Q(D \ltimes t')$.  On the other side, since $t \in V_s(R_e)$, there also exists a tuple $t_e$ in $\pi_{\y_e}  (\Join_{e' \in \T_e} R_{e'})$.  $t_e$ can join with $t''$ as $t$ can join with $t'$ and $\pi_{e} t_e = t, \pi_{p_{e}} t'' = t'$.  Hence, $t_e \Join t'' \in Q(D)$, indicates that $t \in V_l(R_e)$.        
\end{proof}

}

\begin{proof}[Proof of Lemma~\ref{lem:witness}]
W.l.o.g, we assume that $t$ is inserted. The case that $t$ is deleted follows the same argument. 

\paragraph{Direction {\em $\supseteq$}}  We show that each result in $Q(D \ltimes t')$ also appears in $\Delta Q(D , t)$, for every witness tuple $t'$ of $t$.
	Wlog, consider a query result $q \in Q(D \ltimes t')$ for some witness tuple $t'$, where either $t' \in R_e$ for $e \subseteq \y$ or $t' \in \pi_\y R_e$ for $e \cap \y - p_{e} \neq \emptyset$.
	
	First, $Q(D \ltimes t') \subseteq Q(D+t)$ since we have $t' \in \Delta \left(\pi_{\y} V_s(R_e)\right)$ and $\pi_{\key(e)} t' \in V_p(R_e)$ after the insertion of $t$. Hence, all results witnessed by $t'$ appear in $Q(D+t)$ after the insertion of $t$, i.e., $q \in Q(D+t)$. 
	We next show $q \notin Q(D)$. Now let's go back to the timestamp before the insertion of $t$. Implied by the definition of witness tuple, $t' \notin \pi_\y V_s(R_e)$ then. We distinguish two more cases.
	\begin{itemize}
	    \item Case 1: $e \subseteq \y$, $q \notin Q(D)$ since $\pi_e q = t'$ but $t' \notin \pi_{e} Q(D)$ before the insertion of $t$.  This further indicates $q \notin Q$. 
	    \item Case 2: $e - \y \neq \emptyset$ and $e \cap \y - p_{e} \neq \emptyset$, $t' \notin \pi_{\y} V_s(R_e)$ before the insertion of $t'$. This way, $t' \notin \pi_{\y} Q(D)$, thus $q \notin Q$.
	\end{itemize}

	Combining the analysis above, we have 
$q \in Q(D+t), \textrm{and } q \notin Q(D)$
	i.e., $q \in \Delta Q(D,t)$. So, $\biguplus_{t': \textrm{a witness of $t$}} Q(D \ltimes t') \subseteq \Delta Q(D \ltimes t)$.
	
	\paragraph{Direction {\em $\subseteq$}} We next show that every result in $\Delta Q(D,t)$ belongs to $Q(D+t)\ltimes t'$ for some witness tuple $t'$ of $t$. Consider an arbitrary query result $q \in Q(D+t) - Q(D)$. 
	
	It suffices to show that there exists at least one node $e \in \T$ such that tuple $t'= \pi_{e} q$ if $e \subseteq \y$, or tuple $t' \in \pi_\y R_{e}$ with $t' = \pi_{e \cap \y} q$ if $e \cap \y - p_{e} \neq \emptyset$,
	 must be a witness.  An important observation is that $t'$ now belongs to $\Delta \pi_{\y}  V_s(R_e)$; otherwise, $q \in Q(D)$, coming to a contradiction. Now consider the highest node $e_1$ such that $t_1 = \pi_{e_1 \cap \y} q$ and $t_1 \in \Delta \left(\pi_{\y}  V_s(R_{e_1})\right)$.  If $e_1$ is the root, $t_1$ must be a witness of $t$, implied by the Definition~\ref{def:witness}.
	Otherwise, $e_1$ is not the root.  Consider $t_2 = \pi_{p_{e_1} \cap \y} q$.  As $t_2 \notin \Delta V_s(R_{e_2})$, $t_2$ must in $V_s(R_{e_2})$ and $t_2 \in \pi_{e_2} Q(D)$ before the insertion of $t$, which indicates the $\pi_{\key(e_1)} t_1 = \pi_{\key(e_1)} t_2 \in V_p(R_{e_1})$, and $\pi_{\key(e_1)} t_1 \in \pi_{\key(e_1)} Q(D)$.  In this way, $t_1$ is a witness of $t$ by definition. %Let $e_2$ be the parent node of $e_1$, and $t_2 = \pi_{e_2 \cap \y} q$. As $t_2$ is locally live before the insertion of $t$, all tuples in $\{t_3 \in R_{e_2}: \pi_{\key(e_1)} t_3 = \pi_{\key(e_1)} t_2 = \pi_{\key(e_1)} t_1\}$ do not change their status after the insertion of $t$. Moreover, $t_1$ becomes globally live since it participates in $q$. In this way, $t_1$ is an anchor tuple of $t$. 
	
	\paragraph{Critical Property: $\Delta Q(D,t_1) \cap \Delta Q(D,t_2) = \emptyset$ holds for any pair of witness tuples $t_1, t_2$}
 It remains to show that there is no duplicate results in $\bigcup_{t': \textrm{a witness of $t$}} \Delta Q(D \ltimes t')$. By contradiction, assume that there exists a query result $q$ with at least two witness tuples. Wlog, let $t_1, t_2$ be two distinct witness tuples in $q$, where $t_1 \in R_{e_1}$ for some $e_1 \subseteq \y$ or $e_1 \cap \y -p_{e_1}\neq \emptyset$, and some $e_2 \subseteq \y$ or $e_2 \cap \y -p_{e_2}\neq \emptyset$.
 First, $e_1 \neq e_2$, as $q$ contains at most one tuple in each relation.  %otherwise, $\pi_{e_1 \cap \y} t_1 = \pi_{e_2 \cap \y} t_2$, coming to a contradiction that  If $t_1$ and $R''$ represent the same relation, then there cannot exists a output tuple $t_o$, such that $t_o \in \Delta Q(D, t')$ and $t_o \in \Delta Q(D, t'')$, because $t_o(var(R'))$ cannot equal to both $t'$ and $t''$.
	Note that the insertion of $t \in R_e$ can only change the status of tuples in the ancestors of $e$. Without loss of generality, let $e_1$ be the ancestor of $e_2$. Let $e_3$ be parent node of $e_2$ (it could be the case that $e_1 = e_3$). Let $t_3 = \pi_{e_3 \cap \y} q$. Implied by the definition of witness tuples, $t_3 \in \pi_\y V_s(R_{e_3})$ before the insertion of $t$. Implied by $t_3 \notin \Delta \pi_\y V_s(R_{e_3})$, $t_1 \in \pi_\y V_s(R_{e_1})$ before the insertion, contradicting the fact that $t_1$ is a witness tuple. This way, each result in $\Delta Q(D,t)$ corresponds to one witness tuple, thus there is no duplicates across the extended query results over different witness tuples. 
	%If $R'$ and $R''$ represent different relation, as the bottom-up update will only visit the relations in the path between $R$ and root, then one node will be the ancestor of the other.  Assume  $R' \in D(R'')$ and there exists an output tuple $t_o$ such that $t_o \in \Delta Q(D, t')$ and $t_o \in \Delta Q(D, t'')$, which implies that $t'$ still can perform the bottom-up update, because there exists some tuples (for example, $t''$) above $t'$ that is not up-to-date, which contradict to the fact that $t'$ is the $\anchor$ tuple.
	%
	%Thus, for each output tuple $t_o = t_0 \Join \cdots \Join t_m$, there can be at most one $\anchor$ tuple in $\{t_i|0 \le t \le m\}$, which indicates that there is no duplicate output tuples in $\bigcup_{t' \in \anchor(t)} \Delta Q(D, t')$.
\end{proof}

\begin{proof}[Proof of Lemma~\ref{lem:delta_runningtime}]
We first show the correctness of Algorithm~\ref{alg:delta-enumerate}. 
Consider an arbitrary witness tuple $t' \in R_{e_1}$. Denote the nodes lying on the path from $e_1$ to $r$ as $e_1, e_2, \cdots, e_k (r)$ sequentially. We can first expand $Q(D \ltimes t')$ as follows:
    \begin{equation}
        \label{eq:delta-enumerate}
         t' \Join \left(\Join_{i=1}^k V_r(e_i)\right)
         \Join Q_{\T_{e_1}} \Join \left( \Join_{i=2}^k Q_{\T_{e_i} - \T_{e_{i-1}}}\right)
    \end{equation}
    where $Q_\T$ represents the query defined over relations in $\T$. Implied by the join operator and the properties of free-connex join tree, we can further rewrite (\ref{eq:delta-enumerate}) =: 
    \begin{align*} 
    & \bigcup_{S \in t' \Join \left(\Join_{i=1}^k V_r(e_i) \right)} S \Join Q_{\T_{e_1}} \Join \left(\Join_{i=2}^k Q_{\T_{e_i} - \T_{e_{i-1}}}\right)\\
    = &  \bigcup_{S \in t' \Join \left(\Join_{i=1}^k V_r(e_i) \right)} S \times (Q_{\T_{e_1}} \ltimes \{S\}) \times \left(\Join_{i=2}^k (Q_{\T_{e_i} - \T_{e_{i-1}}} \ltimes \{S\})\right)
    \end{align*}
    which is exactly followed by Algorithm~\ref{alg:delta-enumerate}. Together with Lemma~\ref{lem:witness}, all results of $\Delta Q(D,t)$ are enumerated without duplication. 
    
    We next analyze the time complexity. As all witness tuples can be stored in a data structure (e.g., a linked list) supporting constant-delay enumeration, every $t'$ (line 1) can be retrieved in $O(1)$ delay.  It then suffices to show that $Q(D \ltimes t')$ can be enumerated with $O(1)$ delay for every $t'$. Note that subquery $t' \Join \left(\Join_{i=1}^k V_r(e_i)\right)$ (line 4) can be done in $O(1)$ delay with our hashing index.  
    For the remaining subquery $Q_{\T_{e_1}} \ltimes \{S\}$ or $Q_{\T_{e_i} - \T_{e_{i-1}}} \ltimes \{S\}$, we invoke the procedure {\sc FullEnum} (line 6-8) and all query results can be enumerated with $O(1)$ delay, proved by Lemma~\ref{lem:full_runningtime}. Combing those subqueries in a form of Cartesian product can yield query results with $O(1)$ delay, thus completing the whole proof.
	%For each for-loop, the algorithms need to find all the tuples $t_i \in M(R_i)$, such that $t_i$ can join with the current partial output $t$.  As the index support enumerating all tuples whose value on key is $v$ with constant delay, the algorithm only need to use $O(1)$ time to find such tuples, for each output tuple, the algorithm takes $O(m)$ time, where $m = |\yR|$, which is a constant.  Thus, the algorithm can enumerate the $\Delta Q(D, t)$ in constant delay, as it can find the first output tuple in constant time, and constant time to find the next output tuple. 
	%
	%In addition, when the algorithm visit $t_i$, it will also check whether the state of $t_i$ need to be updated, which only takes $O(1)$ time per tuple.  As each output tuple associates with $m$ such tuple, so the total update time is $m \times O(1)\times |\Delta Q(D, t)| = O(|\Delta Q(D, t)|)$ time. 
\end{proof}

\section{Missing Materials in section~\ref{sec:Analysis}}
\label{appendix:analysis}

\begin{proof}[Proof of Theorem~\ref{the:lb-non-weak-hierarchical-full}]
    Given an instance of OuMv, we encode the matrix by $R_3$ and vectors $(v_i, u_i)$ by $R_2$ and $R_4$ separately.  We construct an update sequence $S$ for $Q$ as follows: 
    %\iffalse
    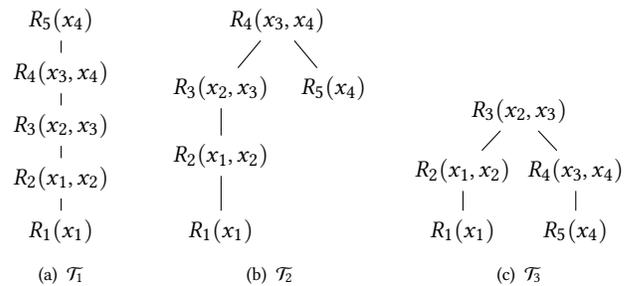
\begin{figure}[h]
    \subfigure[$\T_1$]{
    \centering
    \begin{tikzpicture}      
        \node (p1) at (-2, -2.2) {$R_5(x_4)$ }; 
        \node (p2) at (-2, -2.9) {$R_4(x_3,x_4)$};
        \node (p3) at (-2, -3.6) {$R_3(x_2, x_3)$};
        \node (p4) at (-2, -4.3) {$R_2(x_1,x_2)$};
        \node (p5) at (-2, -5) {$R_1(x_1)$};
  
        \begin{scope}[every path/.style={-}]
            \draw (p1) -- (p2);
            \draw (p2) -- (p3);
            \draw (p3) -- (p4);
            \draw (p4) -- (p5);
            \end{scope} 
        \end{tikzpicture}
        \label{fig:qc_T1}
    }
    \hfill
    %    \subfigure[$\T_2$]{
    %\centering
    %\begin{tikzpicture}      
    %    \node (p1) at ( -1, -2.2) {$R_2(x_1, x_2)$ }; 
    %    \node (p2) at ( -0.25, -4) {$R_4(x_3,x_4)$};
    %    \node (p3) at ( -1.75, -3.1) {$R_1(x_1)$};
    %    \node (p4) at ( -0.25, -3.1) {$R_3(x_2,x_3)$};
    %    \node (p5) at ( -0.25, -5) {$R_5(x_4)$};
    %
    %    \begin{scope}[every path/.style={-}]
    %        \draw (p1) -- (p3);
    %        \draw (p2) -- (p4);
    %        \draw (p5) -- (p2);
    %        \draw (p4) -- (p1);
    %    \end{scope} 
    %\end{tikzpicture}\label{fig:qc_T2}
    %}
    \hfill
    \subfigure[$\T_2$]{
    \centering
    \begin{tikzpicture}      
        \node (p1) at ( -1, -2.2) {$R_4(x_3, x_4)$ }; 
        \node (p2) at ( -1.75, -4) {$R_2(x_1,x_2)$};
        \node (p3) at ( -1.75, -3.1) {$R_3(x_2, x_3)$};
        \node (p4) at ( -0.25, -3.1) {$R_5(x_4)$};
        \node (p5) at ( -1.75, -5) {$R_1(x_1)$};
  
        \begin{scope}[every path/.style={-}]
            \draw (p1) -- (p3);
            \draw (p2) -- (p3);
            \draw (p5) -- (p2);
            \draw (p4) -- (p1);
        \end{scope} 
    \end{tikzpicture}\label{fig:qc_T3}
    }
    \hfill
    \subfigure[$\T_3$]{
    \centering
    \begin{tikzpicture}      
        \node (p1) at ( -0.25, -3) {$R_4(x_3, x_4)$ }; 
        \node (p2) at ( -1.75, -3) {$R_2(x_1,x_2)$};
        \node (p3) at ( -1.0, -2.2) {$R_3(x_2, x_3)$};
        \node (p4) at ( -0.25, -3.8) {$R_5(x_4)$};
        \node (p5) at ( -1.75, -3.8) {$R_1(x_1)$};
  
        \begin{scope}[every path/.style={-}]
            \draw (p1) -- (p3);
            \draw (p2) -- (p3);
            \draw (p5) -- (p2);
            \draw (p4) -- (p1);
        \end{scope} 
    \end{tikzpicture}
    \label{fig:qc_T4}
    }
    \caption{Join trees for $Q = R_1(x_1) \Join R_2(x_1, x_2) \Join R_3(x_2, x_3) \Join R_4(x_3, x_3) \Join R_5(x_4)$.}
    \label{fig:join_tree_core}
\end{figure}
%\fi
(1) we add a tuple $t = (i,j)$ with lifespan $I(t) =[-k, k]$, for each pair $(i,j) \in [n] \times [n]$ if $M_{ij} \neq 0$; (2) we add a tuple $t = (i)$ with lifespan $I(t) = [i-2k,i]$ into $R_1$ and $R_5$; (3) for each pair of vectors $(v_i, u_i)$, we add a tuple $t = (i, v_{ij})$ with lifespan $I(t) = [i,i+2k]$ to $R_2$ if $v_{ij}\neq 0$, and add a tuple $t = (u_{ij}, i)$ with lifespan $I(t) = [i,i+2k]$ to $R_4$ if $u_{ij} \neq 0$;  
(4) if  a query result is enumerated, we output true for $v_i^{T} M u_i$, and false otherwise;
(5) we repeat (3)-(4) for the next pair $(v_{i+1}, u_{i+1})$, until $n$ pairs of vectors are all processed. Each tuple in $S$ has the same lifespan as $2k$, thus it is a FIFO sequence. 

We note that in any free-connex join tree $\T$ of $Q$, there always exists a subtree in which either $R_1 -R_2-R_3$ or $R_5-R_4-R_3$ is a leaf-to-root path. 
Wlog, assume $R_1-R_2-R_3$ is a leaf-to-root path. First, for each tuple $t \in R_1$, $\lambda(t) =1$ as $R_1$ is a leaf node. For $t= (i,v_{ij}) \in R_2$, we observe that $\tilde{I}(t) = [i,i]$ as $I(t) = [i,i+2k]$ and $I(t') = [i-2k,i]$ for some tuple $t' \in R_1$. But in this case, $\lambda(t) = 1$ still holds, as there exists no tuple $t' \in R_1$ with $\tilde{I}(t') \subseteq [i,i]$. However, for each tuple $t \in R_3$, $\lambda(t) = n$ as 
there exists a tuple $t' \in R_2$ such that $\tilde{I}(t') = [i,i]$ for every $i \in [n]$. Hence, the enclosureness of $S$ on every free-connex join tree is $\lambda = \frac{n^2\cdot n+n^2 \cdot 1}{n^2} = n$.

The correctness of this simulation is obvious. 
This way, if there is a data structure that can be updated in $O(\lambda^{1-\epsilon})$ time while supporting $O(\lambda^{2-\epsilon})$-delay enumeration for $Q$ over any FIFO sequence, then the OuMv problem can be solved in $O(n^2\cdot\lambda^{1-\epsilon} + n\cdot\lambda^{2-\epsilon}) = O(n^{3-\epsilon})$ time.  Note that the construction above requires a database of size at least $n^2 = \lambda^2$, thus $\lambda \le \sqrt{|D|}$.
\end{proof}

\begin{proof}[Proof of Theorem~\ref{thm:main}]
We next turn to the update cost of our indexes. As mentioned at the beginning of Section~\ref{sec:Analysis}, the total update cost of the entire sequence is asymptotically dominated by that of P-Update, which is further bounded by the number of times all the counters $\textsf{count}[t]$ can change.  The following lemma connects this quantity with the enclosureness of the update sequence.
\begin{lemma}
  \label{lem:UpdateCountDAG}
  For any tuple $t$, $\textsf{count}[t]$ changes $O(\lambda(t))$ times.
\end{lemma}

\begin{proof}
	The status change of tuple $t \in R_e$ falls into one of the following two cases: (1) tuple $t$ is being inserted or deleted; (2) some tuple $t'\in R_{e'}$ for $e' \in \T_e$ is inserted or deleted, and this update propagates to $t$. Note that tuple $t$ can be inserted and deleted once in its lifespan, thus bounded by $O(1)$ and the cost is reflected in R-Update. Then, we will focus on the second case.% when a tuple $t' \in R_{e'}$ for $e' \in \T_e$ is inserted or deleted, which triggers P-Update and affects $t$. 
	%
	%More specifically, $t$ changes from locally non-alive to locally alive only when a tuple $t' \in R_{e'}$ for some $e' \in \T_e$ is inserted, and from locally alive to locally non-alive only when a tuple $t' \in R_{e'}$ for some $e' \in \T_e$ is deleted. 
	 
	 We start with the case that $e$ has one child node in $\T$. In this case, $\textsf{count}[t]$ has its value changed between $0$ and $1$.
	 Note that if an insertion changes $t$ from $R_e/V_s(R_e)$ to $V_s(R_e)$, subsequent insertions won't change the status of $t$ unless a deletion occurs. Consider a set of $k$ disjoint intervals $\tilde{I}_1, \tilde{I}_2, \cdots, \tilde{I}_k$ in ordering, such that $\tilde{I}_j \in \widetilde{\mathcal{I}}_e$, $\tilde{I}_j \subseteq I(t)$ for each $j \in [k]$, and there exists no additional interval $\tilde{I}$ such that $\tilde{I} \subsetneq \tilde{I}_j$ or $\tilde{I}' \subseteq [\tilde{I}^+_j, \tilde{I}_{j+1}^-]$ for any $j \in \{1,2,\cdots,k\}$. %\qichen{Note that there could be multiple such sets of intervals. We denote such a set as {\em minimal}, if there does not exist any $\tilde{I}(t')$ such that $\tilde{I}(t') \subsetneq \tilde{I}(t_i)$ for any ${t_i}$.} 
	 Each of the $k$ intervals can change the status of $t$ at most twice, so they together can change the status of $t$ at most $O(k)$ times. The effective lifespan of $t$ exactly captures such a quantity.  %i.e., the status of tuple $t$ can change at most %the maximum times minimal time gap that the tuple will keep its state.  If a tuple $t$ is inserted and it should be locally alive, then the live state will be kept at least during the whole {\em effective lifespan}, until the first deletion happens in $D$ or the tuple is deleted.  
	
	We next consider a case when $e$ has two child nodes $e_1, e_2 \in \T$. Similarly, consider a set of $k$ disjoint intervals $\tilde{I}_1, \tilde{I}_2, \cdots, \tilde{I}_k$ in ordering, such that $\tilde{I}_j \in \widetilde{\mathcal{I}}_e$, $\tilde{I}_j \subseteq I(t)$ for each $j \in [k]$, and there exists no additional interval $\tilde{I}$ such that $\tilde{I} \subsetneq \tilde{I}_j$ or $\tilde{I} \subseteq [\tilde{I}^+_j, \tilde{I}_{j+1}^-]$ for any $j \in \{1,2,\cdots,k\}$. We can make the following two observations:
	\begin{enumerate}
	    \item[(1)] For any $\tilde{I}_j$, $\textsf{count}[t]$ can change at most $2$ times within $\tilde{I}$.
	    \item[(2)] For any two adjacent intervals $\tilde{I}_j$ and 
	    $\tilde{I}_{j+1}$, $\textsf{count}[t]$ can change at most $4$ times in their gap.
	\end{enumerate}
    Together, we can conclude that $\textsf{count}[t]$ can change at most $6 \cdot k = O(k)$ times when there are two child nodes. We next go into details of (1) and (2) separately. 
    
	For (1), we assume $\tilde{I}_j \in \widetilde{\mathcal{I}}_{e_1}$ without loss of generality. By the definition of effective lifespan, there cannot be any insertion or deletion in any node of $\T_{e_1}$ within $\tilde{I}_j$. %, otherwise, the new update can form a tighten effective lifespan and violates the requirements. 
	Nevertheless, updates may still exist within $\tilde{I}_j$ on some node of $\T_{e_2}$, which might further change $\textsf{count}[t]$. We distinguish two more cases.
	%
	%\begin{center}
	%    \includegraphics[width=0.8\linewidth]{twocase.pdf}
	%\end{center}
	%
	If $\textsf{count}[t]$ changes from $1$ to $2$, due to an insertion from $\T_{e_2}$, a deletion must not exist within $\tilde{I}_j$ on any node of $\T_{e_2}$, implied by the fact that there exists no $\tilde{I}$ such that $\tilde{I} \subsetneq \tilde{I}_j$. Hence, $\textsf{count}[t]$ can change at most once in $\tilde{I}_j$ for this case.
	Otherwise, $\textsf{count}[t]$ changes from $2$ to $1$, after a deletion from $\T_{e_2}$. We then go into the first case and $\textsf{count}[t]$ can change at most one more time. In total, $\textsf{count}[t]$ can change at most twice.
	
    For (2), it is clear that at the right endpoint of $\tilde{I}_j$ and the left endpoint of $\tilde{I}_{j+1}$, $\textsf{count}[t]$ can change once as the deletion and insertion of an effective lifespan.  In the meantime, there does not exist another effective lifespan within their gap, %interval $[\tilde{t}_j^-, \tilde{t}_{j+1}^+]$, 
    %as the definition of enclosureness, 
    so for any $e_i \in \{e_1, e_2\}$, there exists no deletion on $\T_{e_i}$ in the gap following an insertion in $\T_{e_i}$. %, which causes $\textsf{count}[t]$ to be monotonically decreasing until the first insertion occurs, and monotonically increasing until $\tilde{t}_{j+1}^+$, hence, 
    This way, $\textsf{count}[t]$ can change at most four times (i.e. $2 \to 1 \to 0 \to 1 \to 2$) within their gap.

	At last, we consider the general case when $e$ has multiple child nodes in $\T$. In this case, $\textsf{count}[t]$ has its value changed among $0, 1, \cdots, j$, where $j$ is the number of child nodes of $e$.  By extending the previous two observations, we conclude that $\textsf{count}[t]$ can change at most $3j \cdot k$ times, where $j$ can be considered as a constant. With respect to all possible choices of $k$, we observe that 
	\[
    	k \le \max_{\substack{\mathcal{J} \subseteq \widetilde{{\mathcal{I}}}_e \\ \forall I(t_1) \in\mathcal{J}, I(t_1) \subseteq I(t) \\ \forall I(t_2), I(t_3) \in \mathcal{J}, I(t_2) \cap I(t_3) = \emptyset}} 1+|\mathcal{J}| = \lambda(t), 
	\]
	thus $\textsf{count}[t]$ can change at most $O(\lambda(t))$
	times. 
\end{proof}

The time cost of Algorithm~\ref{alg:P-update} is determined by the number of iterations of for-loop (line 2 or 7).  One can easily observe that  $\textsf{count}[t]$ will be changed for some tuple $t$ once in each iteration, therefore the running time can be bounded by the number of changes to $\textsf{count}[t]$ over all tuples $t$. Now consider an update sequence $S$ with enclosureness $\lambda$. Implied by Lemma~\ref{lem:UpdateCountDAG} the total update cost is $O\left(\sum_{t \in \I} \lambda(t)\right)$, which is $O\left(\frac{\sum_{t\in \I} \lambda(t)}{|\I|}\right) = O(\lambda)$ amortized. 

%\update{On the other hand, if $e$ is a generalized relation, the time cost of Algorithm~\ref{alg:P-update} will be constant, as $e = \key(e_i)$ for any child node $e_i$ of $e$, hence the for-loop in line 2 or 7 contains only one possible tuple, therefore the number of changes to $\textsf{count}[t]$ for any conceptual tuples $\hat{t} \in e$ is bounded by 
%\[
%    \sum_{\hat{t} \in e} \textsf{count}[\hat{t}] \le \sum_{e' \in \C_e} \sum_{t' \in e'} \textsf{count}[t'],
%\]
%meaning the update cost of a generalized relation can be bounded by the update cost of all relations.  Since we consider the data complexity and the number of generalized relations as a constant, we can ignore the cost of maintaining any generalized relation since it only adds a constant factor to the final complexity.
%}

Putting everything together, we have completed the proof for Theorem~\ref{thm:main}.
\end{proof}

\begin{proof}[Proof of Lemma~\ref{lem:q-hierarchical}]
    In~\cite{idris17:_dynam}, it has been proved that a CQ is q-hierarchical if and only if there is a ``simple'' generalized join tree, such that all original relations are leaf nodes of $\T$, and every internal node $e'$, which corresponds to a generalized relation, must have $e' \subseteq e$ for every its child $e$. We note that such a simple generalized join tree is essentially a height-1 free-connex join tree as defined in Section~\ref{sec:preliminaries}.
\end{proof}

\begin{proof}[Proof of Lemma~\ref{lem:weak-hierarchical}]
    Given a height-2 free-connex join tree $\T$ and consider an arbitrary tuple $t \in R_e$. If $e$ is a leaf node, $\tilde{I}(t) = [t^+, t^-]$ and $\lambda_\T(t) = 1$. If $e$ is an internal node, $\tilde{I}(t) \subseteq [t^+, t^-]$. But here, as the join tree is a height-2 free-connex join tree, %\update{from lemma~\ref{lem:height-2-generalized} we can learn that} 
    every $e$'s child node must be a leaf node, hence every tuple $t_1 \in R_{e'}$ for $e' \in \T_e$ has $\tilde{I}(t_1) = [t^+_1, t^-_1]$, and there exists no tuple $t_2$ such that $t^+ < t^+_2$ and $t^-_2 < t^-$. As each tuple $t$ has $\lambda_\T(t)=1$, by definition, $\lambda_\T(S) = 1$.
\end{proof}

\begin{proof}[Proof of Lemma~\ref{lem:insertion-only}]
    As there is no deletion for every tuple $t$, $\tilde{I}(t) = [t^+, +\infty]$.  Hence, for every $t$, by definition, $\lambda_\T (t) = 1$. 
\end{proof}

\section{SQL Queries}
\label{appendix:queries}
\paragraph{3-Hop Full Join Query}
\begin{lstlisting}[ language=SQL,
	deletekeywords={IDENTITY},
	deletekeywords={[2]INT},
	morekeywords={clustered},
	mathescape=true,
	xleftmargin=-1pt,
	framexleftmargin=-1pt,
	frame=tb,
	framerule=0pt ]
SELECT G1.src as A, G2.src as B, G3.src as C, G3.dst as D
FROM G G1, G G2, G G3
WHERE G1.dst = G2.src AND G2.dst = G3.src
AND FILTER OVER (G3.dst)
\end{lstlisting}

\paragraph{4-Hop Full Join Query}
\begin{lstlisting}[ language=SQL,
	deletekeywords={IDENTITY},
	deletekeywords={[2]INT},
	morekeywords={clustered},
	mathescape=true,
	xleftmargin=-1pt,
	framexleftmargin=-1pt,
	frame=tb,
	framerule=0pt ]
SELECT G1.src as A, G2.src as B, G3.src as C, G3.dst as D, G4.dst as E
FROM G G1, G G2, G G3, G G4
WHERE G1.dst = G2.src AND G2.dst = G3.src AND G3.dst = G4.src AND FILTER OVER (G4.dst)
\end{lstlisting}

\paragraph{3-Hop Join-Project Query}
\begin{lstlisting}[ language=SQL,
	deletekeywords={IDENTITY},
	deletekeywords={[2]INT},
	morekeywords={clustered},
	mathescape=true,
	xleftmargin=-1pt,
	framexleftmargin=-1pt,
	frame=tb,
	framerule=0pt ]
SELECT G2.src as B, G3.src as C
FROM G G1, G G2, G G3
WHERE G1.dst = G2.src AND G2.dst = G3.src
\end{lstlisting}

\paragraph{4-Hop Join-Project Query}
\begin{lstlisting}[ language=SQL,
	deletekeywords={IDENTITY},
	deletekeywords={[2]INT},
	morekeywords={clustered},
	mathescape=true,
	xleftmargin=-1pt,
	framexleftmargin=-1pt,
	frame=tb,
	framerule=0pt ]
SELECT G2.src as B, G3.src as C, G3.dst as D
FROM G G1, G G2, G G3, G G4
WHERE G1.dst = G2.src AND G2.dst = G3.src AND G3.dst = G4.src
AND FILTER OVER (G4.dst)
\end{lstlisting}

\paragraph{2-Comb Query}
\begin{lstlisting}[ language=SQL,
	deletekeywords={IDENTITY},
	deletekeywords={[2]INT},
	morekeywords={clustered},
	mathescape=true,
	xleftmargin=-1pt,
	framexleftmargin=-1pt,
	frame=tb,
	framerule=0pt ]
SELECT G1.src as A, G2.src as B, G3.src as C, G3.dst as D
FROM G G1, G G2, G G3, V1, V2
WHERE G1.dst = G2.src AND G2.dst = G3.src AND V1.v = G1.src and V2.v = G3.dst
\end{lstlisting}

\paragraph{Star Query}
\begin{lstlisting}[ language=SQL,
	deletekeywords={IDENTITY},
	deletekeywords={[2]INT},
	morekeywords={clustered},
	mathescape=true,
	xleftmargin=-1pt,
	framexleftmargin=-1pt,
	frame=tb,
	framerule=0pt ]
SELECT G1.src, COUNT(G1.dst, G2.dst, G3.dst, G4.dst)
FROM G G1, G G2, G G3, G G4
WHERE G1.src = G2.src AND G1.src = G3.src AND G1.src = G4.src
GROUP BY G1.src;
\end{lstlisting}

\paragraph{Dumbbell Full Join Query}
\begin{lstlisting}[ language=SQL,
	deletekeywords={IDENTITY},
	deletekeywords={[2]INT},
	morekeywords={clustered},
	mathescape=true,
	xleftmargin=-1pt,
	framexleftmargin=-1pt,
	frame=tb,
	framerule=0pt ]
SELECT *
FROM G G1, G G2, G G3, G G4, G G5, G G6, G G6
WHERE G1.dst = G2.src
    AND G2.dst = G3.src
    AND G3.dst = G1.src
    AND G5.dst = G6.src
    AND G6.dst = G7.src
    AND G7.dst = G5.src
    AND G4.src = G3.dst
    AND G4.dst = G5.src
\end{lstlisting}

\paragraph{Dumbbell Join-Project Query}
\begin{lstlisting}[ language=SQL,
	deletekeywords={IDENTITY},
	deletekeywords={[2]INT},
	morekeywords={clustered},
	mathescape=true,
	xleftmargin=-1pt,
	framexleftmargin=-1pt,
	frame=tb,
	framerule=0pt ]
SELECT G4.src, G4.dst
FROM G G1, G G2, G G3, G G4, G G5, G G6, G G6
WHERE G1.dst = G2.src
    AND G2.dst = G3.src
    AND G3.dst = G1.src
    AND G5.dst = G6.src
    AND G6.dst = G7.src
    AND G7.dst = G5.src
    AND G4.src = G3.dst
    AND G4.dst = G5.src
\end{lstlisting}

\paragraph{SNB Query 1}
\begin{lstlisting}[ language=SQL,
	deletekeywords={IDENTITY},
	deletekeywords={[2]INT},
	morekeywords={clustered},
	mathescape=true,
	xleftmargin=-1pt,
	framexleftmargin=-1pt,
	frame=tb,
	framerule=0pt ]
SELECT p_personid, p_firstname, p_lastname, m_messageid, k_person1id
FROM person, message, knows
WHERE p_personid = m_creatorid 
  AND k_person2id = p_personid;
\end{lstlisting}

\paragraph{SNB Query 2}
\begin{lstlisting}[ language=SQL,
	deletekeywords={IDENTITY},
	deletekeywords={[2]INT},
	morekeywords={clustered},
	mathescape=true,
	xleftmargin=-1pt,
	framexleftmargin=-1pt,
	frame=tb,
	framerule=0pt ]
SELECT k1.k_person1id, k1.k_person2id, k2.k_person2id, t_tagid, m_messageid
FROM tag, message, message_tag, knows1 k1, knows2 k2
WHERE m_messageid = mt_messageid 
  AND mt_tagid = t_tagid 
  AND k1.k_person2id = k2.k_person1id 
  AND m_creatorid = k2.k_person2id 
  AND m_c_replyof is NULL 
  AND FILTER OVER (k1.k_person1id)
\end{lstlisting}

\paragraph{SNB Query 3}
\begin{lstlisting}[ language=SQL,
	deletekeywords={IDENTITY},
	deletekeywords={[2]INT},
	morekeywords={clustered},
	mathescape=true,
	xleftmargin=-1pt,
	framexleftmargin=-1pt,
	frame=tb,
	framerule=0pt ]
SELECT k1.k_person1id, k1.k_person2id, k2.k_person2id, t_tagid, m_messageid
FROM tag, message, message_tag, knows1 k1, knows2 k2
WHERE m_messageid = mt_messageid 
  AND mt_tagid = t_tagid 
  AND k1.k_person2id = k2.k_person1id
  AND k2.k_person2id <> k1.k_person1id
  AND m_creatorid = k2.k_person2id 
  AND m_c_replyof is NULL 
  AND FILTER OVER (k1.k_person1id)
\end{lstlisting}

\paragraph{SNB Query 4}
\begin{lstlisting}[ language=SQL,
	deletekeywords={IDENTITY},
	deletekeywords={[2]INT},
	morekeywords={clustered},
	mathescape=true,
	xleftmargin=-1pt,
	framexleftmargin=-1pt,
	frame=tb,
	framerule=0pt ]
SELECT t_name, t_tagid, count(distinct m_messageid)
FROM tag, message, message_tag, knows
WHERE m_messageid = mt_messageid 
  AND mt_tagid = t_tagid 
  AND m_creatorid = k_person2id 
  AND m_c_replyof is NULL 
  AND FILTER OVER (k_person1id)
GROUP BY t_name, t_tagid
\end{lstlisting}

\end{document}